\documentclass{jfm}

\usepackage{graphicx}
\usepackage{natbib}
\usepackage{subfig}
\usepackage[amssymb]{SIunits}
\usepackage{amsmath}
\usepackage{color}
\usepackage{rotating}

\ifCUPmtlplainloaded \else
  \checkfont{eurm10}
  \iffontfound
    \IfFileExists{upmath.sty}
      {\typeout{^^JFound AMS Euler Roman fonts on the system,
                   using the 'upmath' package.^^J}%
       \usepackage{upmath}}
      {\typeout{^^JFound AMS Euler Roman fonts on the system, but you
                   dont seem to have the}%
       \typeout{'upmath' package installed. JFM.cls can take advantage
                 of these fonts,^^Jif you use 'upmath' package.^^J}%
      }
  \else
  \fi
\fi

\ifCUPmtlplainloaded \else
  \checkfont{msam10}
  \iffontfound
    \IfFileExists{amssymb.sty}
      {\typeout{^^JFound AMS Symbol fonts on the system, using the
                'amssymb' package.^^J}%
       \usepackage{amssymb}%
       \let\le=\leqslant  
       \let\ge=\geqslant  
      }{}
  \fi
\fi

\ifCUPmtlplainloaded \else
  \IfFileExists{amsbsy.sty}
    {\typeout{^^JFound the 'amsbsy' package on the system, using it.^^J}%
     \usepackage{amsbsy}}
    {\providecommand\boldsymbol[1]{\mbox{\boldmath $##1$}}}
\fi

\newcommand{\nom}{Nu_\omega}

\newcommand{\usro}{Ro^{-1}}
\newcommand{\bu}{\boldsymbol{u}}

\newcommand{\mean}[1]{\left \langle #1 \right \rangle}

\newsavebox{\astrutbox}
\sbox{\astrutbox}{\rule[-5pt]{0pt}{20pt}}

\newcommand{\tttc}{$\text{T}^3\text{C}$}

\title[Optimal Taylor-Couette flow: Radius ratio dependence]{Optimal Taylor-Couette flow: \\ radius ratio dependence}

\author[R. Ostilla M\'onico and others]%
{R\ls O\ls D\ls O\ls L\ls F\ls O\ns O\ls S\ls T\ls I\ls L\ls L\ls A \ns M \ls \'O \ls N\ls I\ls C\ls O$^1$,\break%
S\ls A\ls N\ls D\ls E\ls R\ns G.\ns H\ls U\ls I\ls S\ls M\ls A\ls N$^1$,\break%
T\ls I\ls M\ns J.\ns G.\ns J\ls A\ls N\ls N\ls I\ls N\ls K$^1$,\ns
D\ls E\ls N\ls N\ls I\ls S\ns P.\ns M.\ns V\ls A\ls N\ns G\ls I\ls L\ls S$^{1}$,\break%
R\ls O\ls B\ls E\ls R\ls T\ls O\ns V\ls E\ls R\ls Z\ls I\ls C\ls C\ls O$^{2,1}$,\break%
S\ls I\ls E\ls G\ls F\ls R\ls I\ls E\ls D\ns G\ls R\ls O\ls S\ls S\ls M\ls A\ls N\ls N$^{3}$,\ns%
C\ls H\ls A\ls O\ns S\ls U\ls N$^1$,\break%
\and D\ls E\ls T\ls L\ls E\ls F\ns L\ls O\ls H\ls S\ls E$^1$}

\affiliation{$^1$Physics of Fluids, Mesa+ Institute, University of Twente, P.O. Box 217, 7500 AE Enschede, The Netherlands\\[\affilskip]
$^2$Dipartimento di Ingegneria Meccanica, University of Rome ``Tor Vergata'', Via del Politecnico 1, Roma 00133, Italy\\[\affilskip]
$^3$Department of Physics, University of Marburg, Renthof 6, D-35032 Marburg, Germany}

\date{\today}

\begin{document}
 
\maketitle

\begin{abstract}
Taylor--Couette flow with independently rotating inner (i) \&
outer (o) cylinders
is explored numerically and experimentally to determine
the effects of the radius ratio $\eta$ on the system response. Numerical simulations reach Reynolds numbers
of up to $Re_i=9.5\cdot10^3$ and $Re_o=5\cdot10^3$, corresponding to Taylor numbers of up to $Ta=10^8$
for four different radius ratios $\eta=r_i/r_o$ between $0.5$ and $0.909$. The experiments, performed in
the Twente Turbulent Taylor--Couette ($T^3C$) setup, reach Reynolds numbers of up
to $Re_i=2\cdot10^6$ and $Re_o=1.5\cdot10^6$, corresponding to $Ta=5\cdot10^{12}$ for $\eta=0.714-0.909$.
Effective scaling laws for the torque $J^\omega(Ta)$ are found, which for sufficiently large driving $Ta$
are independent of the radius ratio $\eta$. As previously reported for $\eta=0.714$, optimum transport at a non--zero 
Rossby number $Ro=r_i|\omega_i-\omega_o|/[2(r_o-r_i)\omega_o]$ is found in both experiments and numerics. $Ro_{opt}$
is found to depend on the radius ratio and the driving of the system. At a driving in the range 
between $Ta\sim3\cdot10^{8}$ and  $Ta\sim10^{10}$, $Ro_{opt}$ saturates to an asymptotic 
$\eta$-dependent value. Theoretical predictions
for the asymptotic value of $Ro_{opt}$ are compared to the experimental results, and found
to differ notably. Furthermore, the local angular velocity profiles from experiments
and numerics are compared, and a link between a flat bulk profile and optimum transport for all
radius ratios is reported.

\end{abstract}


\section{Introduction}

\begin{figure}
 \begin{center}
  \includegraphics[width=0.4\textwidth]{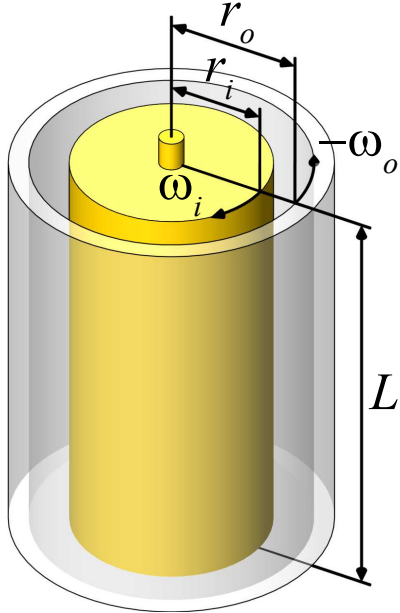}
  \caption{Schematic of the Taylor-Couette system. The system consists of two coaxial cylinders, which have 
an inner cylinder radius of $r_i$ and an outer cylinder radius of $r_o$. Both cylinders are of length $L$. 
The inner cylinder rotates with an angular velocity $\omega_i$ and the outer cylinder rotates with 
an angular velocity of $\omega_o$.}
  \label{fig:RBTCnosplit}
 \end{center}
\end{figure}

Taylor-Couette (TC) flow consists of the flow between two coaxial cylinders which are independently rotating. 
A schematic drawing of the system can be seen in Fig. \ref{fig:RBTCnosplit}. The rotation difference between the 
cylinder shears the flow thus driving the system. This rotation difference has been traditionally expressed by two Reynolds 
numbers, the inner cylinder $Re_i=r_i\omega_id/\nu$, and the outer cylinder 
$Re_o=r_o\omega_od/\nu$ Reynolds numbers, where $r_i$ and $r_o$ are the radii of the inner and outer cylinder, respectively, $\omega_i$
 and $\omega_o$ the inner and outer cylinder angular velocity, $d=r_o-r_i$ the gap width, and $\nu$ the kinematic 
viscosity. The geometry of TC is characterized by two nondimensional parameters, namely
the radius ratio $\eta=r_i/r_o$ and the aspect ratio $\Gamma=L/d$. 

Instead of taking $Re_i$ and $Re_o$, the driving in TC can alternatively be characterized by the Taylor $Ta$ and the rotation rate,
also called the Rossby $Ro$ number. 
The Taylor number can be seen as the non-dimensional forcing (the differential rotation) of the system
defined as 
$Ta = \sigma (r_o - r_i)^2 (r_o + r_i)^2 (\omega_o - \omega_i)^2 / (4 \nu^2)$, or 
\begin{equation}
Ta = (r_a^6 d^2 / r_o^2 r_i^2 \nu^2) (\omega_o - \omega_i)^2. 
\end{equation}
Here $\sigma = r_a^4 / r_g^4$ with $r_a = (r_o + r_i)/2$ the arithmetic and $r_g = \sqrt{r_o r_i}$ the geometric mean radii. The 
Rossby number is defined as:

\begin{equation}
 Ro=\displaystyle\frac{|\omega_i-\omega_o|r_i}{2\omega_od},
 \label{eq:Rosdef}
\end{equation}

\noindent and can be seen as a measure of 
the rotation of the system as a whole. $Ro<0$ corresponds to counterrotating cylinders, and $Ro>0$ to corotating cylinders.

TC is among the most investigated systems in fluid mechanics, mainly owing to its simplicity as an experimental model for shear flows. 
TC is in addition a closed system, so global balances which relate the angular velocity transport 
to the energy dissipation can be obtained. Specifically, in Eckhardt, Grossmann \& Lohse (2007) \nocite{eck07} 
(from now on referred to as EGL 2007), an exact relationship 
between the global parameters and the volume averaged energy dissipation rate was derived. 
This relationship has an analogous form to the one which can be obtained for Rayleigh-B\'enard (RB) flow, i.e. a flow in which heat is transported from a hot bottom plate
to a cold top plate. 

TC and RB flow have been extensively used to explore new concepts in fluid mechanics. 
Instabilities \citep{swi81,pfi81,pfi88,cha81,dra81,bus67}, nonlinear dynamics and chaos \citep{lor63,ahl74,beh85,dom86,str94},
pattern formation \citep{and86, cro93,bod00}, and turbulence \citep{sig94,gro00,kad01,lat92,ahl09,loh10} have been 
studied in both TC and RB and both numerically and experimentally. The main reasons behind the popularity of these systems
are, in addition to the fact that they 
are closed systems, as mentioned previously, their simplicity due to the high amount of symmetries present. 
It is also worth noting that plane 
Couette flow is the limiting case of TC when the radius ratio $\eta=1$.

Experimental investigations of TC have a long history, 
dating back to the initial work in the end of the 1800s by \cite{cou890} in France, who concentrated on 
outer cylinder rotation and developed the viscometer and \cite{mal896} in the UK, who also rotated the inner
cylinder and found indications of turbulence.
Later work by \cite{wen33} and \cite{tay36},
greatly expanded on the system, the former measuring torques and velocities for several radius and rotation ratios in the turbulent
case, and the latter being the first to mathematically describe the cells which form
if the flow is linearly unstable. The subject can be traced back even further to Stokes, and even Newton. 
For a broader historical context, we refer the reader to \cite{don91b}. 

Experimental work continued during the years \citep{smi82,and86,ton90,lat92,lat92a,lew99,gil11a,gil11,pao11,hui12} at low and 
high $Ta$ numbers for different ratios of the rotation frequencies $a = - \omega_o / \omega_i$.
$a$ is positive for counter--rotation and negative for co--rotation. $-a\equiv\mu$, another measure used for the
ratio of rotation frequencies.
This work has been complemented by numerical simulations,
not only in the regime of pure inner cylinder rotation \citep{fas84,cou96,don07,don08,pir08}, but also 
for eigenvalue study \citep{geb93b}, and counter-rotation at fixed $a$ \citep{don08}. Recently \citep{bra12,ost12},
simulations have also explored the effect of the outer cylinder rotation on the system at large Reynolds numbers. 

The recent experiments \citep{gil11a,gil11,pao11,mer12} and simulations \citep{bra12,ost12} have 
shown that at fixed $Ta$ an optimal angular momentum transport is obtained at \emph{non-zero} $a_{opt}$, 
and that the location of this maximum $a_{opt}$ varies with $Ta$. 
However, both experiments and simulations have been restricted to two radius ratios, namely $\eta=0.5$ and
$\eta=0.714$. The same radius ratios were also used for studies carried out on scaling laws of the torque and the ``wind''
of turbulence at highly 
turbulent Taylor numbers \citep{lew99, pao11, gil11, hui12, mer12}. Up to now, it is not clear how the radius
ratio affects the scaling laws of the system response or the recently found phenomena of optimal transport as a function of $Ta$.

Two suggestions were made to account for the radius ratio dependence of optimal transport. Van Gils et. al (2011b) \nocite{gil11}
wondered whether the optimal transport in general lies in or at least close to the Voronoi boundary (meaning a line
of equal distance) of the Esser-Grossmann
stability lines \citep{ess96} in the $(Re_o,Re_i)$ phase space as it does for $\eta=0.714$. However, this bisector value does not
give the correct optimal transport for $\eta=0.5$ \citep{mer12,bra13}. Therefore \cite{bra12} proposed a dynamic extension
of the Esser-Grossmann instability theory. This model correctly gives the observed optimal transport
(within experimental error bars) between $\eta=0.5$ and $\eta=0.714$ for three experimental data
sets \citep{wen33,pao11,gil11} and one numerical data set \citep{bra13}, but it is not clear how it performs outside the
$\eta$-range $[0.5,0.714]$.

In this paper, we study the following questions: how does the radius ratio $\eta$  affect the 
flow? How are the scaling laws of the angular momentum transport affected? What is the role of the geometric
parameter called pseudo-Prandtl number $\sigma$ introduced in EGL2007? Can the effect of the radius 
ratio be interpreted as a kind of non-Oberbeck-Boussinesq effect, analogous to this effect in Rayleigh-B\'enard flow? 
Finally, are the predictions and insights of \cite{gil11}, \cite{ost12} and \cite{bra13} on the 
optimal transport also valid for other values of $\eta$?

In order to answer these questions, both direct numerical simulations (DNS) and experiments have been undertaken. Numerical simulations, 
with periodic axial boundary conditions, have been performed using the finite--difference code previously used in \cite{ost12}. 
In these simulations, three more values of $\eta$ have been investigated: one in which the gap 
is larger ($\eta=0.5$), and two in which the gap is smaller ($\eta=0.833$ and $0.909$). With the previous 
simulations from \cite{ost12} at $\eta=0.714$, a total of four radius ratios has been analyzed. 

In both experiments and numerics, only one aspect ratio $\Gamma$ has been studied for
every radius ratio. Since the work of \cite{ben78} it is known that 
multiple flow states with a different amount of vortex pairs can coexist in TC for the same non-dimensional
flow parameters. However, with increased driving,
the bifurcations become less important and many branches do not survive. Indeed, \cite{lew99} found that for pure inner cylinder rotation only one 
branch with 8 vortices (for $\Gamma=11.4$ and $\eta=0.714$) remains when $Re_i$ is increased above $2\cdot10^4$. 
As the Reynolds numbers reached in the experiments greatly exceed this value we do not expect to see the effect of 
multiple states in the current experimental results.

For the numerical simulations, axially periodic boundary conditions have been taken. \cite{bra12} already studied the effect of 
the axial periodicity length on the system, and found that for a fixed vortical wavelength, the number of vortices does not affect the overall 
flow behaviour. It was also found that, in analogy to experiments, the effect of vortical wavelength, and hence of multiple states,
becomes smaller with increased driving.

\begin{figure}
 \begin{center}
  \subfloat{\label{fig:PhaseSpace05}\includegraphics[width=0.49\textwidth,trim = 0mm 0mm 9mm 0mm, clip]{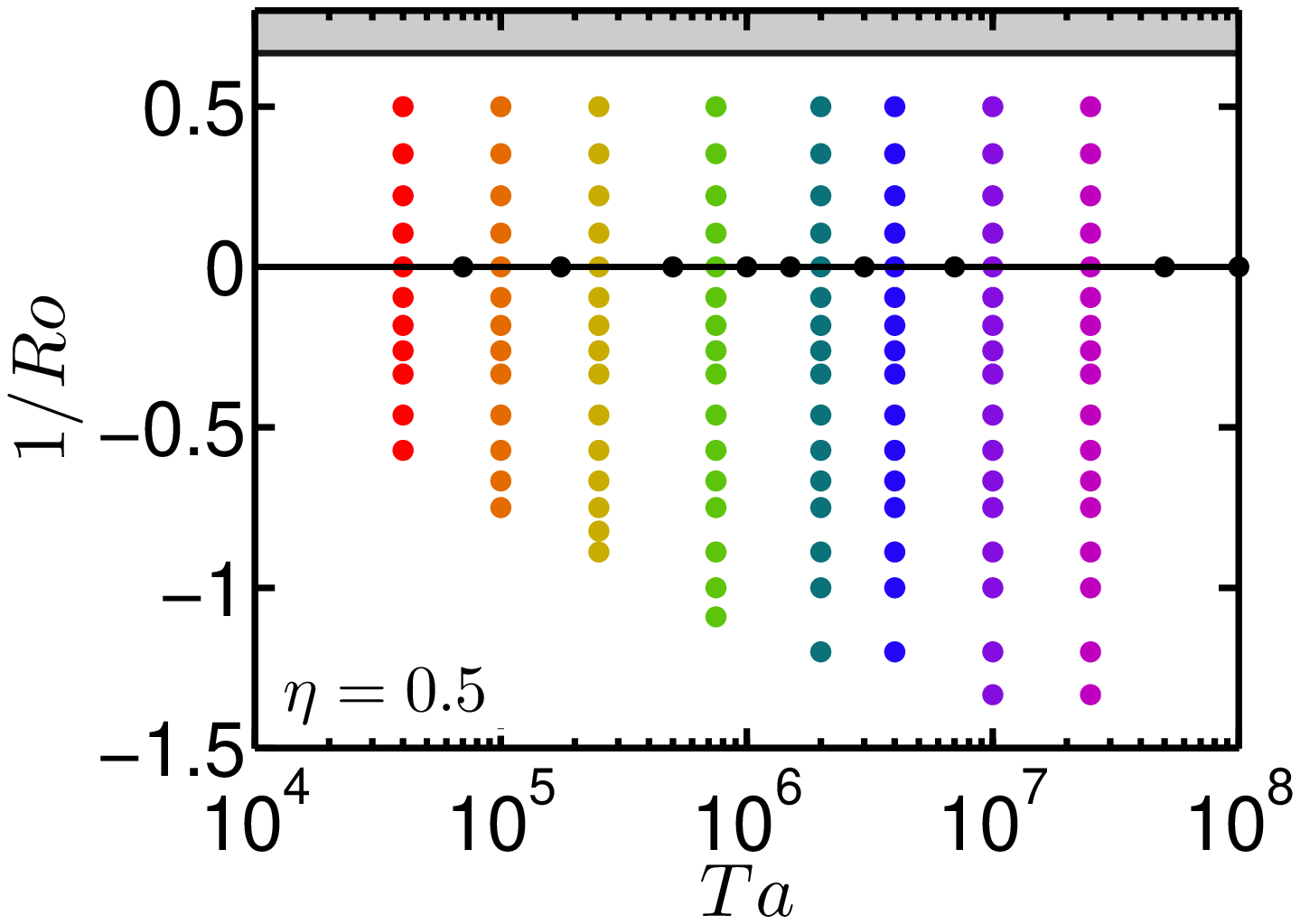}}                
  \subfloat{\label{fig:PhaseSpace0714}\includegraphics[width=0.49\textwidth,trim = 0mm 0mm 9mm 0mm, clip]{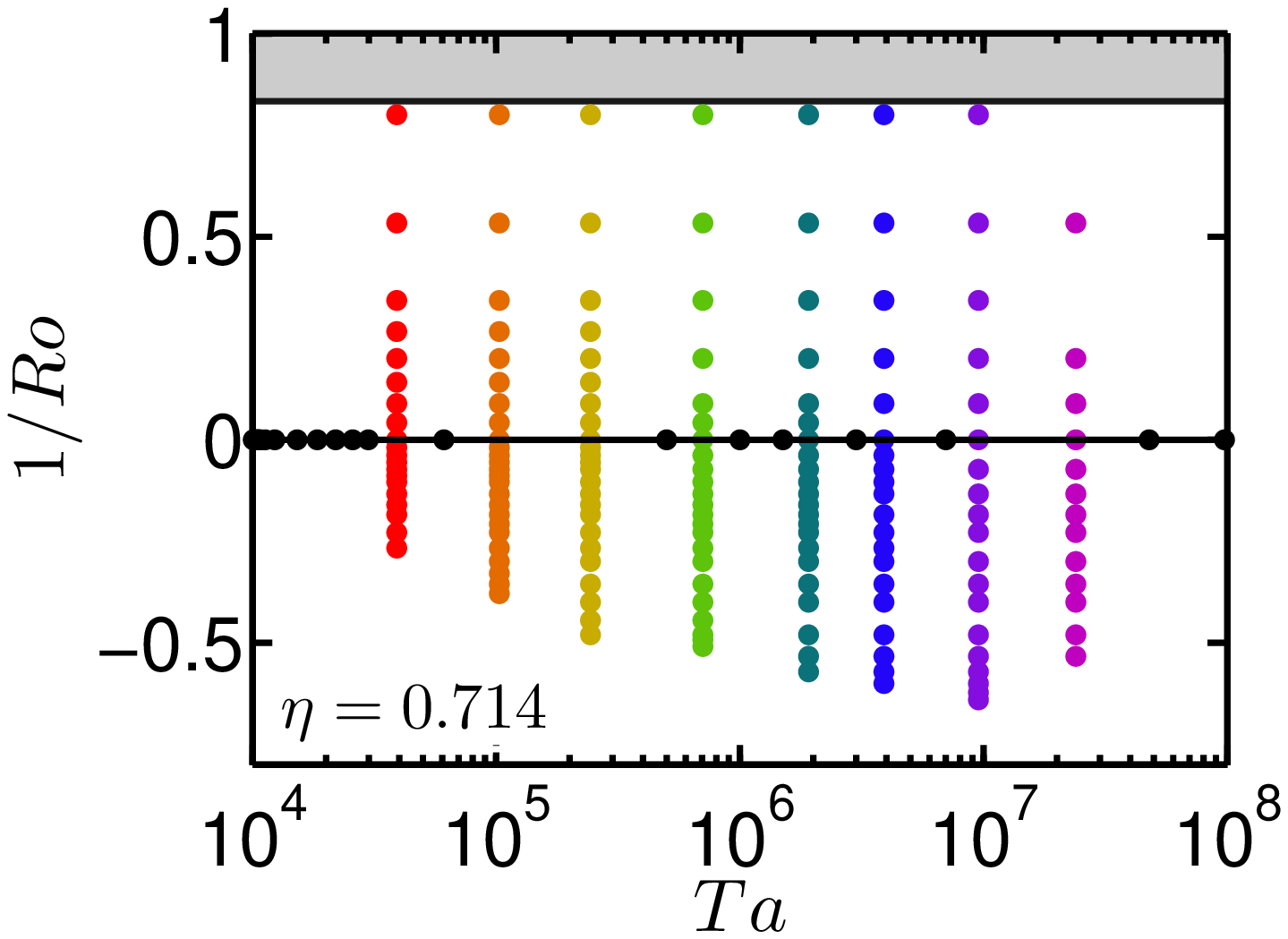}}\\
  \subfloat{\label{fig:PhaseSpace0833}\includegraphics[width=0.49\textwidth,trim = 0mm 0mm 9mm 0mm, clip]{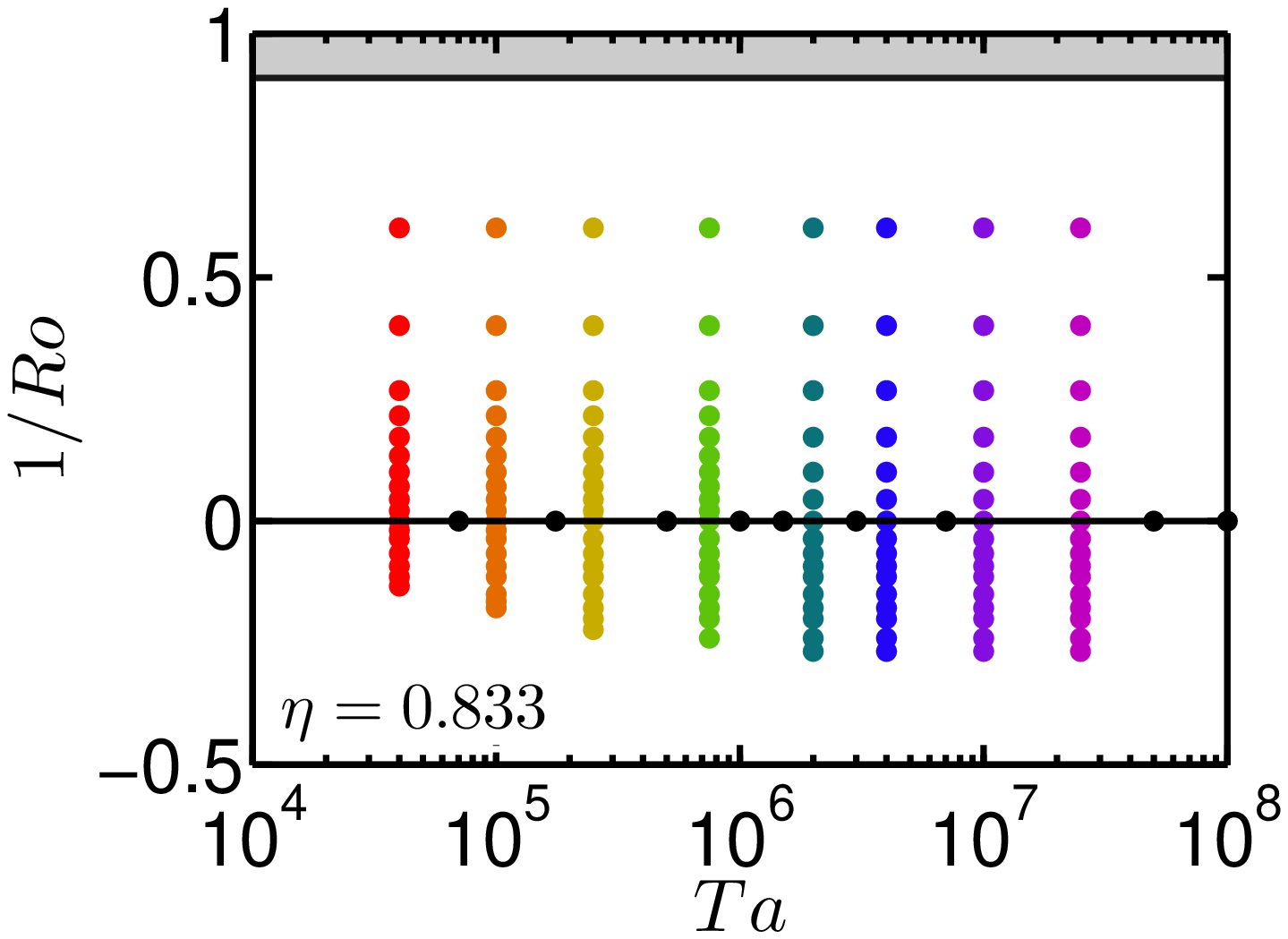}}   
  \subfloat{\label{fig:PhaseSpace0909}\includegraphics[width=0.49\textwidth,trim = 0mm 0mm 9mm 0mm, clip]{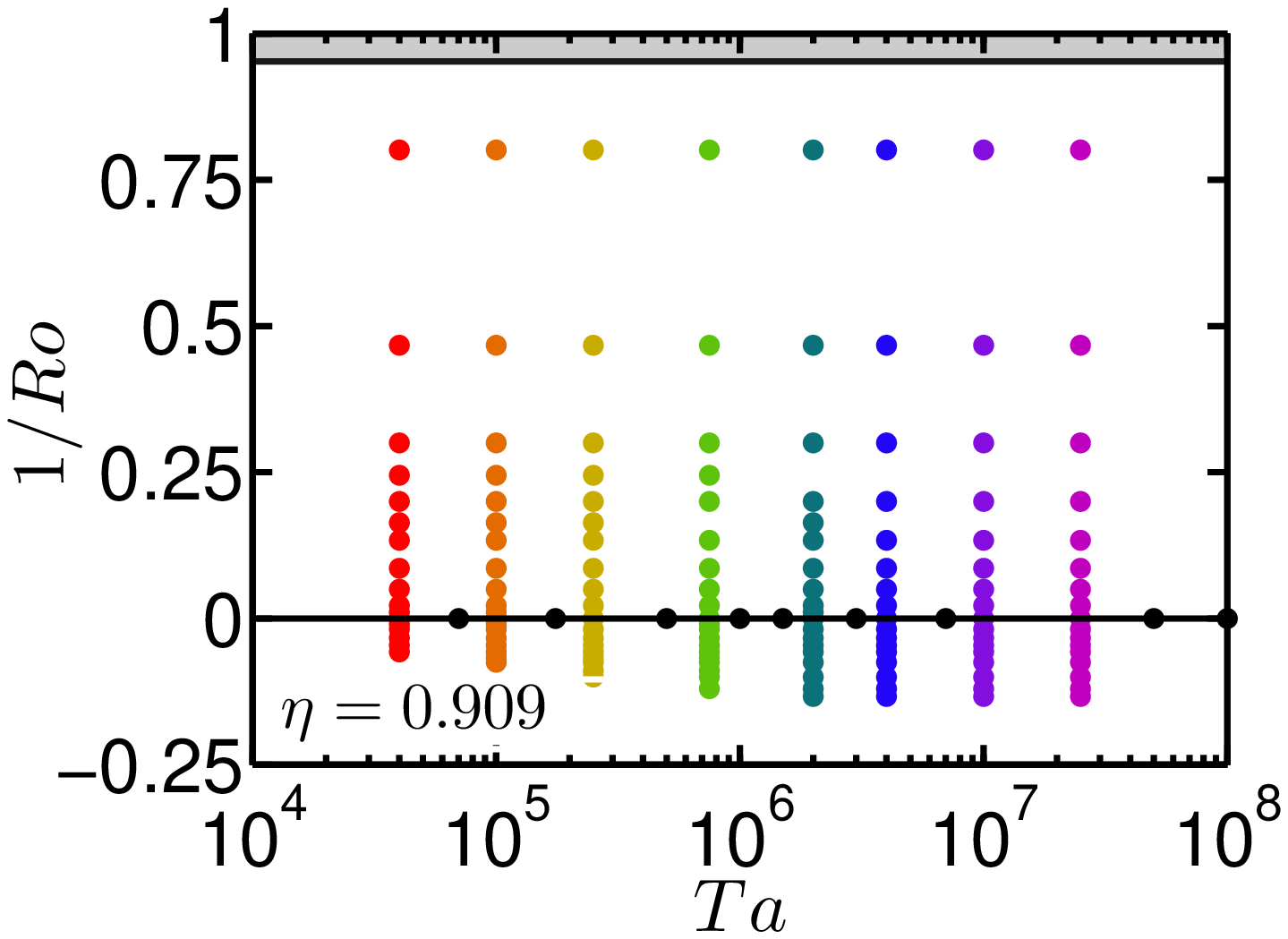}}                 
  \caption{Control parameter phase space which was numerically explored in this paper in the ($Ta$, $1/Ro$) representation. From top-left to 
bottom-right: $\eta=0.5$, $0.714$, $0.833$ and $0.909$. $\Gamma=2\pi$ was fixed, and axial periodicity was employed. 
   The grey-shaded area signals boundary conditions for which the angular momentum $L=r^2\omega$ of the outer cylinder ($L_o$) 
is larger than the angular momentum of the inner cylinder ($L_i$). This causes the flow to have an overall transport 
of angular momentum towards the inner cylinder. In this region, the Rayleigh stability criterium applies, which states 
that if $dL^2/dr > 0$ the flow is linearly stable to axisymmetric perturbations.
   }
  \label{fig:PhaseSpace}
 \end{center} 
\end{figure}

Figure \ref{fig:PhaseSpace} shows the ($Ta$,$1/Ro$) parameter space explored in the simulations 
for the four selected values of the radius ratio $\eta$.
A higher density of points has been used in places where the global response ($\nom$, $Re_w$) of the flow 
shows more variation with the control parameters $Ta$ and $1/Ro$. A fixed aspect ratio of $\Gamma=2\pi$ has been taken for all simulations, 
and axially periodic boundary conditions have been used. These simulations have the same upper 
bounds of $Ta$ (or $Re_i$) as the ones of \cite{ost12}.

\begin{figure}
 \begin{center}
  \subfloat{\label{fig:PhaseSpaceExp0716}\includegraphics[width=0.47\textwidth,trim = 0mm 0mm 0mm 0mm, clip]{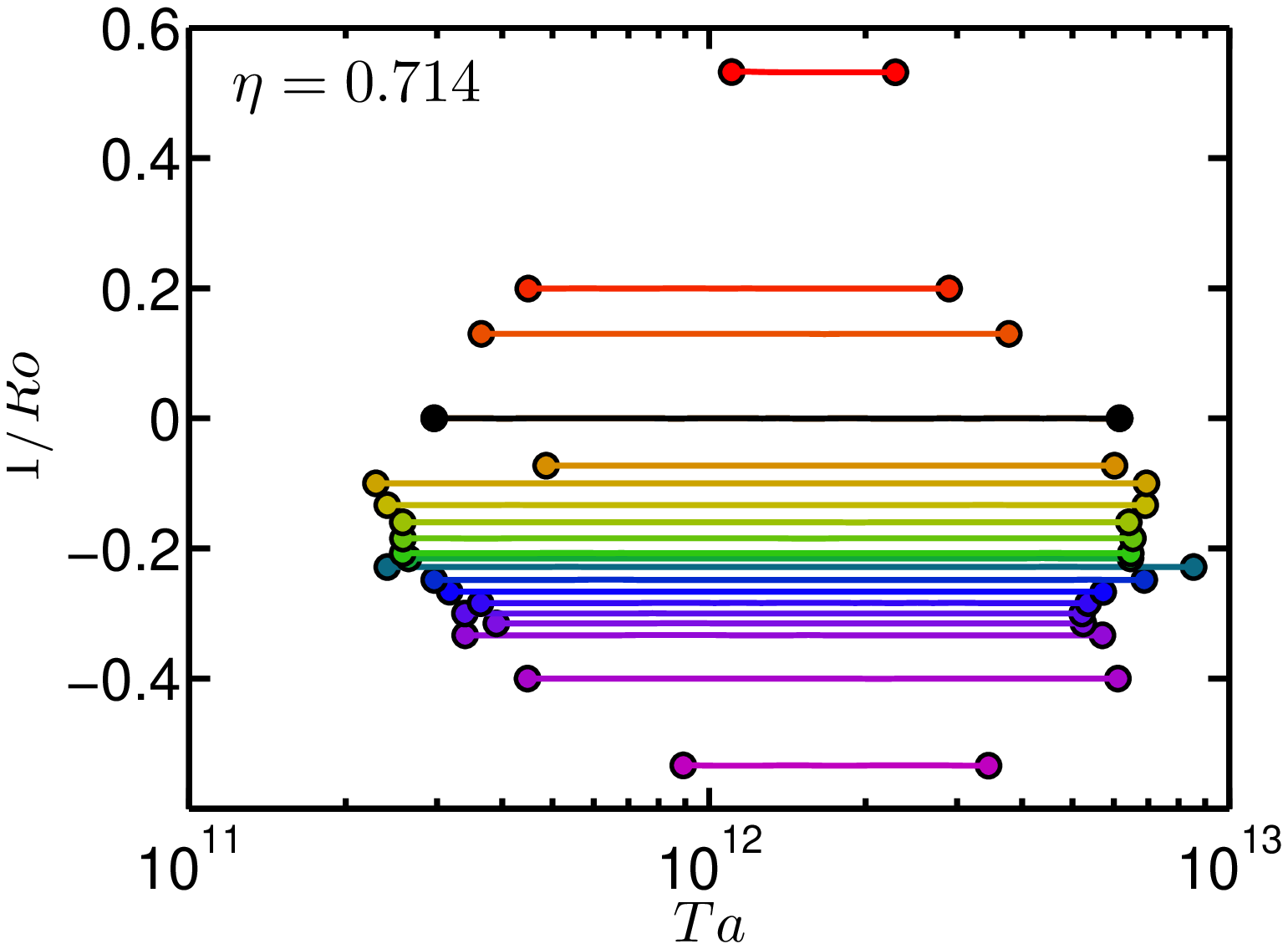}}                
  \subfloat{\label{fig:PhaseSpaceExp0769}\includegraphics[width=0.47\textwidth,trim = 0mm 0mm 0mm 0mm, clip]{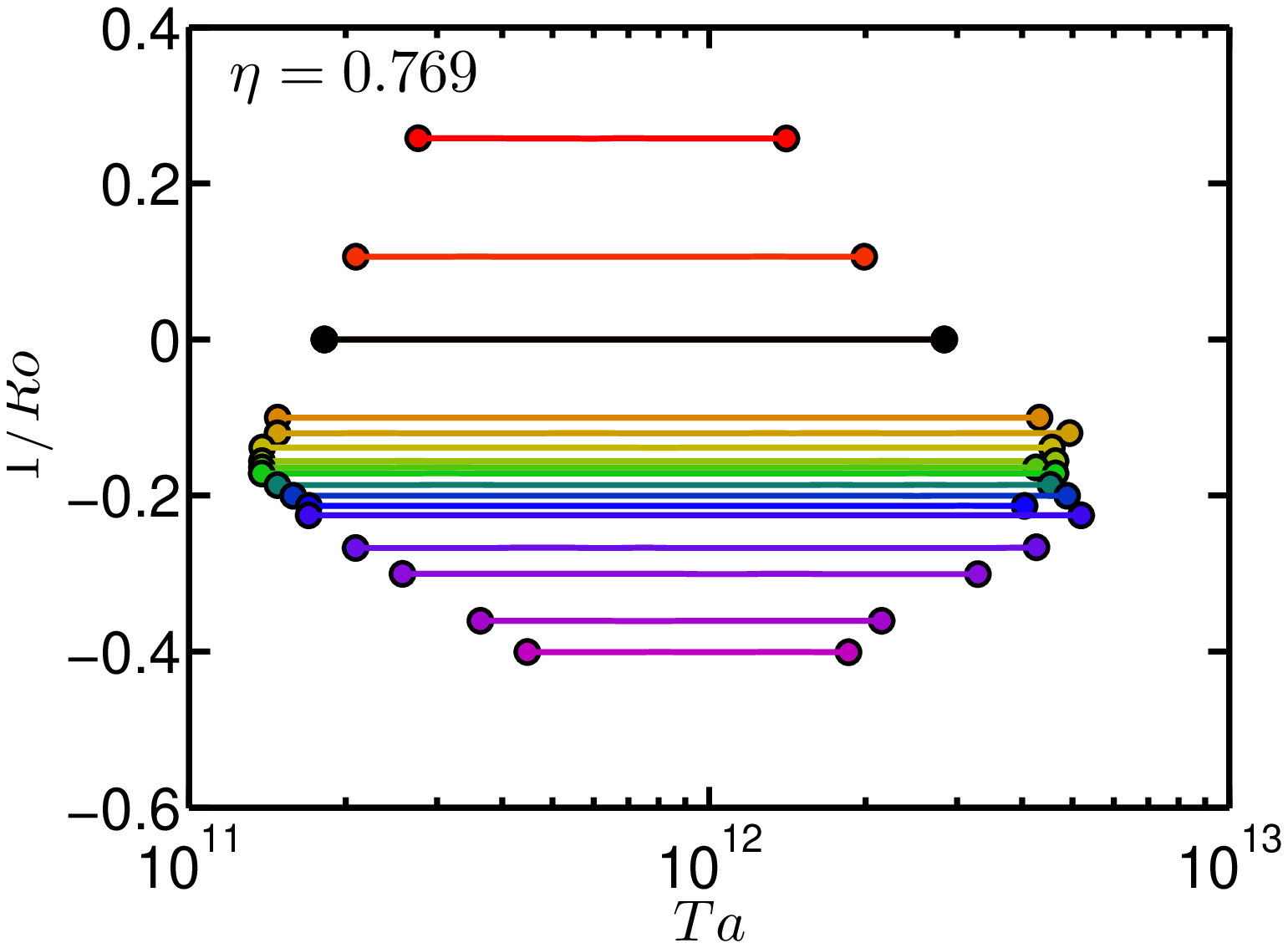}}\\                
  \subfloat{\label{fig:PhaseSpaceExp0833}\includegraphics[width=0.47\textwidth,trim = 0mm 0mm 0mm 0mm, clip]{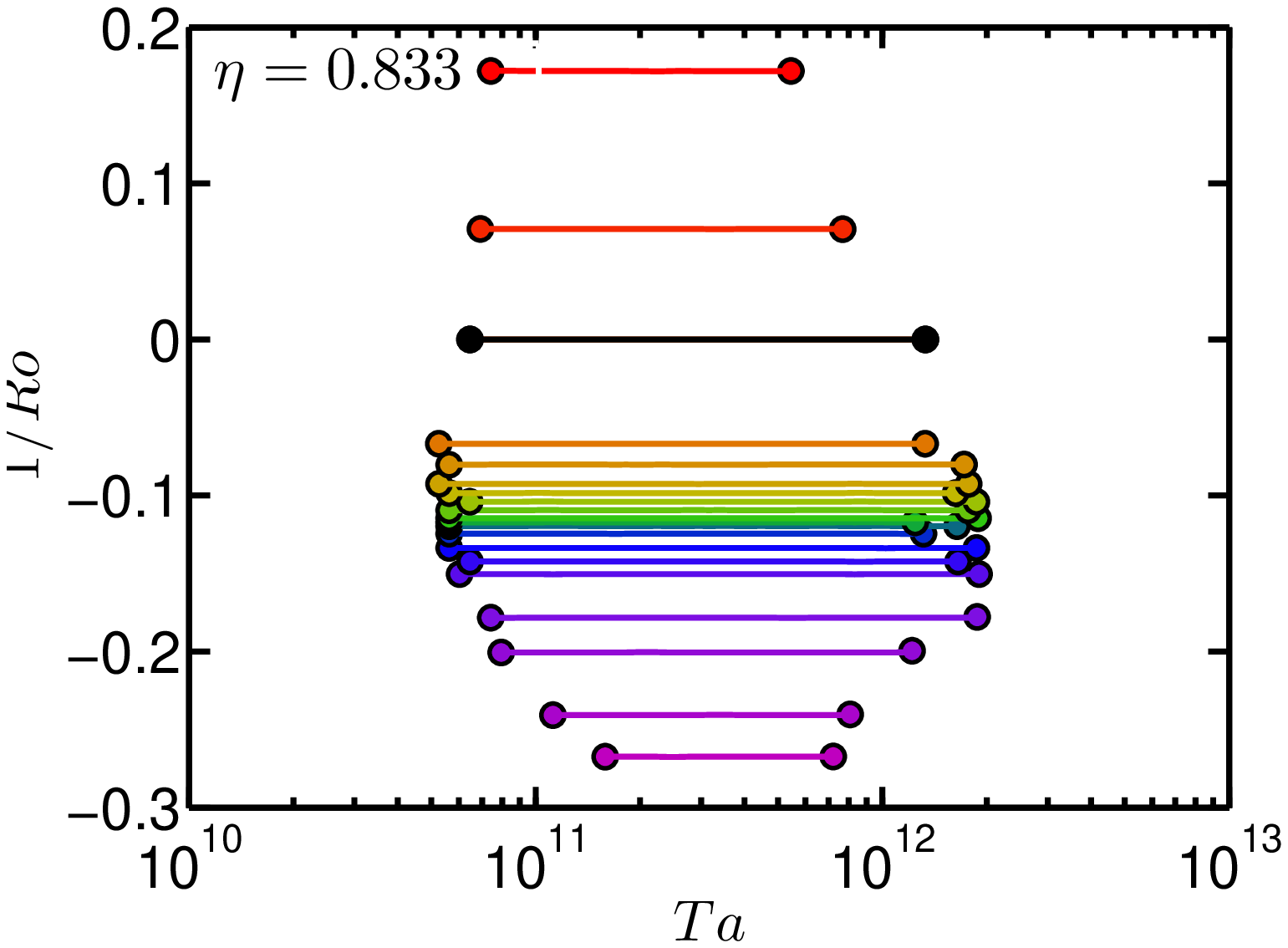}}                
  \subfloat{\label{fig:PhaseSpaceExp0909}\includegraphics[width=0.47\textwidth,trim = 0mm 0mm 0mm 0mm, clip]{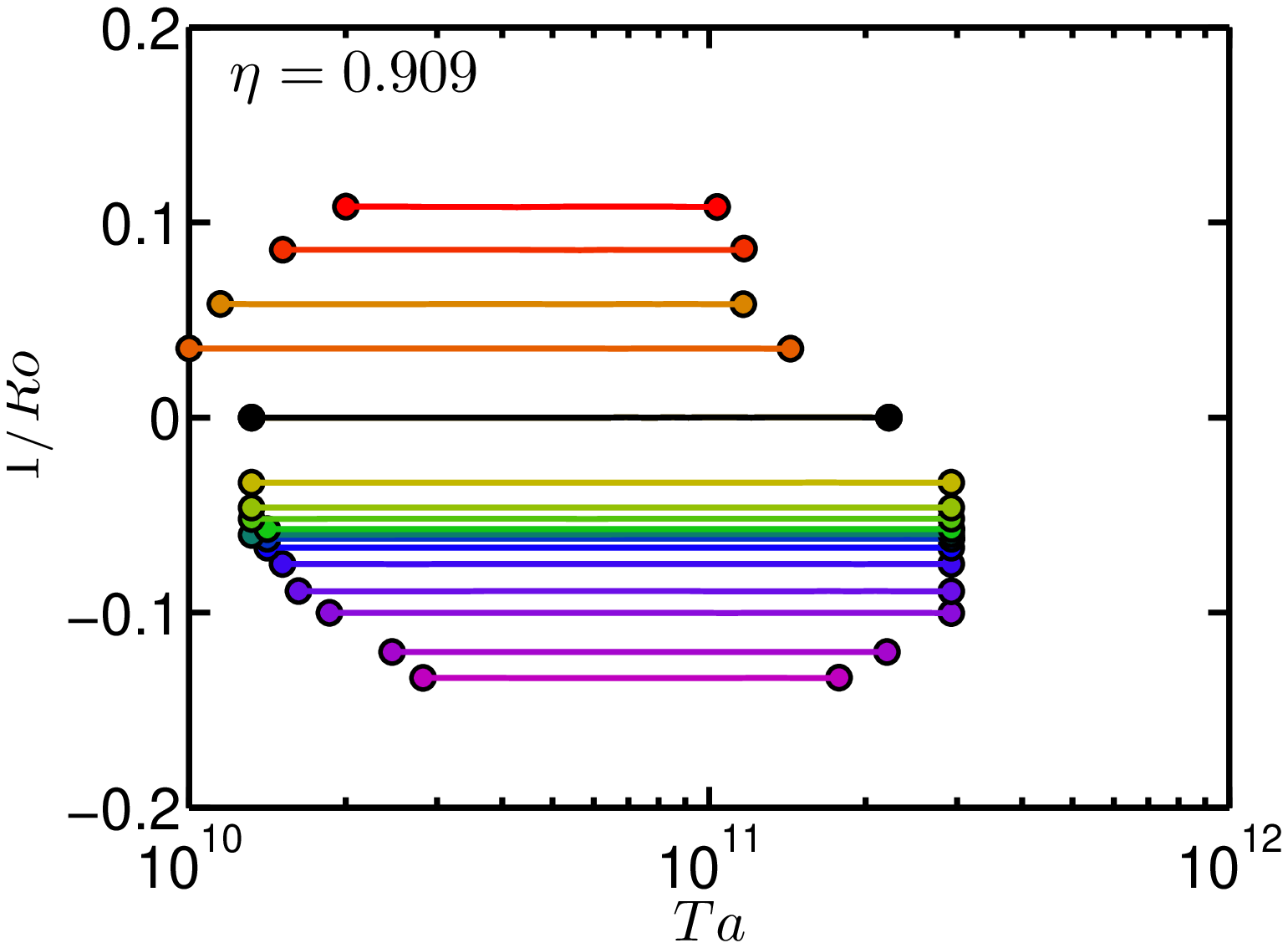}}                
  \caption{Control parameter phase space which was explored in experiments in the ($Ta$, $\usro$) representation for $\eta=0.716$ (top-left), $\eta=0.769$ 
(top-right), $\eta=0.833$ (bottom-left) and $\eta=0.909$ (bottom-right). } 

  \label{fig:PhaseSpaceExp}
 \end{center} 
\end{figure}

In addition to these simulations, experiments have been performed with the Twente Turbulent Taylor-Couette 
($T^3C$) facility, with which we achieve larger Ta numbers. Details of the
setup are given in \cite{gil11a}. Once again, four values of $\eta$ have
 been investigated, but, due to experimental constraints, we have been limited to investigate
only smaller gap widths, i.e. values $\eta\ge 0.714$, namely 
$\eta=0.714$, $0.769$, $0.833$ and $0.909$. The experimentally explored parameter space are shown in Fig. \ref{fig:PhaseSpaceExp}.

The structure of the paper is as follows. 
In sections 2 and 3, we start by describing the 
numerical code and the experimental setup, respectively. 
In section 4, the global response of the system, quantified by the non-dimensionalized
angular velocity current $Nu_\omega$ is analyzed. 
To understand the global response, we analyze the local data which 
can be obtained from the DNS simulations in section 5. Angular velocity profiles in the bulk 
and in the boundary layers are analyzed and related to the global angular velocity optimal transport. We finish in section 6
 with a discussion of the results and an outlook for further investigations.

\section{Numerical method}

In this section, the used numerical method is explained in some detail. 
The rotating frame in which the Navier-Stokes equations are solved and the employed non-dimensionalizations 
are introduced in the first section. This is followed by detailing the spatial 
resolution checks which have been performed.

\subsection{Code description}
\label{sec:codedesc}

The employed code is a finite difference code, which solves the Navier-Stokes equations in 
cylindrical coordinates. A second--order spatial discretization is used, and the equations are 
advanced in time by a fractional time integration method. This code is based on the so-called Verzicco-code,
whose numerical algorithms are detailed in \cite{ver96}. A combination of MPI and OpenMP directives are used to 
achieve large scale parallelization. This code has been extensively
used for Rayleigh-B\'enard flow; for recent simulations see \cite{ste10,ste11}.  In the context of TC flow,
\cite{ost12} already validated the code for $\eta=0.714$.

The flow was simulated in a rotating frame, which was chosen to rotate with $\Omega=\omega_o \boldsymbol{e}_z$. 
This was done in order to simplify the boundary conditions. In that frame the outer cylinder is stationary 
for any value of $a$, while the inner cylinder has an azimuthal velocity of 
$u_\theta(r=r_i)=r_i(\omega^\ell_i-\omega^\ell_o)$, where the $\ell$ superscript denotes variables
in the lab frame, while no superscript denotes variables in the rotating frame. 
We then choose the inner cylinder rotation rate in the rotating frame as
the characteristic velocity of the system $U \equiv |u_\theta(r_i)| = r_i|\omega_i-\omega_o|$ and the 
characteristic length scale $d$ to non-dimensionalize the equations and boundary conditions.


Using this non-dimensionalization, the inner cylinder velocity boundary condition simplifies to: 
$\hat{u}_\theta(r=r_i)=\text{sgn}(\omega_i-\omega_o)$. In this paper, $\omega_i-\omega_o$ 
is always positive. Thus, in this rotating frame the flow geometry is simplified to a pure inner cylinder rotation with the 
boundary condition  $\hat{u}_{\theta}(r_i) = 1$. The outer cylinder's effect on the flow is felt as a 
Coriolis force in this rotating frame of reference. The Navier-Stokes equations then read:

\begin{equation}
 \displaystyle\frac{\partial \hat{\bu}}{\partial \hat{t}} + \hat{\bu}\cdot\hat{\nabla}\hat{\bu} = -\hat{\nabla} \hat{p} +  
\left (\displaystyle\frac{f(\eta)}{Ta}\right)^{1/2} \hat{\nabla}^2\hat{\bu} + Ro^{-1} {\boldsymbol{e}_z}\times\hat{\bu}~,
\label{eq:rotatingTC}
\end{equation}
where $Ro$ was defined previously in Eq. \ref{eq:Rosdef}, and $f(\eta)$ is
\begin{equation}
 f(\eta) = \displaystyle\frac{(1+\eta)^3}{8\eta^2}~.
\end{equation}

It is useful to continue the non-dimensionalization by defining the normalized radius $\tilde{r} = (r-r_i)/d$ 
and the normalized height $\tilde{z}=z/d$. We define the time- and azimuthally-averaged velocity field as:

\begin{equation}
 \hat{\bar{\bu}}(r,z)=\langle \hat{\bu}(\theta,r,z,t) \rangle_{\theta,t}~,
\end{equation}

\noindent where $\langle \phi(x_1,x_2,...,x_n) \rangle_{x_i}$ indicates averaging of the field $\phi$ with respect to $x_i$. 

To quantify the torque in the system, we first note that the angular velocity current

\begin{equation}
 J^\omega = r^3(\langle u_r^\ell \omega^\ell\rangle_{\theta,z,t} - \nu \partial_r\langle\omega^\ell\rangle_{\theta,z,t})
 \label{eq:jomega}
\end{equation}

\noindent is conserved, i.e. independent on the radius $r$ (EGL 2007). $J^\omega$ represents the current
of angular velocity from the inner cylinder to the outer cylinder (or vice versa). The first term is the convective
contribution to the transport, while the second term is the diffusive contribution.

In the state with the lowest driving, 
and ignoring end plate effects, a laminar, time independent velocity field which is purely azimuthal, $u^\ell_\theta(r) = Ar+B/r$, with
$u_r=u_z=0$, is induced by the rotating cylinders.  This laminar flow produces an angular velocity current $J_0^\omega$, 
which can be used to nondimensionalize the angular velocity current and a non-zero dissipation rate $\epsilon_{u,0}$.

\begin{equation}
 \nom=\displaystyle\frac{J^\omega}{J_0^\omega}.
\end{equation}

\noindent $\nom$ can be seen as an angular velocity ``Nusselt'' number.

When $J^\omega$, and therefore $\nom$ are calculated numerically, the values will depend on the radial position,
due to finite time averaging.
We can define $\Delta_J$ to quantify this radial dependence as:

\begin{equation}
 \Delta_J = \displaystyle\frac{\max(J^\omega(r))-\min(J^\omega(r))}{\langle J^\omega(r)\rangle_r} 
=\displaystyle\frac{\max(\nom(r))-\min(\nom(r))}{\langle\nom(r)\rangle_r}
\end{equation}

\noindent which analytically equals zero but will deviate when calculated numerically. 

The convective dissipation per unit mass can be calculated either from its definition 
as a volume average of the local dissipation rate for an incompressible fluid,

\begin{equation}
 \varepsilon_u=\epsilon^\ell_{u} = \displaystyle\frac{\nu}{2} \mean{(\partial_i^\ell u_j + \partial_j^\ell u_i)^2}_{V,t},
 \label{eq:epsilonlocal}
\end{equation}

\noindent or a global balance can also be used. The exact relationship (EGL 2007) 

\begin{equation}
 \epsilon^\ell_{u} - \epsilon^\ell_{u,0} = \displaystyle\frac{\nu^3}{d^4}\sigma^{-2}Ta(\nom-1)~,
 \label{eq:epsiloneck}
\end{equation}

\noindent where $\epsilon_{u,0}$ is the volume averaged dissipation rate in the purely 
azimuthal laminar flow, links the volume averaged dissipation to the global driving $Ta$ and response $\nom$. 

This link can be and has been used for code validation and for checking spatial resolution adequateness. 
The volume averaged dissipation can be calculated from both (\ref{eq:epsilonlocal}) and (\ref{eq:epsiloneck}) 
and later checked for sufficient agreement. We define the quantity $\Delta_{\epsilon}$ as the relative 
difference between the two ways of numerically calculating the dissipation, namely either via $\nom$ with eq.
\ref{eq:epsiloneck} or directly from the velocity gradients, eq. \ref{eq:epsilonlocal}:

\begin{equation}
 \Delta_{\epsilon} = \displaystyle\frac{\nu^3d^{-4}\sigma^{-2}Ta(\nom-1)+\epsilon_{u,0} - \frac{\nu}{2} \mean{(\partial_i u^\ell_j + \partial^\ell_j u_i)^2}_{V,t} }{ \frac{\nu}{2} \mean{(\partial^\ell_i u_j + \partial^\ell_j u_i)^2}_{V,t} }~.
 \label{eq:epsilondifference}
\end{equation}

\noindent $\Delta_{\epsilon}$ is equal to $0$ analytically, but will deviate
when calculated numerically.  The deviation of $\Delta_J$ and $\Delta_{\epsilon}$ from zero
is an indication of the adequateness of the resolution.

We would like to emphasize that the requirement for $\Delta_J<0.01$ is much stricter 
than torque balance, which can simply be expressed
as $\nom(r_i)=\nom(r_o)$.  As analyzed in \cite{ost12},
a value of less than $1\%$ for $\Delta_J$ and about $1\%$ for $\Delta_{\epsilon}$ is linked to grid adequateness 
at the Taylor number simulated.
To ensure convergence in time, the time-averages of the Nusselt number and the energy dissipation calculated locally
(equation \ref{eq:epsilonlocal}) were also checked to converge in time within $1\%$.

\subsection{Resolution checks}

Spatial resolution checks were performed in two ways. First, as mentioned previously, 
the values of $\Delta_J$ and $\Delta_{\epsilon}$ were checked. As an additional check, 
simulations at selected values of $Ta$ were performed at a higher resolution.
As the explored parameter space is large, these checks were performed only for the highest value of $Ta$ 
simulated for the grid size. A lower driving of the flow
for the same grid size is expected to have a smaller error due to spatial discretization, as 
spatial discretization errors increase with increased $Re$, and thus increased $Ta$. 

Concerning the temporal resolution there are numerical and physical constraints; the former requires a time step
small enough to keep the integration scheme stable and this is achieved 
by using an adaptive time step based on a Courant--Frederich--Lewy ($CFL$) criterium. 
The $3^{\rm rd}$--order Runge--Kutta time--marching algorithm allows for a $CFL$ of up to
$\sqrt{3}$, but this can be reduced due to the implicit factorization of the
viscous terms. For safety, the maximum $CFL$ has been taken as $1.4$. 
From the physical point of view, the time step size must also be small enough to properly describe the
fast dynamics of the smallest flow scale which is the Kolmogorov scale.
Although the time step size should be determined by the most restrictive among the two criteria above, our experience
suggests that as long as the $CFL$ number criterion is satisfied, which guarantees numerical stability, 
the results become insensitive to the time step size and all the flow scales are adequately described temporally.
Direct confirmation of this statement can be found in \cite{ost12}.

The results for $\eta=0.5$, $0.833$, and $0.909$ are presented in Table \ref{tbl:restests}. Uniform
discretization was used in azimuthal and axial directions. In the radial direction, points were
clustered near the walls by using hyperbolic tangent-type clustering, or a clipped Chebychev type
clustering for higher values of $Ta$. A table including the results for the spatial 
resolution tests at $\eta=0.714$ can be found in \cite{ost12}.

\begin{table}
  \begin{center} 
  \def~{\hphantom{0}}
  \begin{tabular}{cccccccc}
  $\eta$  & $Ta$ & $N_\theta$ x $N_r$ x $N_z$ & $\nom$ & 100$\Delta_{J}$ & 100$\Delta_{\epsilon}$ & Case \\
  $0.5$   & $2.5\cdot10^5$ & 100x100x100 & 2.03372 & 0.30 & 1.11 & R \\
  $0.5$   & $2.5\cdot10^5$ & 150x150x150 & 2.03648 & 0.76 & 0.89 & E \\
  $0.5$   & $7.5\cdot10^5$ & 150x150x150 & 2.56183 & 0.47 & 0.92 & R \\
  $0.5$   & $7.5\cdot10^5$ & 256x256x256 & 2.55673 & 0.74 & 0.31 & E \\
  $0.5$   & $1\cdot10^7$   & 300x300x300 & 4.23128 & 0.33 & 1.07 & R \\
  $0.5$   & $1\cdot10^7$   & 400x400x400 & 4.22574 & 0.97 & 1.06 & E \\ 
  $0.5$   & $2.5\cdot10^7$ & 350x350x350 & 5.07899 & 0.85 & 1.12 & R \\ 
  $0.5$   & $2.5\cdot10^7$ & 512x512x512 & 5.08193 & 0.87 & 1.98 & E \\ 
  $0.5$   & $5\cdot10^7$   & 768x512x1536 & 6.08284 & 0.45 & 1.56 & R \\  
  $0.5$   & $1\cdot10^8$   & 768x512x1536 & 7.48561 & 1.46 & 0.88 & R \\ 

  $0.833$ & $2.5\cdot10^5$ & 180x120x120 & 2.72293 & 0.21 & 0.76 & R \\
  $0.833$ & $2.5\cdot10^5$ & 300x180x180 & 2.72452 & 0.29 & 0.29 & E \\
  $0.833$ & $1\cdot10^7$   & 384x264x264 & 7.07487 & 0.29 & 0.61 & R \\
  $0.833$ & $1\cdot10^7$   & 512x384x384 & 7.17245 & 0.13 & 1.16 & E \\ 
  $0.833$ & $2.5\cdot10^7$ & 512x384x384 & 8.62497 & 0.71 & 1.05 & R \\ 
  $0.833$ & $2.5\cdot10^7$ & 768x576x576 & 8.51678 & 0.90 & 1.26 & E \\ 
  $0.833$ & $5\cdot10^7$   & 512x384x384 & 9.68437 & 0.26 & 2.92 & R \\ 
  $0.833$ & $1\cdot10^8$   & 768x576x576 & 11.4536 & 0.89 & 2.29 & R \\ 

  $0.909$ & $2.5\cdot10^5$ & 180x120x120 & 2.31902 & 0.16 & 0.91 & R \\ 
  $0.909$ & $2.5\cdot10^5$ & 300x200x200 & 2.30810 & 0.07 & 0.17 & E \\
  $0.909$ & $2\cdot10^6$   & 256x180x180 & 3.76826 & 0.55 & 0.49 & R \\
  $0.909$ & $2\cdot10^6$   & 384x256x256 & 3.77532 & 0.39 & 0.21 & E \\
  $0.909$ & $2.5\cdot10^7$ & 384x256x256 & 7.83190 & 0.43 & 3.15 & R \\ 
  $0.909$ & $2.5\cdot10^7$ & 450x320x320 & 7.86819 & 0.81 & 2.07 & E \\ 
  $0.909$ & $5\cdot10^7$   & 2305x400x1536 & 9.74268 & 0.46 & 1.02 & R \\ 
  $0.909$ & $1\cdot10^8$   & 2305x400x1536 & 11.3373 & 0.57 & 1.06 & R \\ 

  

 \end{tabular}
 \caption{Resolution tests for $\Gamma=2\pi$ and $\eta=0.5$, $0.833$ and $0.909$. The first column
 displays the radius ratio, the second column displays the Taylor number, the third 
column displays the resolution employed, the fourth column the calculated $\nom$, the 
fifth column and sixth colums the relative discrepancies $\Delta_{J}$ and $\Delta_{\epsilon}$, 
and the last column the 'case': (R)esolved and (E)rror reference. $\Delta_{\epsilon}$ is positive,
and exceeds the $1\%$ threshold reported in \cite{ost12} for some cases at the largest $\eta$,
but even so resolution appears to be sufficient as variations of $\nom$ are small.}
 \label{tbl:restests}
\end{center}
\end{table}

\section{Experimental setup}

The Twente Turbulent Taylor-Couette (\tttc) apparatus has been built to obtain high $Ta$ numbers. 
It has been described in detail in \cite{gil11a} and \cite{gil12}. The inner cylinder with outside radius 
$r_i=\unit{0.200}{\meter}$ consists of three sections. The total height of those axially 
stacked sections is $L=\unit{0.927}{\meter}$. We measure the torque only on the middle 
section of the inner cylinder, which has a height of $L_\text{mid} = \unit{0.536}{\meter}$, 
to reduce the effect of the torque losses at the end-plates in our measurements. This
approach has already been validated in \cite{gil12}. 
The transparent outer cylinder is made 
of acrylic and has an inside radius of $r_o  = \unit{0.279}{\meter}$. We vary the radius ratio 
by reducing the diameter of the outer cylinder by adding a `filler' that 
is fixed to the outer cylinder and sits between the inner and the outer cylinder, effectively 
reducing $r_o$ while keeping $r_i$ fixed. We have 3 fillers giving us 4 possible outer 
radii: $r_o = \unit{0.279}{\meter}$ (without any filler), $ \unit{0.26}{\meter}$, $\unit{0.24}{\meter}$, 
and $\unit{0.22}{\meter}$, 
giving experimental access to $\eta=0.716$, $0.769$, $0.833$, and $0.909$, respectively. Note that by 
reducing the outer radius, we not only change $\eta$, but also change $\Gamma=L/(r_o-r_i)$ 
from $\Gamma(\eta=0.716) = 11.68$ to $\Gamma(\eta=0.909) = 46.35$. 

For high $Ta$ the heating up of the system becomes apparent and it has to be 
actively cooled in order to keep the temperature constant. We cool the working fluid (water) 
from the top and bottom end plates and maintain a constant temperature within
$\pm 0.5K$ throught both the spatial extent and the time run of the experiment. 
The setup has been constructed in such a way that we can rotate both cylinders independently 
while keeping the setup cooled. 

As said before, we measure the torque on the middle inner 
cylinder. We do this by measuring the torque that is transferred 
from the axis to the cylinder by using a load-cell that is inside 
the aforementioned cylinder. Torque measurements are performed using a 
fixed procedure: the inner cylinder is spun up to its maximum rotational 
frequency of \unit{20}{\hertz} and kept there for several minutes. Then 
the system is brought to rest. The cylinders are then brought to their initial rotational 
velocities (with the chosen $1/Ro$), corresponding to a velocity for which the torque is 
accurate enough; generally of order 2--\unit{3}{\hertz}. We then slowly increase both velocities
over 3-6 hours to their final velocities while maintaining $1/Ro$ fixed during the 
entire experiment. During this velocity ramp we continuously acquire the torque of 
this quasi-stationary state. The calibration of the system is done in a similar way; 
first we apply the maximum load on the system, going back to zero load, and then 
gradually adding weight while recording the torque. These procedures ensure that 
hysteresis effects are kept to a minimum, and that the system is always brought to 
the same state before measuring. More details about the setup can be found in \cite{gil11a}. 

Local velocity measurements are done by laser Doppler anemometry (LDA). We 
measure the azimuthal velocity component by focusing two beams in the 
radial-azimuthal plane. We correct for curvature effects of the outer 
cylinder by using a ray-tracer, see \cite{hui12b}. The velocities are 
measured at midheight ($z=L/2$) unless specified otherwise. For every 
measurement position we measured long enough such as to have a 
statistically stationary result, for which about $10^5$ samples were required
for every data point. This ensured a statistical convergence of $<1\%$. 

\section{Global response: Torque}

\label{sec:global}

In this section, the global response of the TC system for the four simulated radius ratios is presented. 
This is done by measuring the scaling law(s) of the non-dimensional torque $\nom$ as function(s) of $Ta$. 
The transition between different types of local scaling laws in different $Ta$-ranges is investigated, and related 
to previous simulations \citep{ost12} and experiments \citep{gil12}.

\subsection{Pure inner cylinder rotation}

The global response of the system is quantified by $\nom$.
By definition, for purely azimuthal laminar
flow, $\nom=1$. Once the flow is driven stronger than a certain critical $Ta$, large azimuthal roll structures appear,
which enhance angular transport through a large scale wind. 

\begin{figure}
 \begin{center}
  \subfloat{\label{fig:TaNuEtasa0}\includegraphics[width=0.98\textwidth]{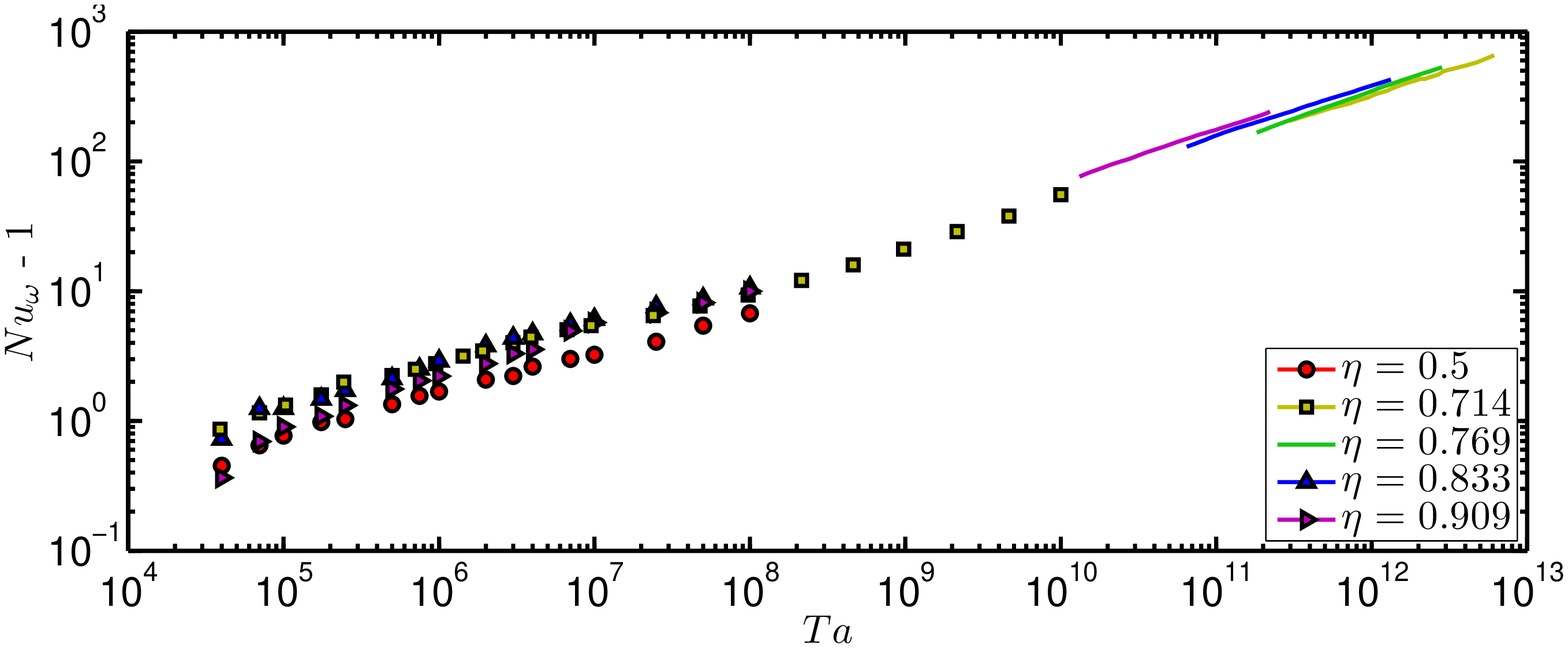}}\\
  \subfloat{\label{fig:TaNuEtasa0C}\includegraphics[width=0.98\textwidth]{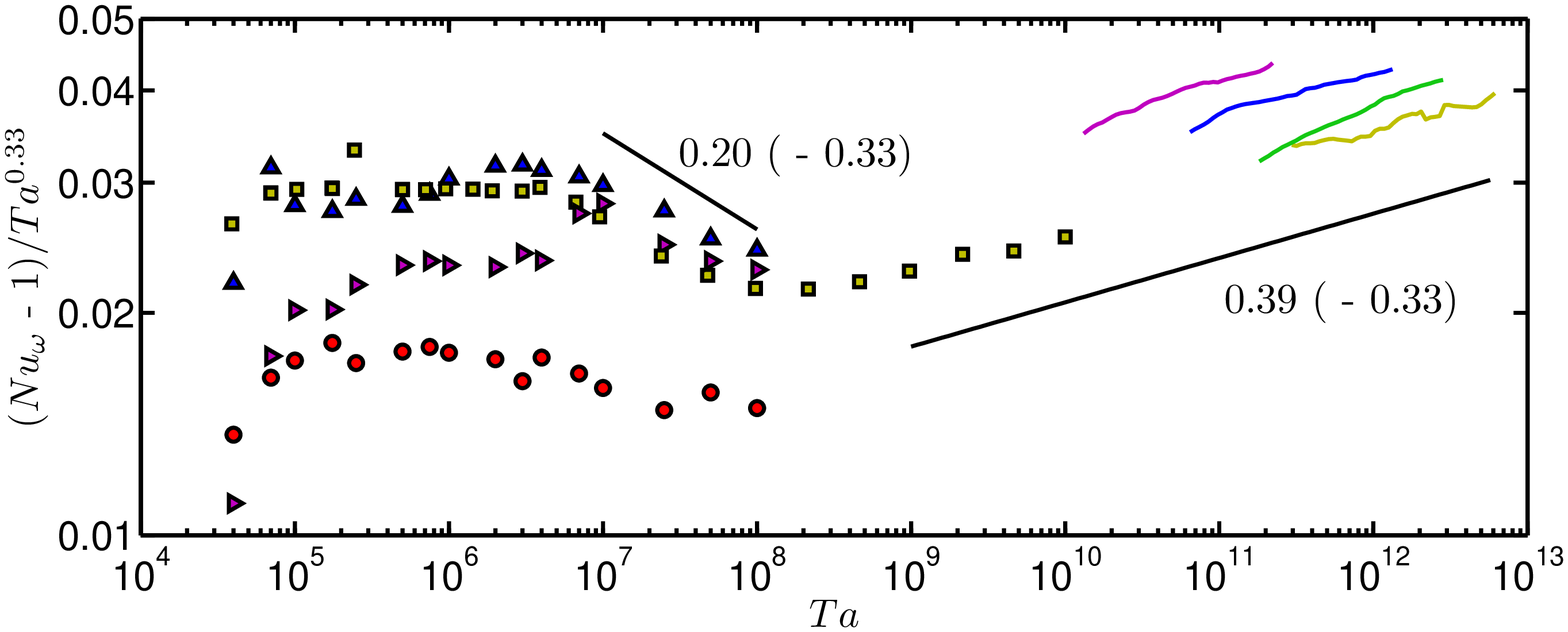}}\\  
  \caption{The global system response for pure inner cylinder rotation as function of the driving $Ta$: 
The top panel shows $\nom-1$ vs $Ta$ for both simulations (points on the left of the graph) and experiments
(lines on the right of the graph). Numerical data from \cite{ost13} for $\eta=0.714$ have been added to these figures.
The bottom panel shows the compensated Nusselt ($(\nom-1)/Ta^{1/3}$) versus $Ta$, with added
lines with scaling law $Ta^{0.20}$ and $Ta^{0.39}$ to guide the eye. }
 \label{fig:GlobalResponsea0}
 \end{center} 
\end{figure}

Figure \ref{fig:GlobalResponsea0} shows the response of the system for increasing $Ta$ in case of pure inner cylinder rotation  for
four values of $\eta$. Experimental and numerical results are shown in the same panels, covering different ranges and thus complementary,
but consistent with each other. Numerical results for $\eta=0.714$ from \cite{ost13} have been added to both panels.

As has already been noticed in \cite{ost12} for $\eta=0.714$, a change in the local scaling law relating 
$Ta$ to $\nom$ occurs at around $Ta\approx 3\cdot10^6$.
We can interpret these changes in the same way as 
\cite{ost12} and relate the transition in the $Ta$-$\nom$ local scaling law to the break up of coherent structures. 

It is also worth mentioning that the exponent in the local scaling laws in the regime before the transition 
depends on the radius ratio. This can be seen in the compensated plot, and explains the curve crossings that we see in the graphs.

For experiments (solid lines of figure \ref{fig:GlobalResponsea0}), a different local scaling law can be seen. In this case 
the experiments are performed at much higher $Ta$ than the simulations. The scaling 
$\nom\sim Ta^{0.38}$ can be related to the so called ``ultimate'' regime, a regime where 
the boundary layers have become completely
turbulent \citep{gro11,gro12,hui13}. As indicated for the case at $\eta = 0.714$ we expect that for increasing $Ta$ also 
the simulations become turbulent enough to reach this scaling law (cf. \cite{ost13}). In this
regime, the local scaling law relating $Ta$ and $\nom$ has no dependence on $\eta$ and thus is universal.

In both experiment and simulation with the largest $Ta$, the value of $\eta$ corresponding to the smallest gap, 
i.e. $\eta=0.909$, has the 
highest angular velocity transport ($\nom$) at a given $Ta$. This can be phrased in terms of the 
pseudo-Prandtl-number $\sigma$, introduced in EGL2007. As
a smaller gap means a smaller $\sigma$, we thus find a decrease of $\nom$ with increasing $\sigma$,
for the drivings explored in experiments,
 similarly as predicted \cite{gro01} and found \cite{xia02} for $Nu(Pr)$ in RB convection for $Pr > 1$.

\subsection{Rossby number dependence}

\label{sec:globalro}

In this subsection, the effect of outer cylinder rotation on angular velocity transport will be studied. 
Previous experimental and numerical work at $\eta=0.714$ \citep{pao11, gil11, ost12, bra12} revealed 
the existence of an optimum transport where, for a given $Ta$, the transport of momentum is highest at a 
Rossby number $\usro_{opt}$, which depends on $Ta$ and saturates 
around $Ta\sim 10^{10}$. In this subsection, this work will be extended to the other values of $\eta$.

\begin{figure}
 \begin{center}
  \subfloat{\label{fig:RoNuNNEta05}\includegraphics[width=0.90\textwidth]{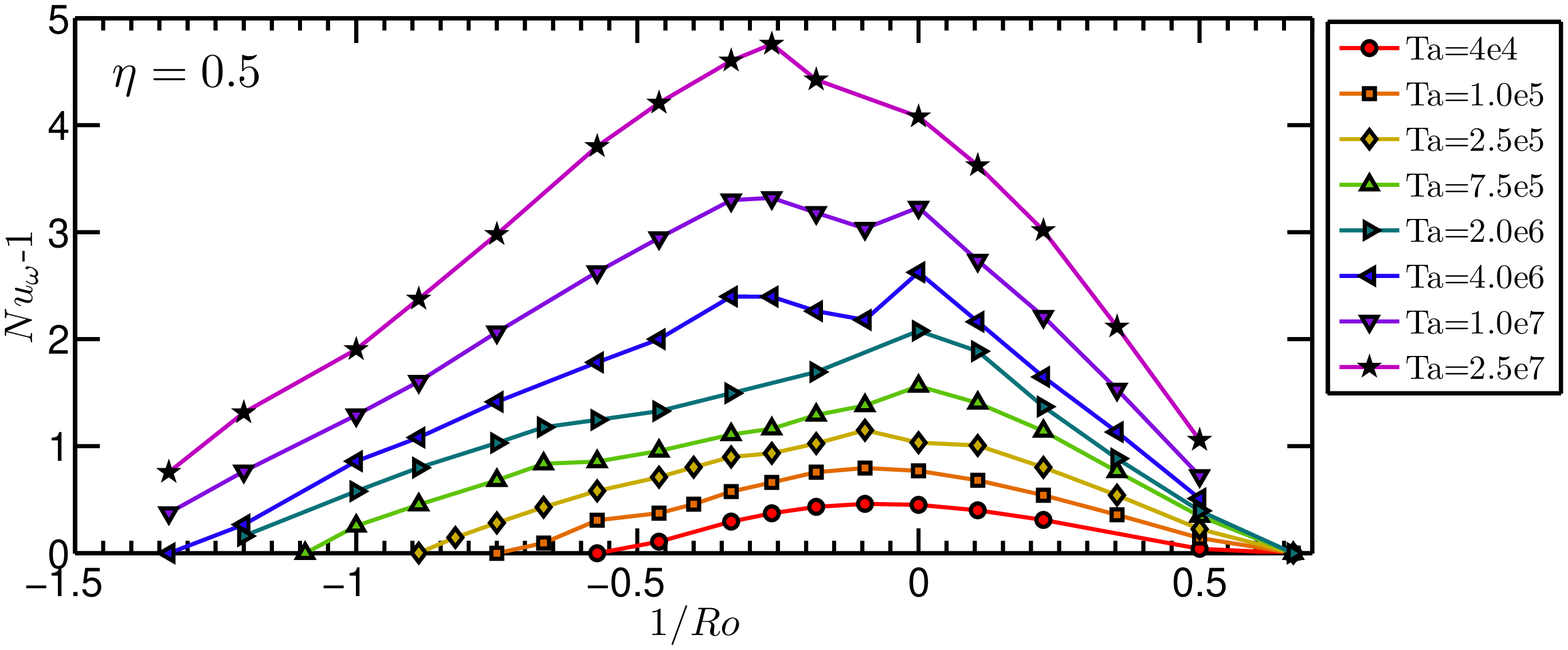}}\\
  \subfloat{\label{fig:RoNuNNEta0714}\includegraphics[width=0.90\textwidth]{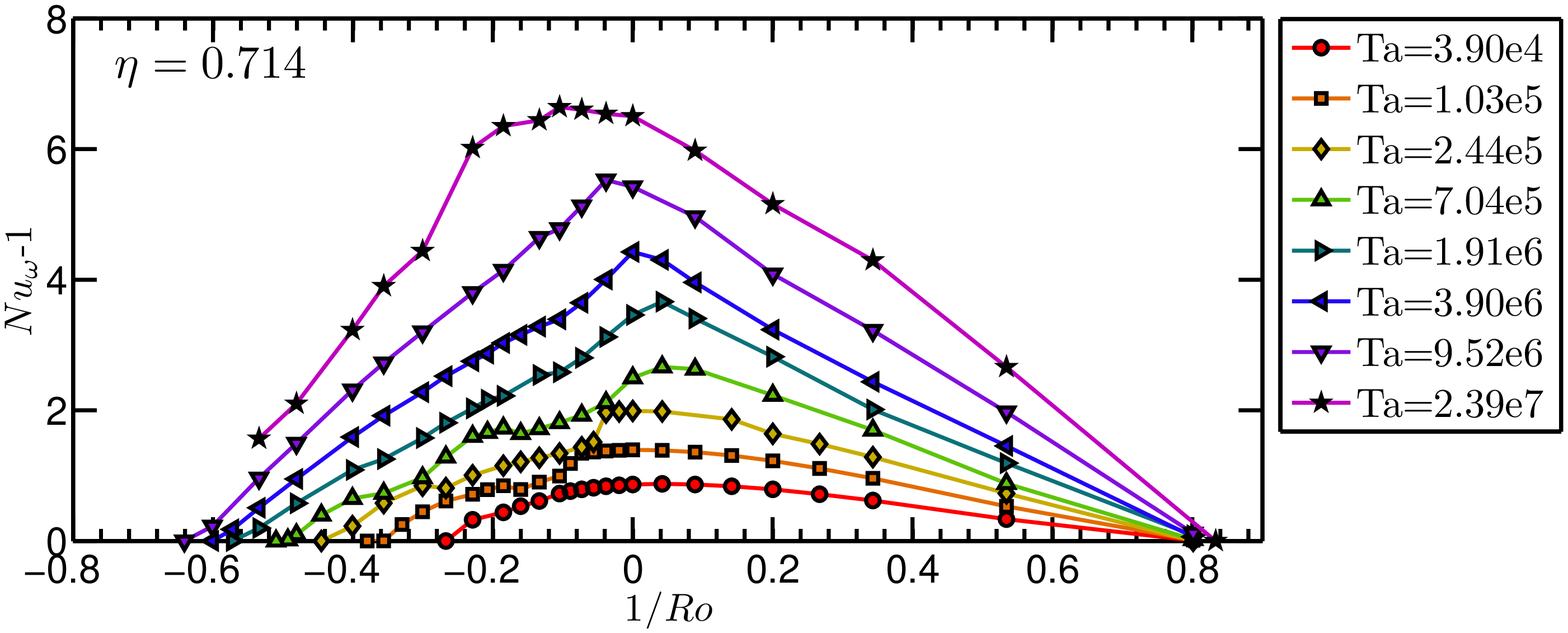}}\\
  \subfloat{\label{fig:RoNuNNEta0833}\includegraphics[width=0.90\textwidth]{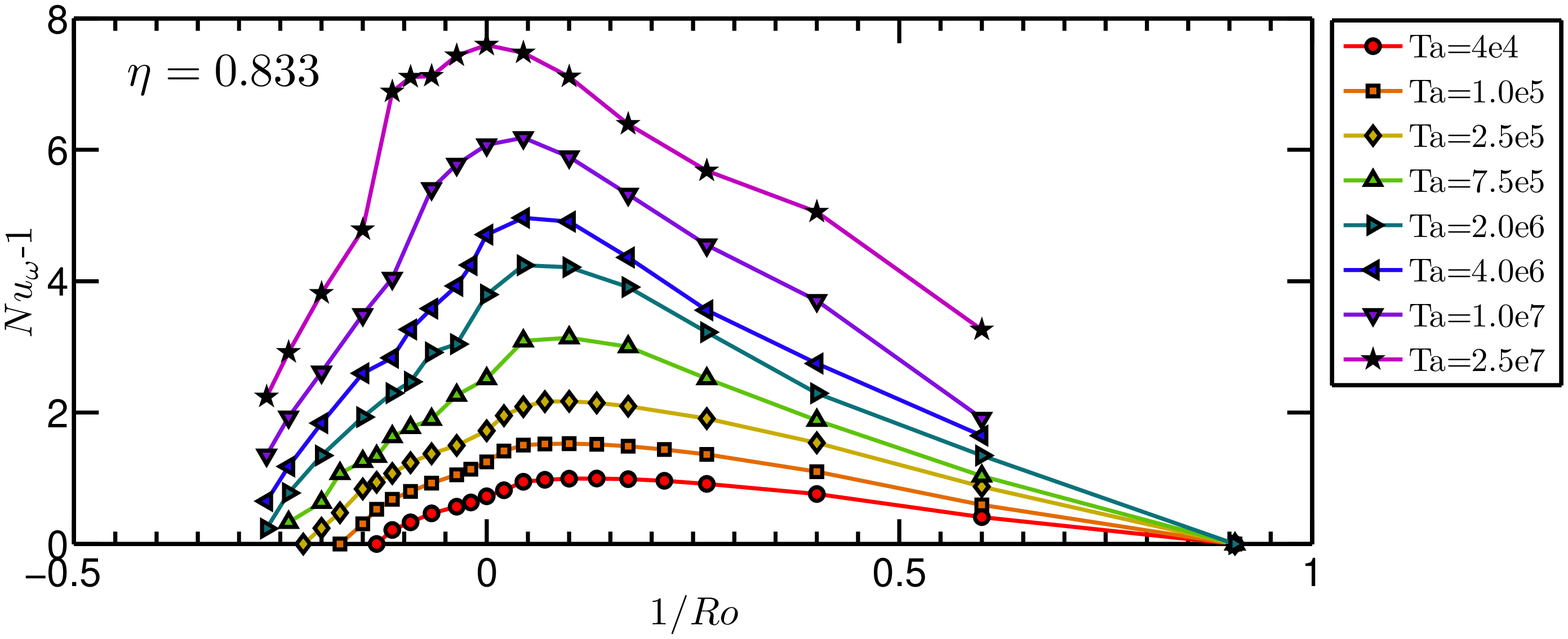}}\\
  \subfloat{\label{fig:RoNuNNEta0909}\includegraphics[width=0.90\textwidth]{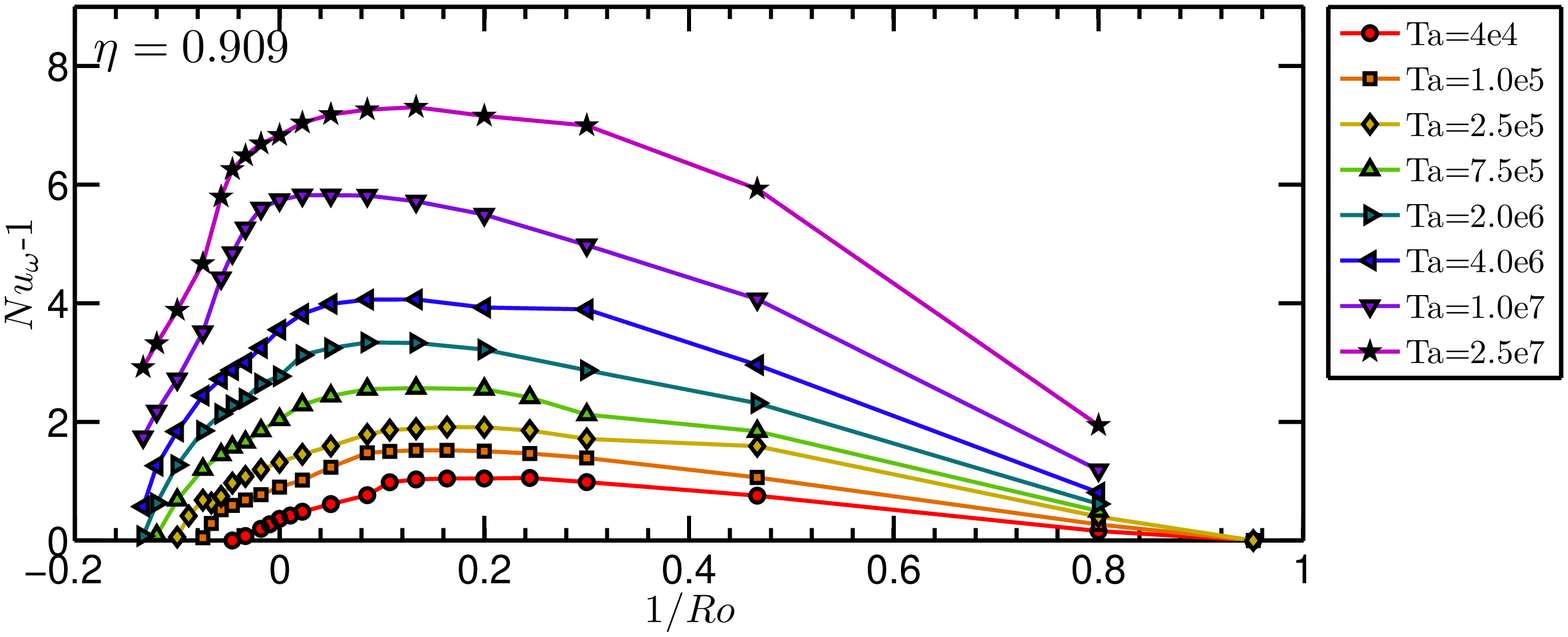}}\\
  \caption{$\nom-1$ versus $\usro$ for the four values of $\eta$ studied numerically, $\eta=0.5$ (top), $0.714$, 
$0.833$ and $0.909$ (bottom). The shape of the curve and the position of the maximum depend very strongly on both $Ta$ and $\eta$. }
 \label{fig:RoNuNN_nums}
 \end{center} 
\end{figure}

Figure \ref{fig:RoNuNN_nums} shows the results of the numerical exploration of the $\usro$ parameter 
space between $Ta=4\cdot10^4$ and $Ta=2.5\cdot 10^7$.
The shape of $\nom = \nom(\usro)$ curves and the position of $\usro_{opt}$ depends very strongly on $\eta$ 
in the $Ta$ range studied in numerics. For the largest gap (i.e., $\eta=0.5$), the optimum 
can be seen to be in the counter--rotating range (i.e., $\usro<0$) as long as $Ta$ is high enough. 
On the other hand, for the smallest gap (i.e., $\eta=0.909$), 
the optimum is at co--rotation (i.e., $\usro>0$) in the whole region studied.
The other values of studied $\eta$ reveal an intermediate behavior.
Optimum transport is located for co--rotation at lower values of $Ta$ and slowly moves towards 
counter--rotation. For all values of $\eta$, when the driving is increased, $\usro_{opt}$ 
tends to shift to more negative values. 

For two values of $Ta$ ($Ta=4\cdot10^6$ and $Ta=10^7$) for a radius ratio $\eta=0.5$,
two distinct peaks can be seen in the $\nom(\usro)$ curve. This can be understood by looking at the flow topology.
For $\usro=0$, three distinct rolls can be seen. However, when decreasing $\usro$, the rolls begin to break up. Some
remnants of large-scale structures can be seen, but these are weaker than the $\usro=0$ case. Having a large-scale
roll helps the transport of angular momentum, leading to the peak in $\nom$ at $\usro=0$. 
Further increasing the driving causes the rolls to also break up for $\usro=0$, and eliminates the anomalous peak.

\begin{figure}
 \begin{center}
  \subfloat{\label{fig:TaNuAllExp}\includegraphics[height=0.3\textwidth]{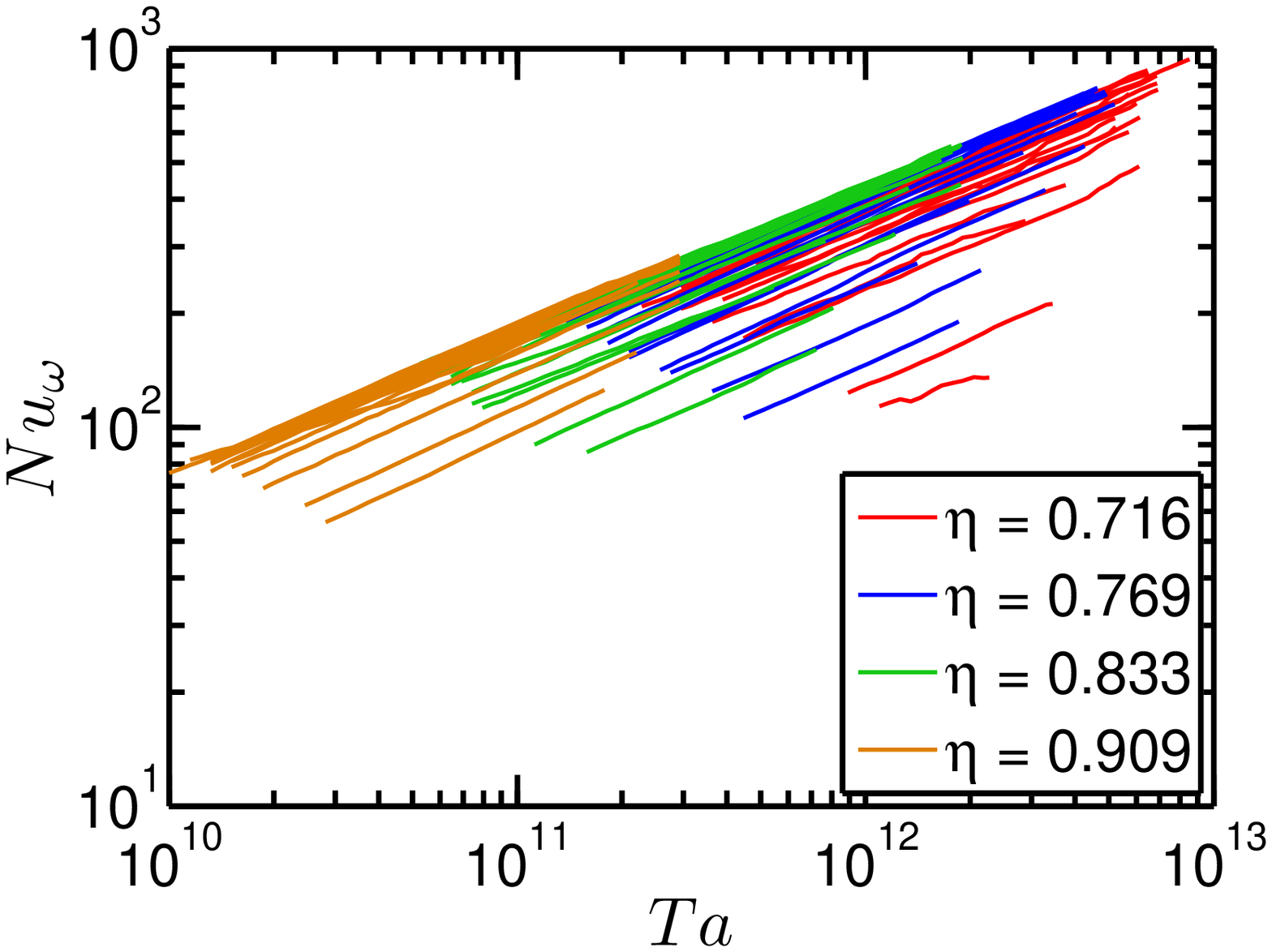}}
  \subfloat{\label{fig:TaNuAllGamma}\includegraphics[height=0.3\textwidth]{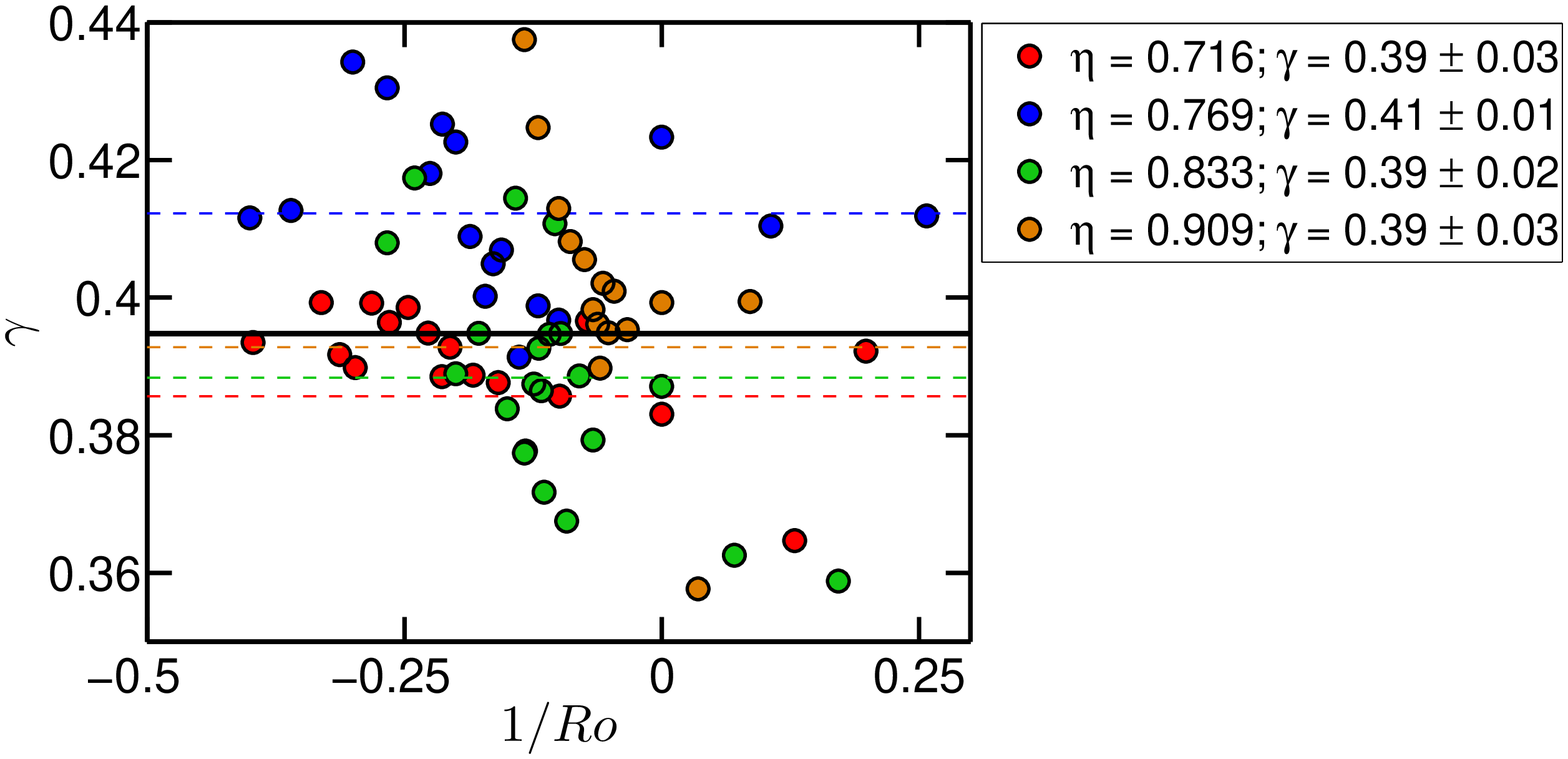}}
  \caption{The left panel shows $\nom$ versus $Ta$ for all values of $\eta$ and 
$\usro$ studied in experiments. The right panel shows the exponent $\gamma$ of the scaling law $\nom \propto Ta^{\gamma}$ 
for various $\usro$, obtained by a least-square linear fit in log-log space. The average value of $\gamma$ 
for each $\eta$ is represented by the dashed lines, while the solid line represents 
the average value of $\gamma=0.39$ for all $\eta$, which will be used for compensating
$\nom$. }
 \label{fig:NuTaAllRosExp}
 \end{center} 
\end{figure}

\begin{figure}
\begin{sideways}
  \centering
  \subfloat{\label{fig:TaNuCompEta0714}\includegraphics[width=0.60\textwidth]{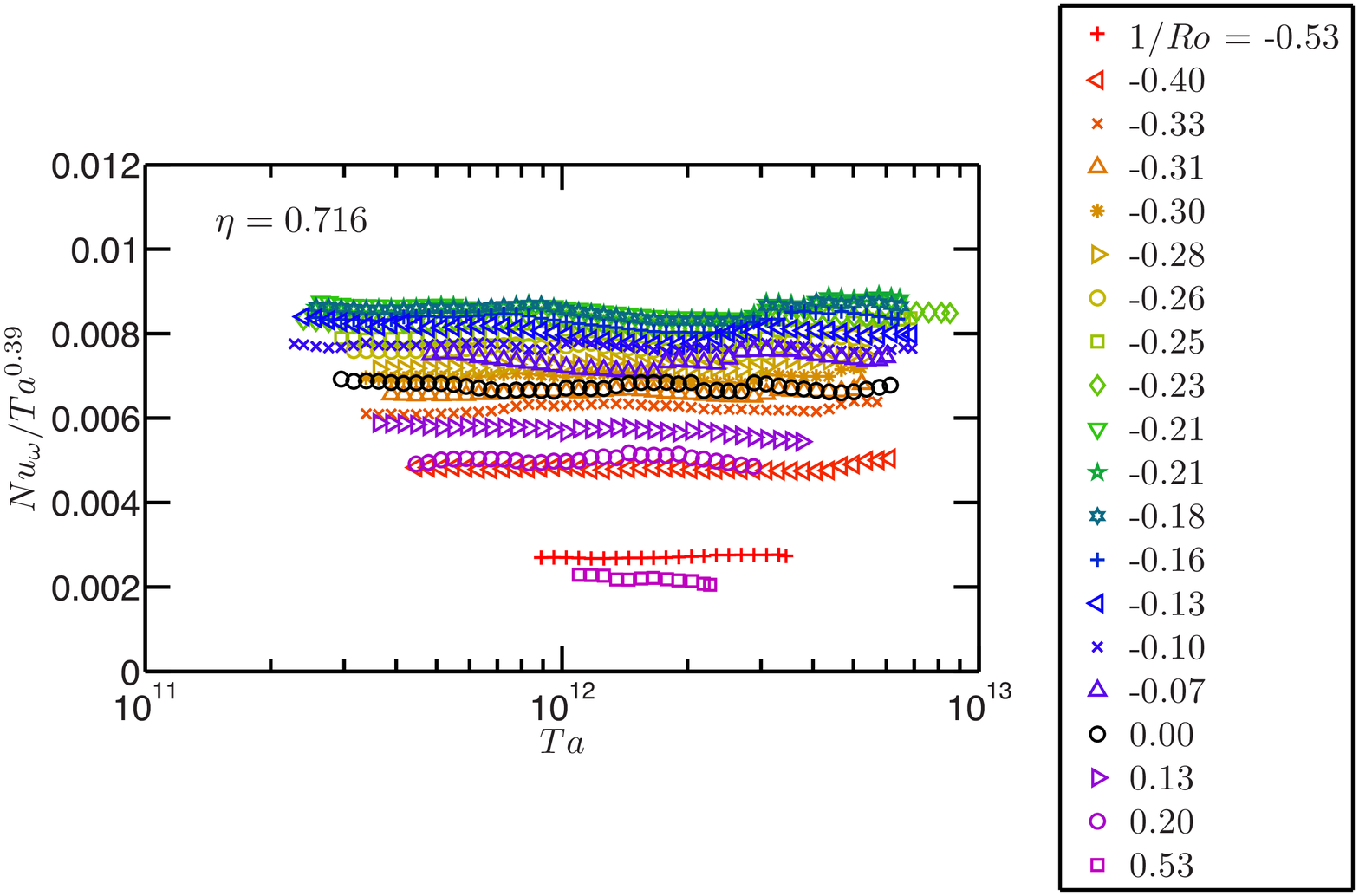}}
  \subfloat{\label{fig:TaNuCompEta0769}\includegraphics[width=0.60\textwidth]{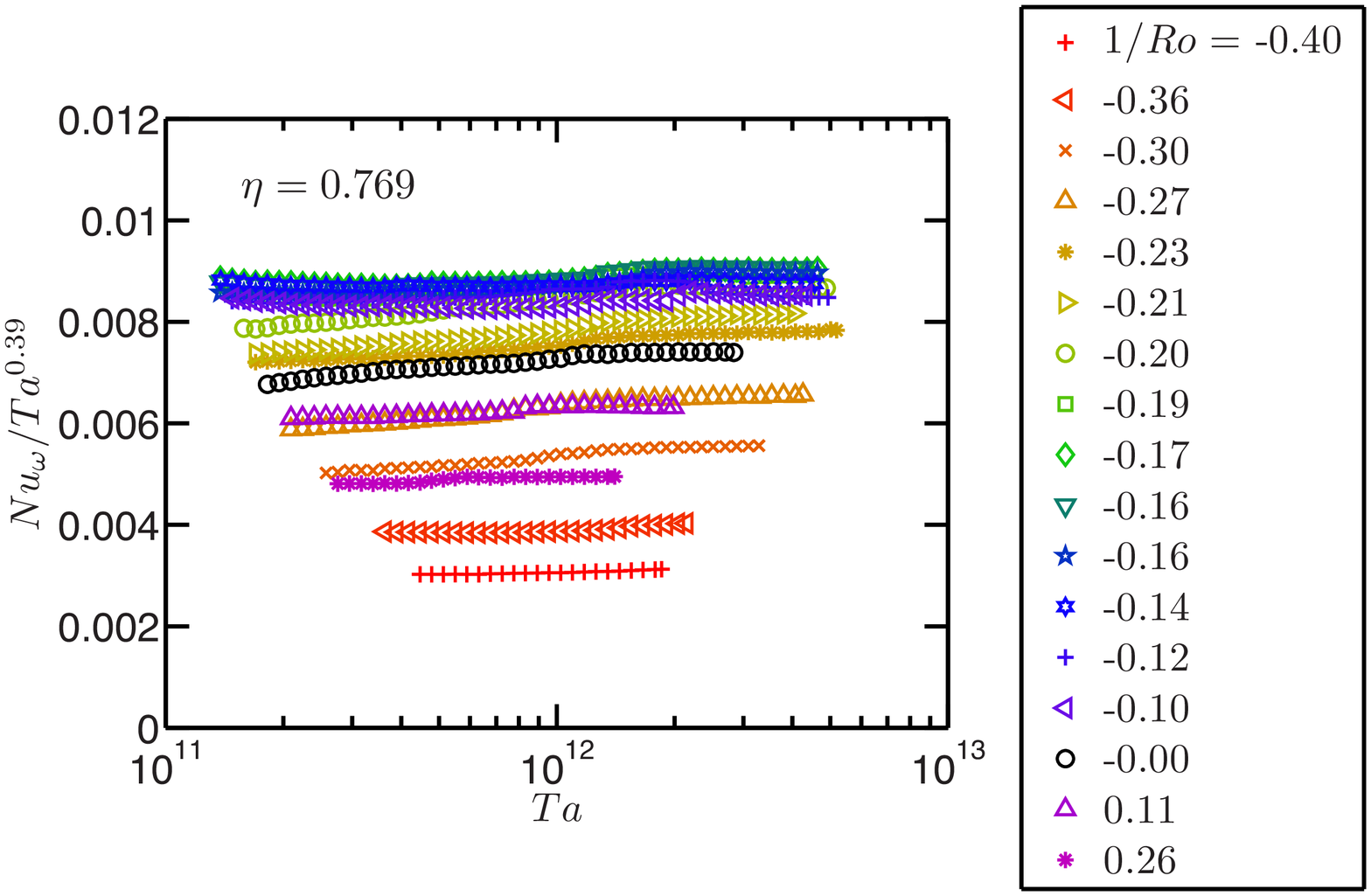}}
\end{sideways}
\mbox{\\}
\begin{sideways}  
  \subfloat{\label{fig:TaNuCompEta0833}\includegraphics[width=0.60\textwidth]{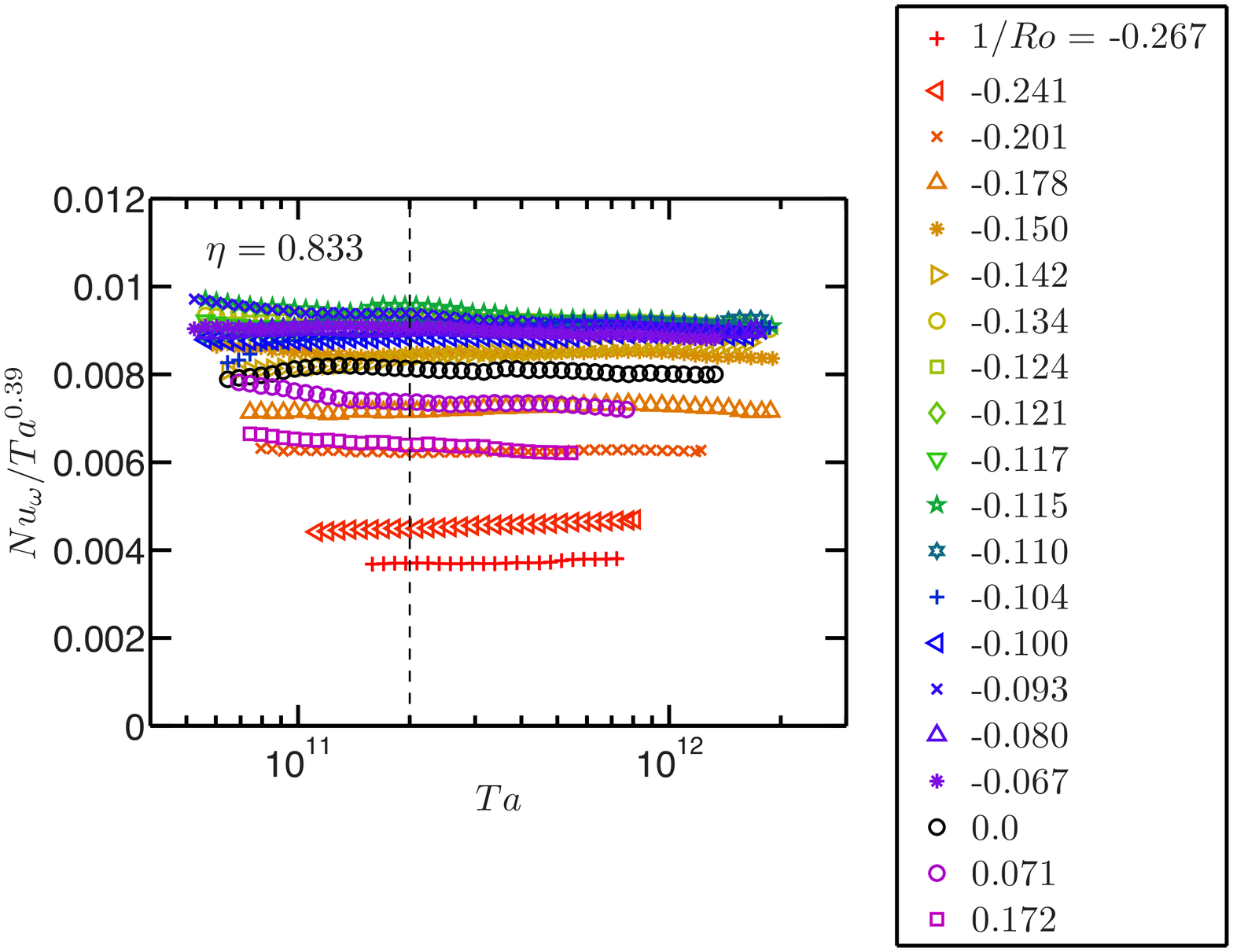}}
  \subfloat{\label{fig:TaNuCompEta0909}\includegraphics[width=0.60\textwidth]{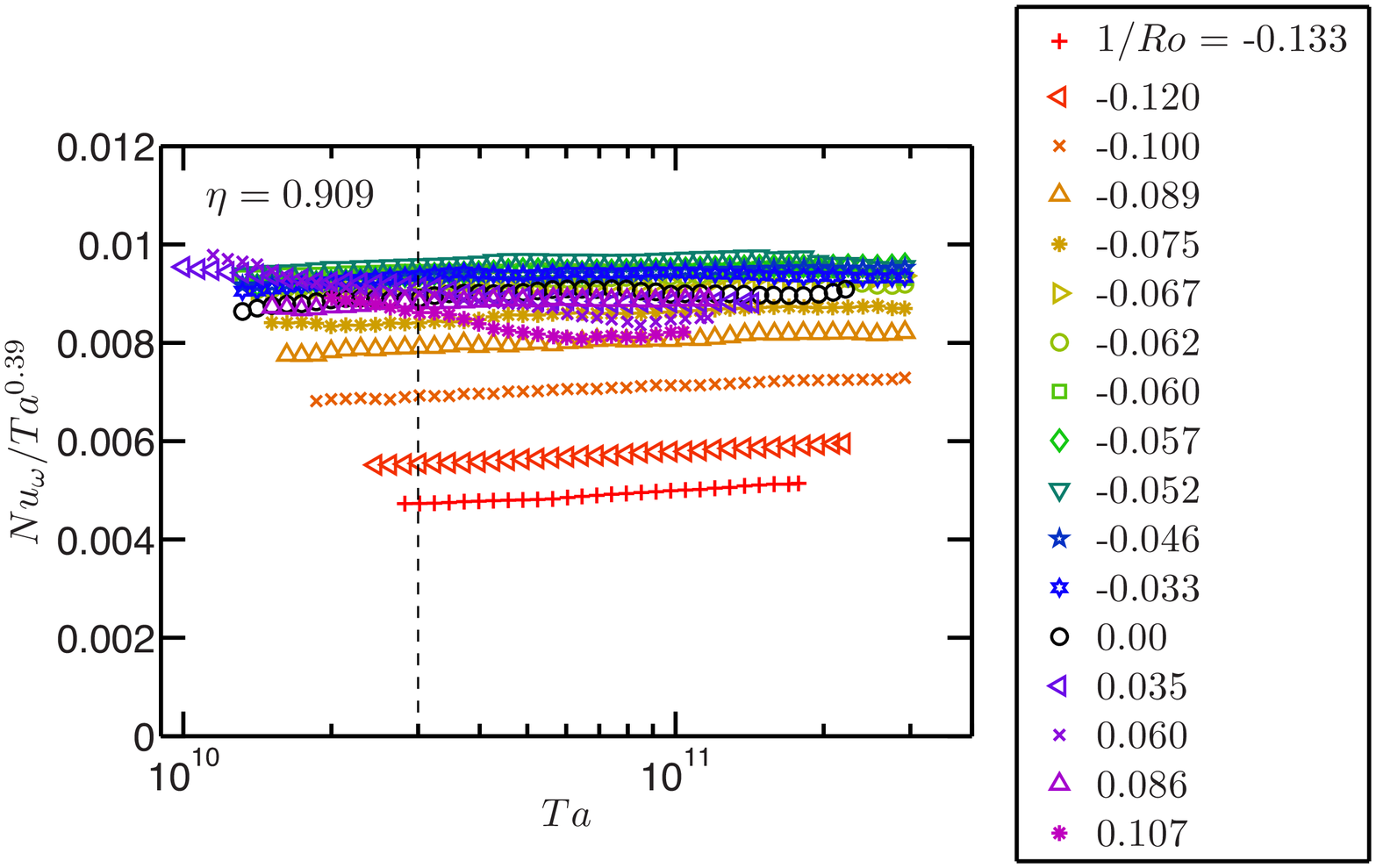}}
\end{sideways}
 \caption{The four panels show $\nom/Ta^{0.39}$ vs. $Ta$ for all
explored values of $\usro$ for the four studied values of $\eta$: $\eta=0.714$ (top-left), $0.769$ (top-right), $0.833$ (bottom-left), and
$0.909$ (bottom-right). The scaling law $\nom\sim Ta^{0.39}$ is 
seen to approximately hold throughout the whole parameter space explored. 
No trends can be appreciated which would lead us to expect further shift of the optimum 
with increased driving. Dashed lines, indicating the cut-off regions used for determining $\langle\nom/Ta^{0.39}\rangle_{Ta}$
have been plotted for $\eta=0.833$ and $\eta=0.909$.}
  \label{fig:NuTaCompAllExp}
\end{figure}

The shift seen in the numerics may or may not continue with increasing $Ta$. The experiments conducted 
explore a parameter space of $10^{10}<Ta<10^{13}$ and thus serve to explore the shift at 
higher driving. Figure \ref{fig:NuTaAllRosExp} presents the obtained results. The left panel 
shows $\nom$ versus $Ta$ for all measurements. The right panel shows the exponent $\gamma$, 
obtained by fitting a least-square linear fit in the log--log plots. Across the $\eta$ and $\usro$ range 
studied, the average exponent is $\gamma\approx0.39$. This value is used in figure
 \ref{fig:NuTaCompAllExp} to compensate $\nom$. The horizontality of all data points reflects the good
scaling and the universality of this ultimate scaling behaviour $\nom \propto Ta^{0.39}$.

 To determine the optimal rotation ratio for the experimental data, a $Ta$-averaged compensated 
Nusselt $\langle\nom/Ta^{0.39}\rangle_{Ta}$ was used. This is defined as:

\begin{equation}
 \langle\nom/Ta^{0.39}\rangle_{Ta} = \displaystyle\frac{1}{Ta_{max}-Ta_{co}} \int^{Ta_{max}}_{Ta_{co}} \nom/Ta^{0.39} dTa,
\end{equation}

\noindent where $Ta_{max}$ is the maximum value of $Ta$ for every ($\eta,\usro$) dataset, and $Ta_{co}$ is a cut-off $Ta$ number used
for the larger $\eta$ ($Ta_{co}=2\cdot10^{11}$  for $\eta=0.833$ and $Ta_{co}=3\cdot10^{10}$ for $\eta=0.909$) to exclude
the initial part of the $\nom/Ta^{0.39}$ data points which seem to have a different scaling for some
of the values of $\usro$ explored. For the smaller values
of $\eta$, $Ta_{co}=Ta_{min}$, the minimum value of $Ta$ for every ($\eta,\usro$) dataset. 
An error bar on this average is estimated as one standard deviation
of the data from the computed average.

\begin{figure}
 \begin{center}
  \subfloat{\label{fig:RoNuTamaxExp}\includegraphics[width=0.96\textwidth]{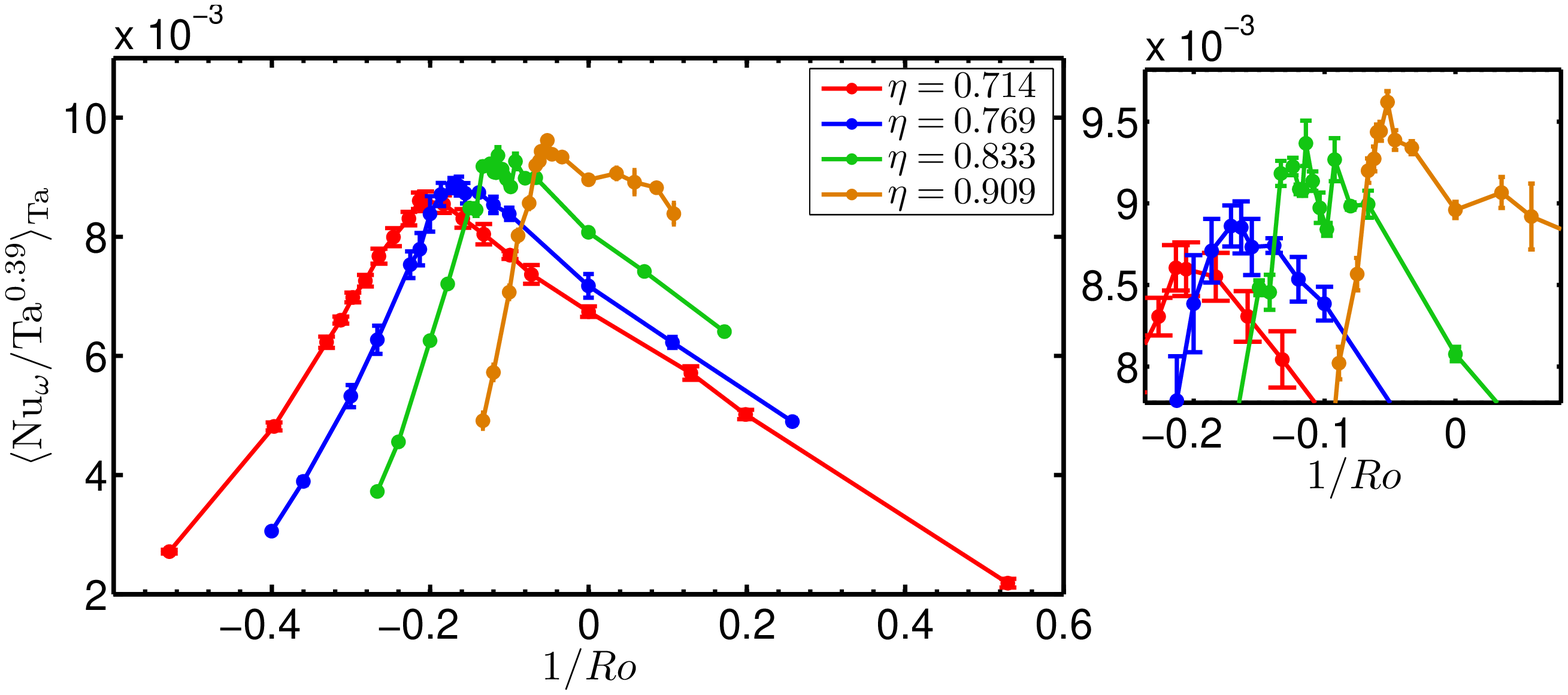}}\\
  \subfloat{\label{fig:aNuTamaxExp}\includegraphics[width=0.96\textwidth]{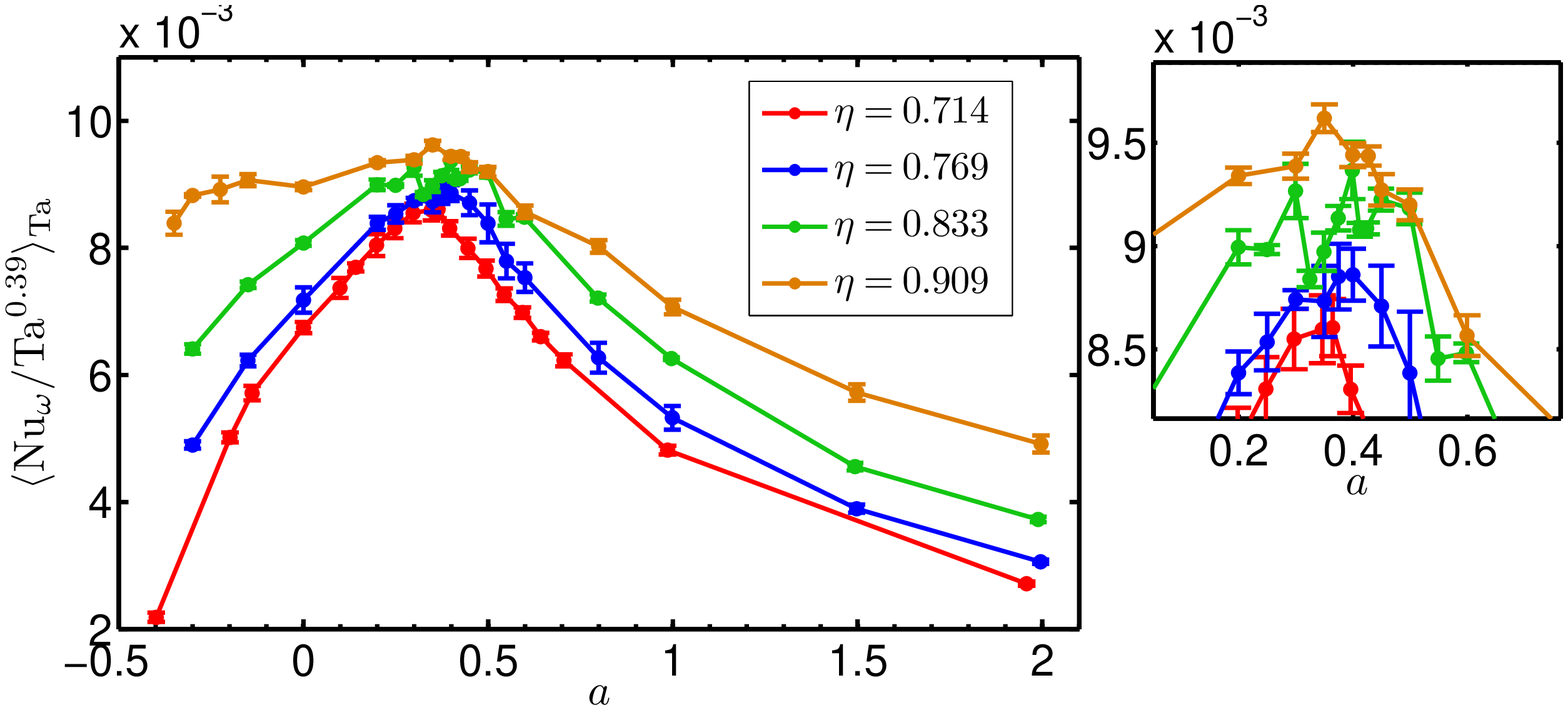}}\\
  \caption{ The panels
show $\langle\nom/Ta^{0.39}\rangle_{Ta}$ versus either $\usro$ (top) or $a$ (bottom) 
at the cut-off region highlighted in figure \ref{fig:NuTaCompAllExp}
for the values of $\eta$ studied experimentally. Insets containing a zoom-in around the optimum have been added for clarity. 
Error bars indicate one standard deviation from the mean value, and are too small to be seen for most data points.
There is a strong $\eta$-dependence of the curve $\nom/Ta^{0.39}$ versus $\usro$, even at the largest
drivings studied in experiments. Optimal transport is located at $\usro_{opt}=-0.20$ for $\eta=0.714$,
$\usro_{opt}=-0.15$ for $\eta=0.769$, $\usro_{opt}=-0.10$ for $\eta=0.833$ and $\usro_{opt}=-0.05$ for
 $\eta=0.909$, corresponding to $a\approx 0.33-0.35$ for all values of $\eta$. In the bottom panel,
the maximum of the graph is less pronounced, i.e. it becomes more flat with increasing 
$\eta$. In the limit $\eta\to 1$, $a$ does not tend to a finite limit, while $\usro$ does. This result
highlights the advantage of using $\usro$ instead of $a$ as a control parameter.}
 \label{fig:RoNuTamax}
 \end{center} 
\end{figure}

The two panels of figure \ref{fig:RoNuTamax} show 
$\langle\nom/Ta^{0.39}\rangle_{Ta}$ as a function of $\usro$ or alternatively of
$a$ for the four values of $\eta$ considered in experiments. 
The increased driving changes the characteristics of the flow. This is reflected in the very
different shapes of the $\usro$-dependence of $\nom$ when comparing figures \ref{fig:RoNuNN_nums} 
and \ref{fig:RoNuTamax}, and in the shift of $\usro_{opt}$. 

\begin{figure}
 \begin{center}

  \subfloat{\label{fig:PeakwidthTasSim}\includegraphics[width=0.47\textwidth]{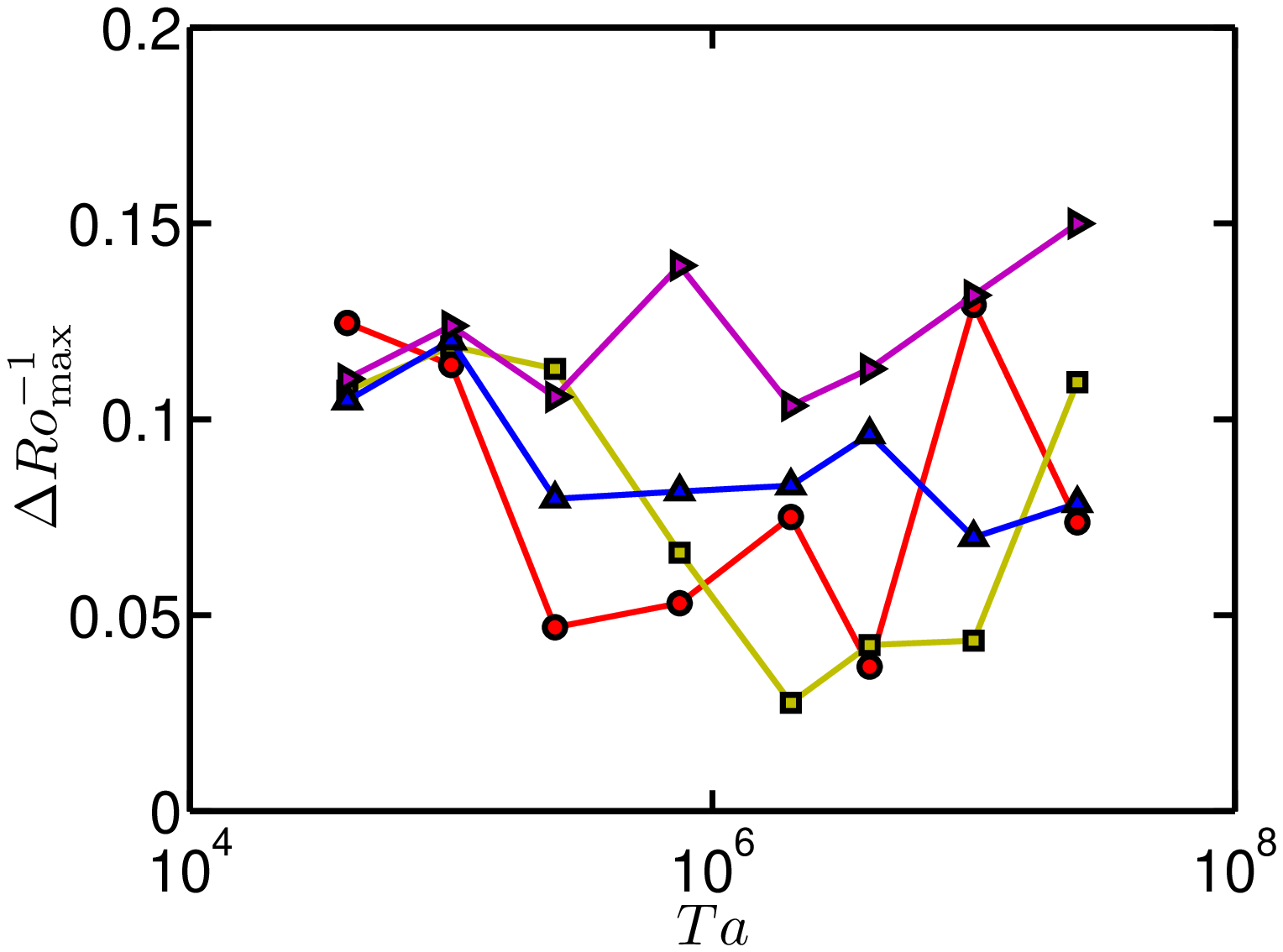}}
  \subfloat{\label{fig:RooptTasSim}\includegraphics[width=0.47\textwidth]{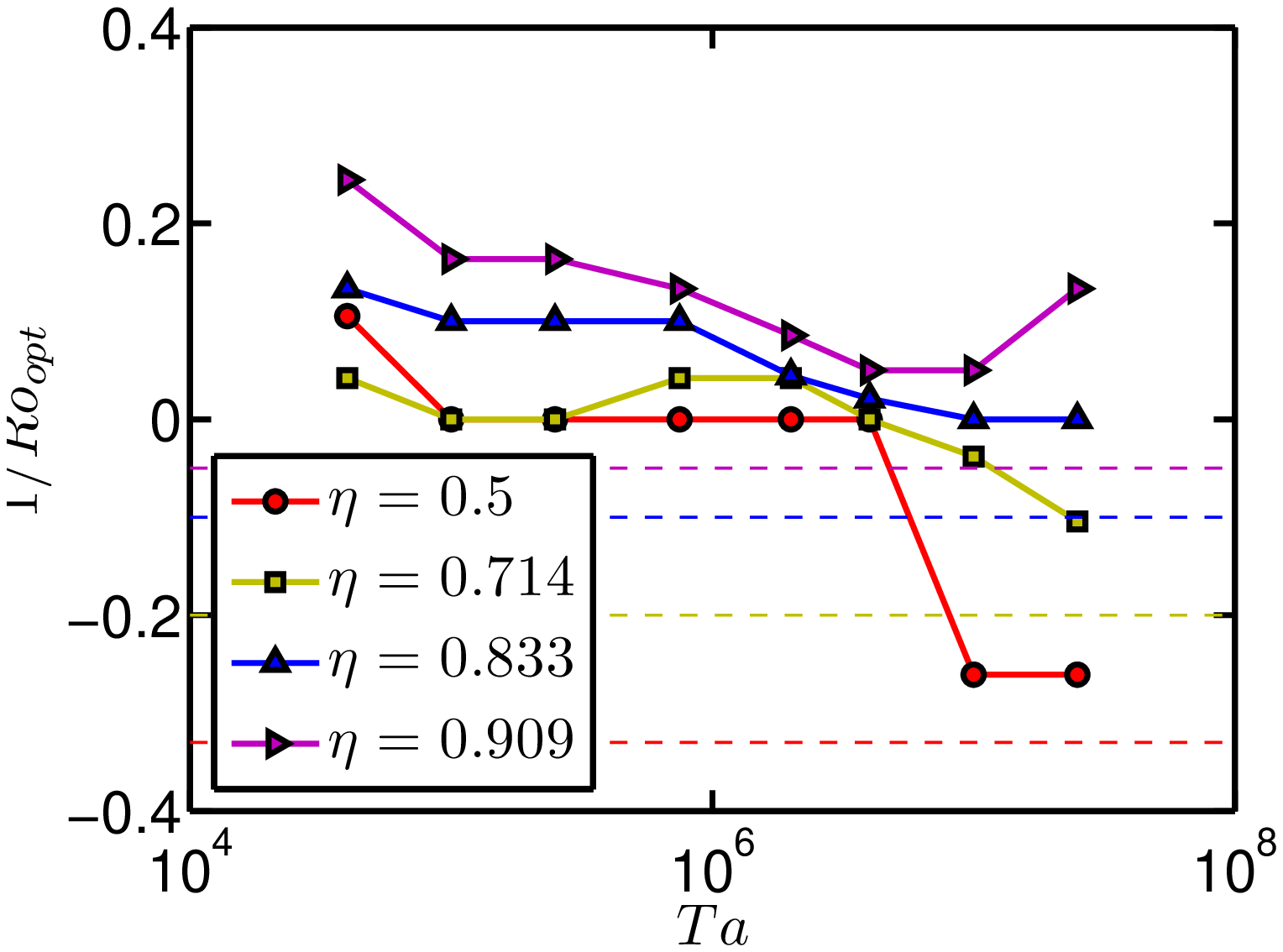}}\\
   \caption{In the left panel, $95\%$ peak width $\Delta\usro_{max}$ vs $Ta$ for the four values of 
$\eta$ analyzed in numerics. The peak width can be seen to vary with driving, and for smaller
gaps is larger for larger values of $Ta$. In the right panel, $\usro_{opt}$ vs $Ta$ for the same four values of $\eta$. 
The location of the optimal transport has a very strong dependence on the driving, especially for the largest 
values of $\eta$. As driving increases beyond the numerically studied range and overlaps with experiments,
 $\usro_{opt}$ should tend to the experimentally found values, represented as dashed lines 
in the figure. The asymptote for $\eta=0.5$ is obtained from \cite{mer12}. 
The trend appears to be less clear for $\eta=0.909$, but this might be understandable from the peak width at the
highest driving $Ta$.}
 \label{fig:RooptTaEtas}
 \end{center} 
\end{figure}

To summarize these effects, figure \ref{fig:RooptTaEtas} presents both the $95\%$ peak width $\Delta\usro_{max}$
and the position of optimal transport $\usro_{opt}$ determined as the realization with the maximum torque
as a function of $Ta$ and $\eta$ obtained
from numerics as well as the asymptotic value from experiments. The peak width $\Delta\usro_{max}$
is defined as:

\begin{equation}
 \Delta\usro_{max} = \displaystyle\frac{\displaystyle\int_{\usro_{-0.95}}^{\usro_{0.95}} \nom(\usro) d\usro }{\mathrm{max}(\nom - 1)}
\end{equation}

\noindent where $\usro_{-0.95}$ and $\usro_{0.95}$ are the values of $\usro$ for which $\nom$ is $95\%$ of the peak value.

The $95\%$ peak width can be seen to vary with driving, reflecting what is seen in figure
\ref{fig:RoNuNN_nums}. The shape of the $\usro$-$\nom$ curve is highly dependent of both $\eta$ and $Ta$.
$\usro_{opt}$ shows a very large
variation across the $Ta$ range studied in numerics. The shift of the $\usro_{opt}$ with $Ta$ is expected to continue until
it reaches the value found in the experiments. This can be seen in the right panel of figure \ref{fig:RoNuNN_nums} 
for $\eta=0.5$ to $\eta=0.833$. For $\eta=0.909$ , the trend seems to change for the last point. However, this is due to the
very large and flat peak of the $\nom(\usro)$ curve- this can also be seen in the left panel and in figure \ref{fig:RoNuNNEta0909}. 

One may also ask the question: 
has the value of $\usro_{opt}$  already
saturated in our experiments? Figure \ref{fig:NuTaCompAllExp} shows the trend for $\nom$ for increasing
$Ta$. This trend does not seem to vary much for different values of $\usro$. 
Therefore, we expect the value of $\usro_{opt}$ to have
already reached saturation in our experiments. 

\begin{figure}
 \begin{center}
   \includegraphics[width=0.80\textwidth]{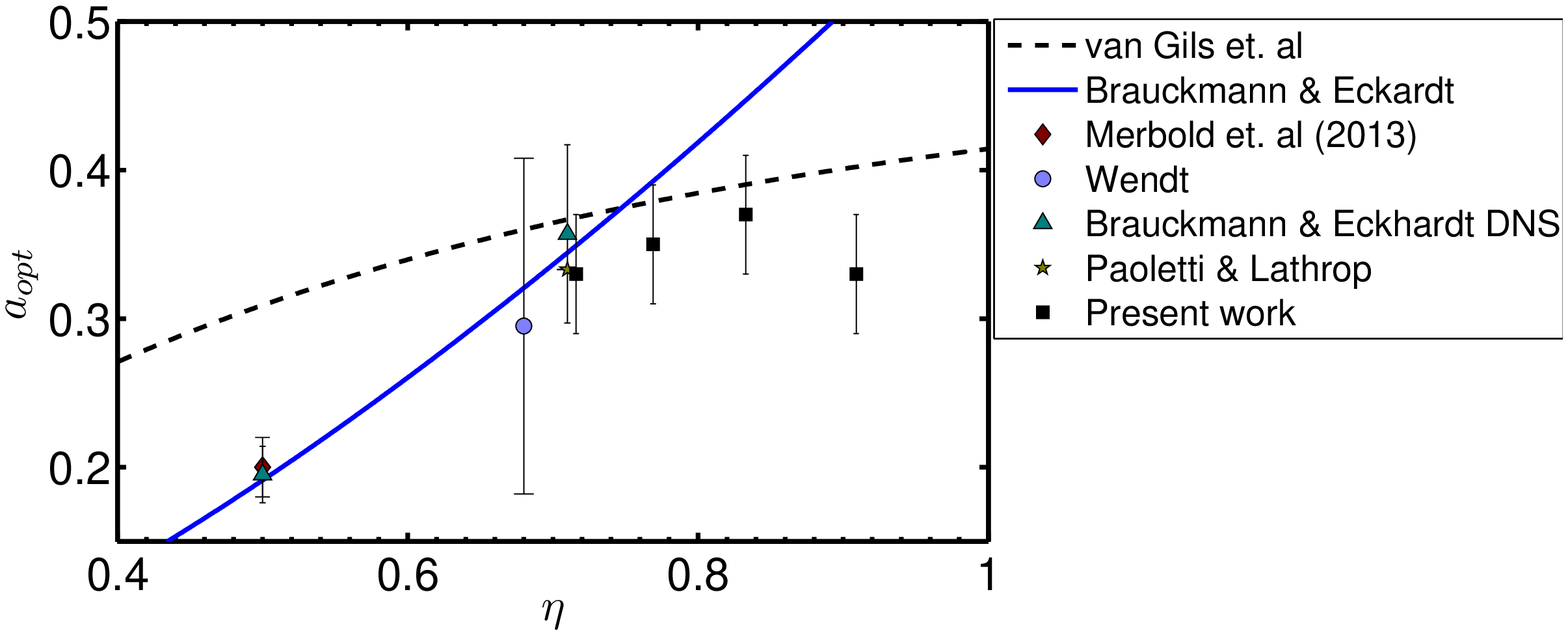}
   \caption{State-of-the-art data for $a_{opt}(\eta)$, from both experiments \citep{wen33,pao11,mer12} 
and numerics \citep{bra13}. The speculation of \cite{gil12} and the
prediction of \cite{bra13} are plotted as lines on the graph. The new experimental 
results deviate up substantially from both predictions, even when taking into account error bars.}
 \label{fig:EtaRooptComparison}
 \end{center} 
\end{figure}

We can compare these new experimental results to the available results from the literature, the speculation
made in \cite{gil12} and the prediction made in \cite{bra13} for the dependence of the saturated $a_{opt}$ on $\eta$. This is shown in
figure \ref{fig:EtaRooptComparison}. Both dependencies are shown to deviate substantially from the experimental
results obtained in the present work. Even if the speculation from \cite{gil12} appears to be better for this $\eta$-range, 
for previous experimental data at $\eta=0.5$, it is in clear difference with the experimentally measured value for 
optimal transport by \cite{mer12}. 

This section has shown that the radius ratio has a very strong effect on the global response 
and especially on optimal transport. Significantly increased
transport for co--rotation has been found at the lowest drivings based on the DNS results. This finding was already reported in 
\cite{ost12} for $\eta=0.714$, but the transport increase was marginal. For $\eta=0.833$ and especially
for $\eta=0.909$ the transport can be increased up to three times. The shift of $\usro_{opt}$ has also been seen to 
be much bigger and to happen in a much slower way for smaller gaps. The reason for this will be studied in 
Section \ref{sec:local}, using the local data obtained from experiments and numerics. 

\section{Local results}
\label{sec:local}

In this section, the local angular velocity profiles will be analyzed. Angular velocity is the transported quantity
in TC flow and shows the interplay between the bulk, where the transport is convection dominated, and the boundary-layers, 
where the transport is diffusion dominated.
Numerical profiles and experimental profiles obtained from LDA will be shown. The angular velocity gradient 
in the bulk will be analyzed and connected to the optimal transport. In addition, the boundary layers will be analyzed and 
compared to the results from the analytical formula from EGL 2007 for the BL thickness ratio in the non-ultimate regime. 

\subsection{Angular velocity profiles}

\begin{figure}
 \begin{center}
  \subfloat{\label{fig:romegaprof_Eta05}\includegraphics[height=0.33\textwidth]{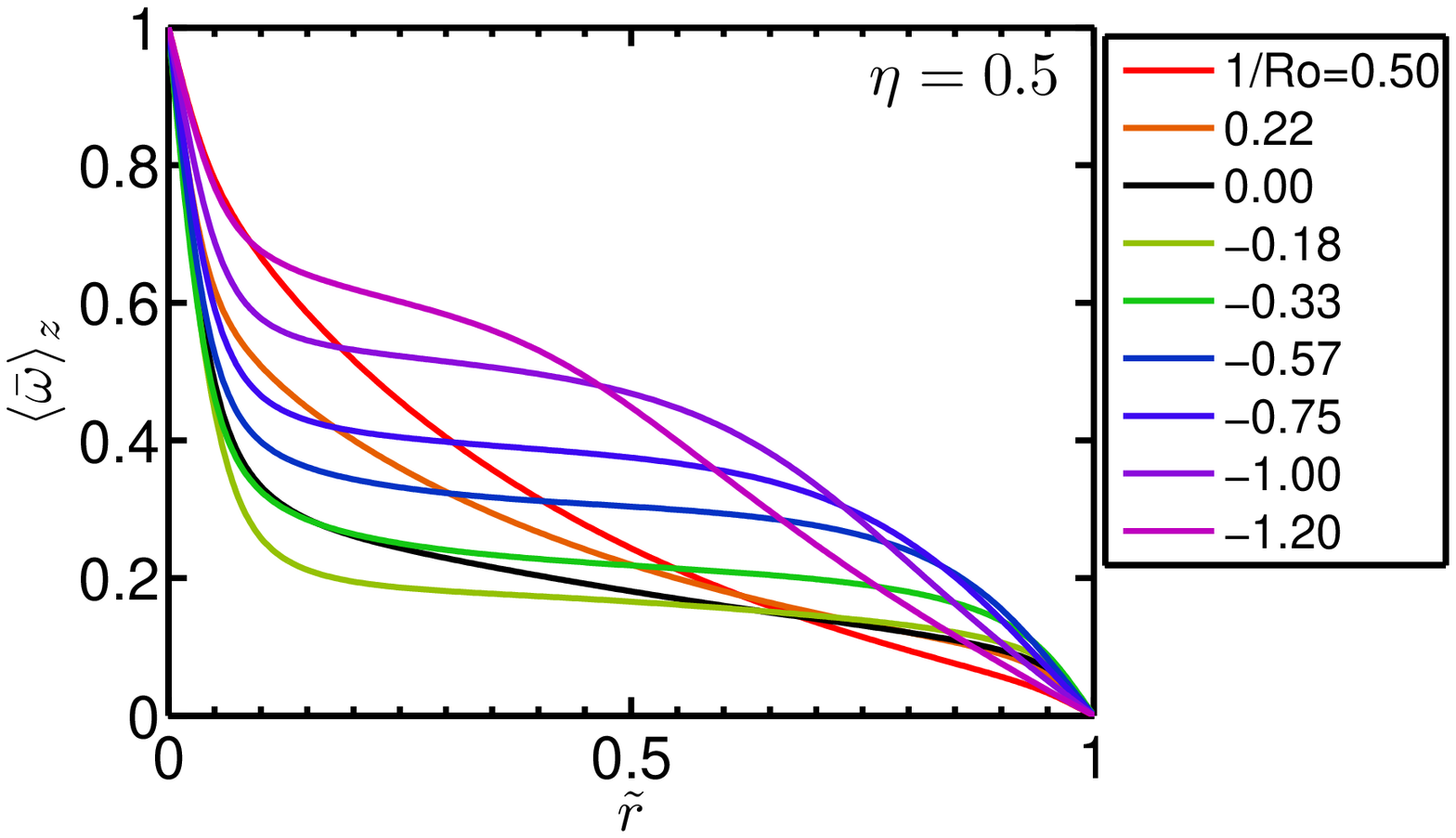}}
  \subfloat{\label{fig:romegaprof_Eta0714}\includegraphics[height=0.33\textwidth]{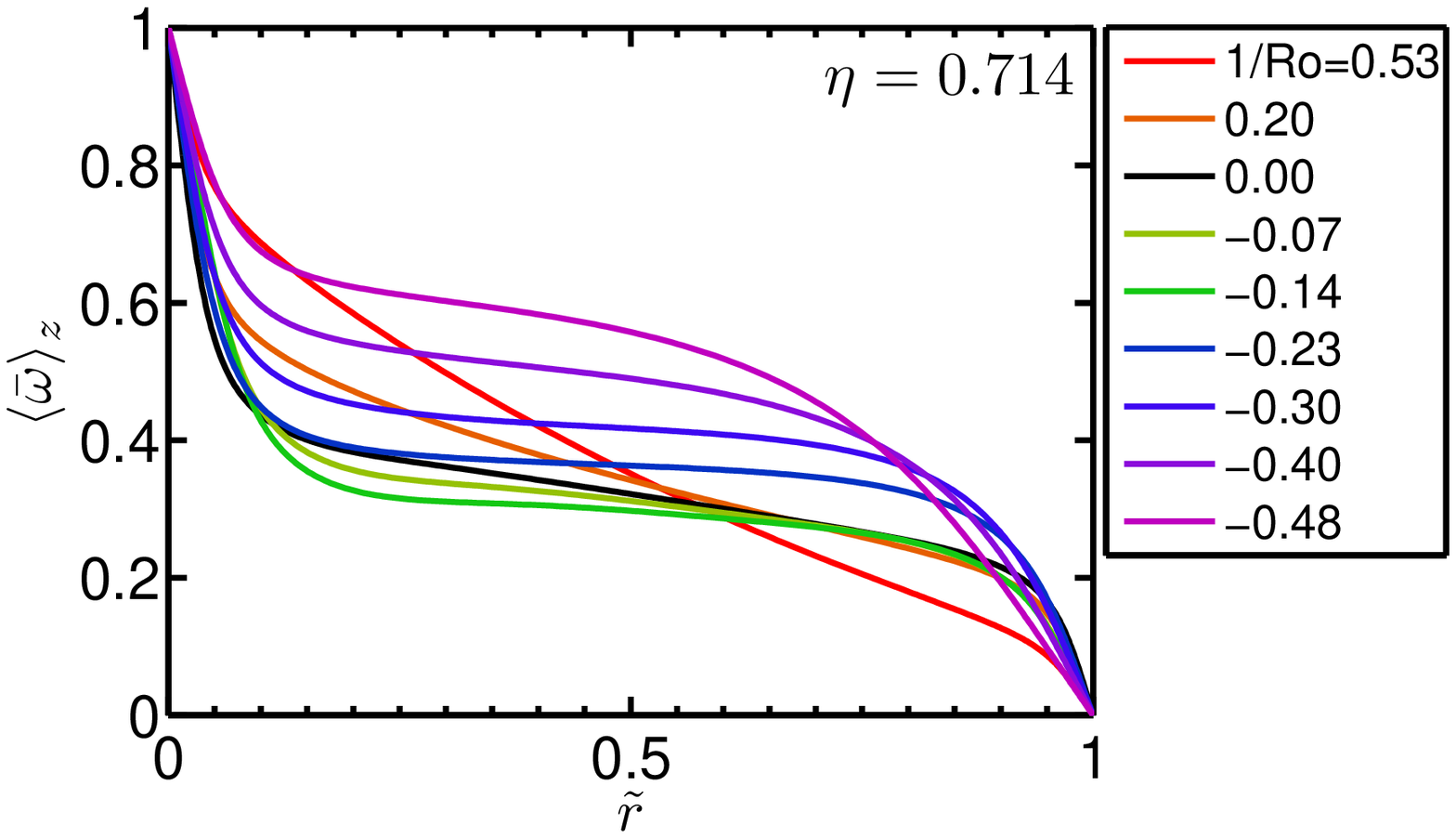}}\\
  \subfloat{\label{fig:romegaprof_Eta0833}\includegraphics[height=0.33\textwidth]{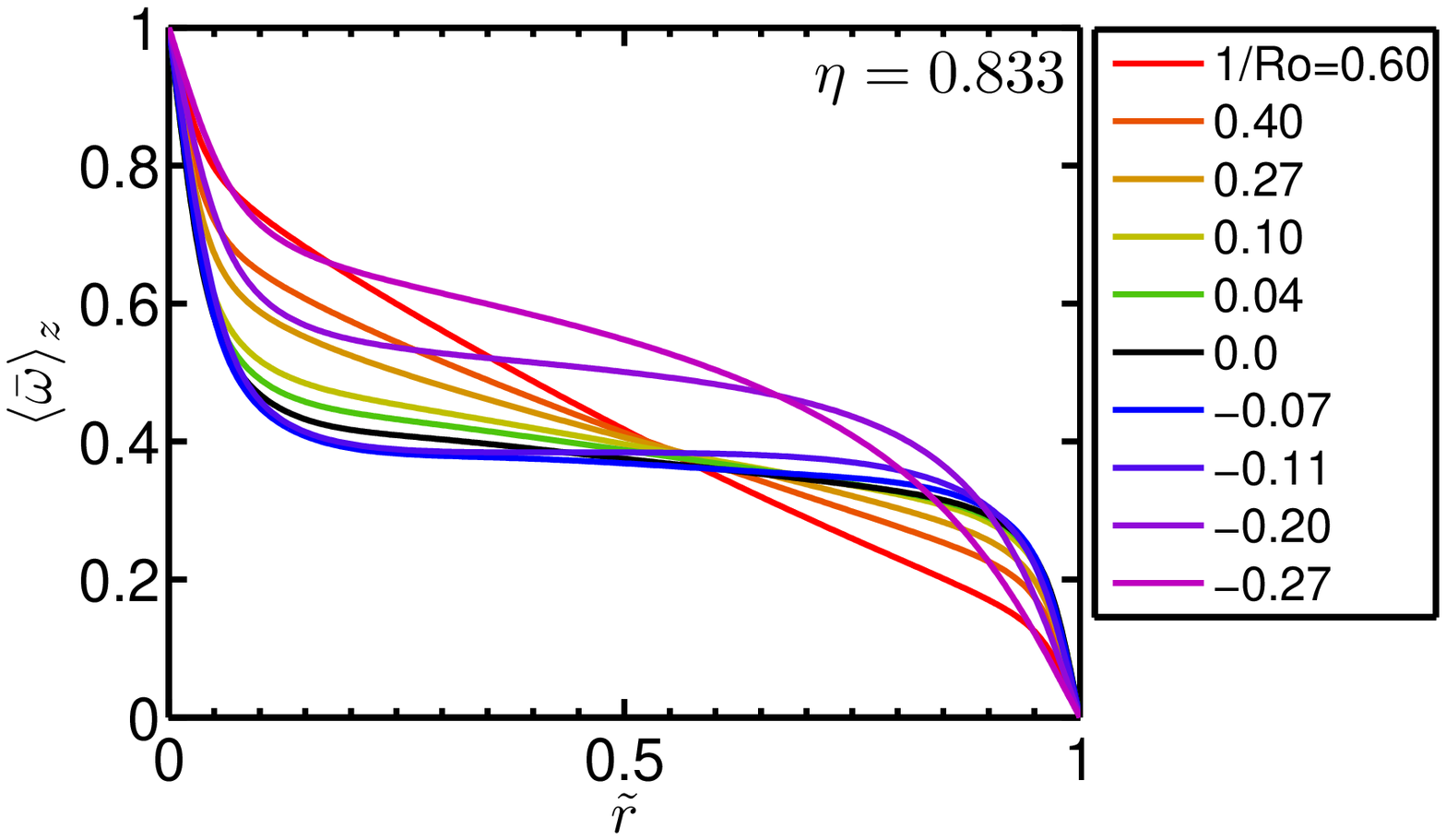}}
  \subfloat{\label{fig:romegaprof_Eta0909}\includegraphics[height=0.33\textwidth]{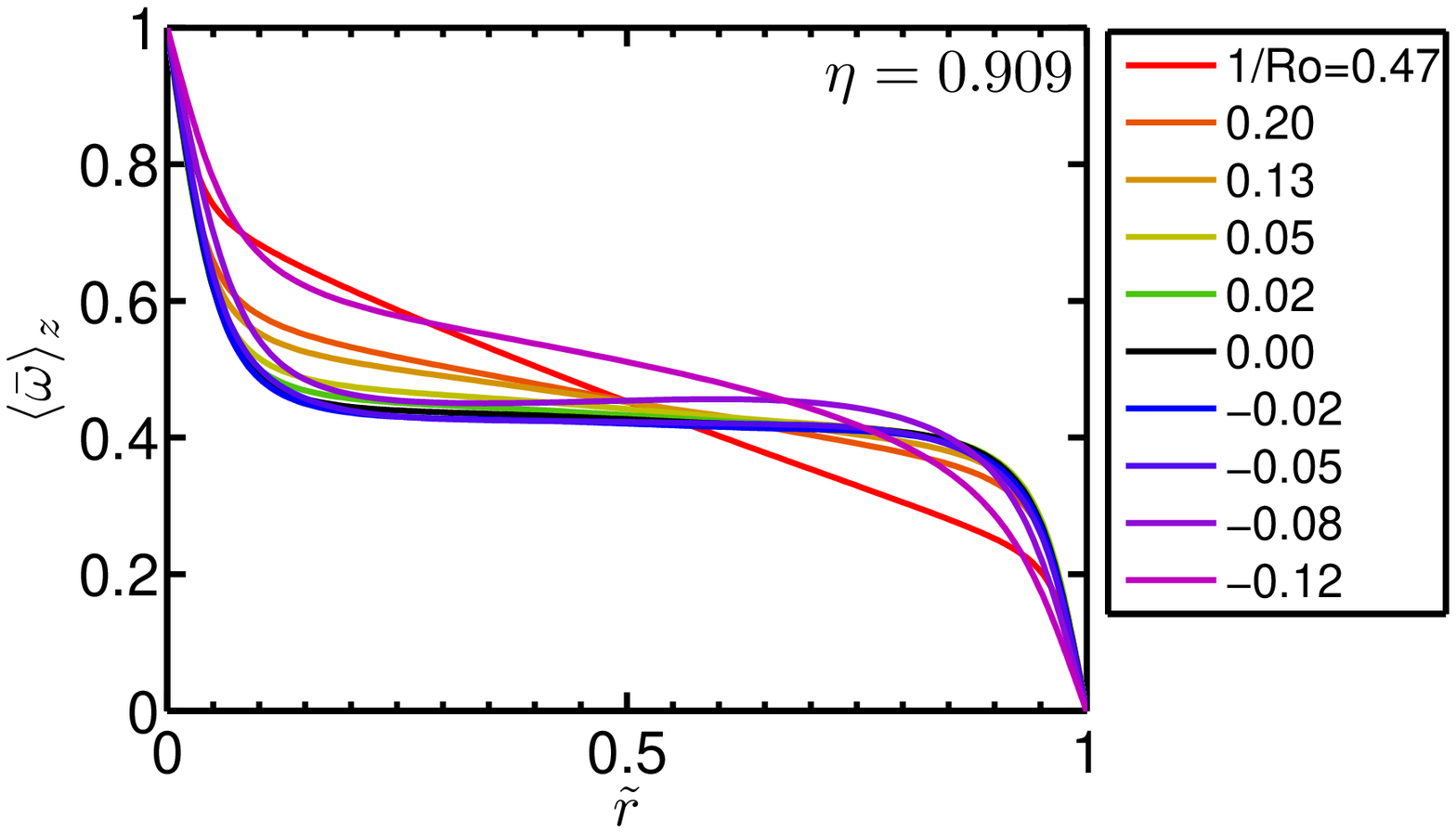}}\\
  \caption{Azimuthally, axially and temporally averaged angular velocity $\langle\bar{\omega}\rangle_z$ versus 
radius $\tilde{r}$ for:  $\eta=0.5$, $\eta=0.714$, $\eta=0.833$, and $\eta=0.909$. 
Data is for $Ta=2.5\cdot10^7$ ($Ta=2.39\cdot10^7$ for $\eta=0.714$) and selected values 
of $\usro$. For smaller $\eta$, the $\omega$-bulk profiles differ more from a straight line, and have, on 
average, a smaller value. } 
  \label{fig:romegaprof_num}
 \end{center} 
\end{figure}

\begin{figure}
 \begin{center}
  \subfloat{\label{fig:lda_Eta0714}\includegraphics[height=0.33\textwidth]{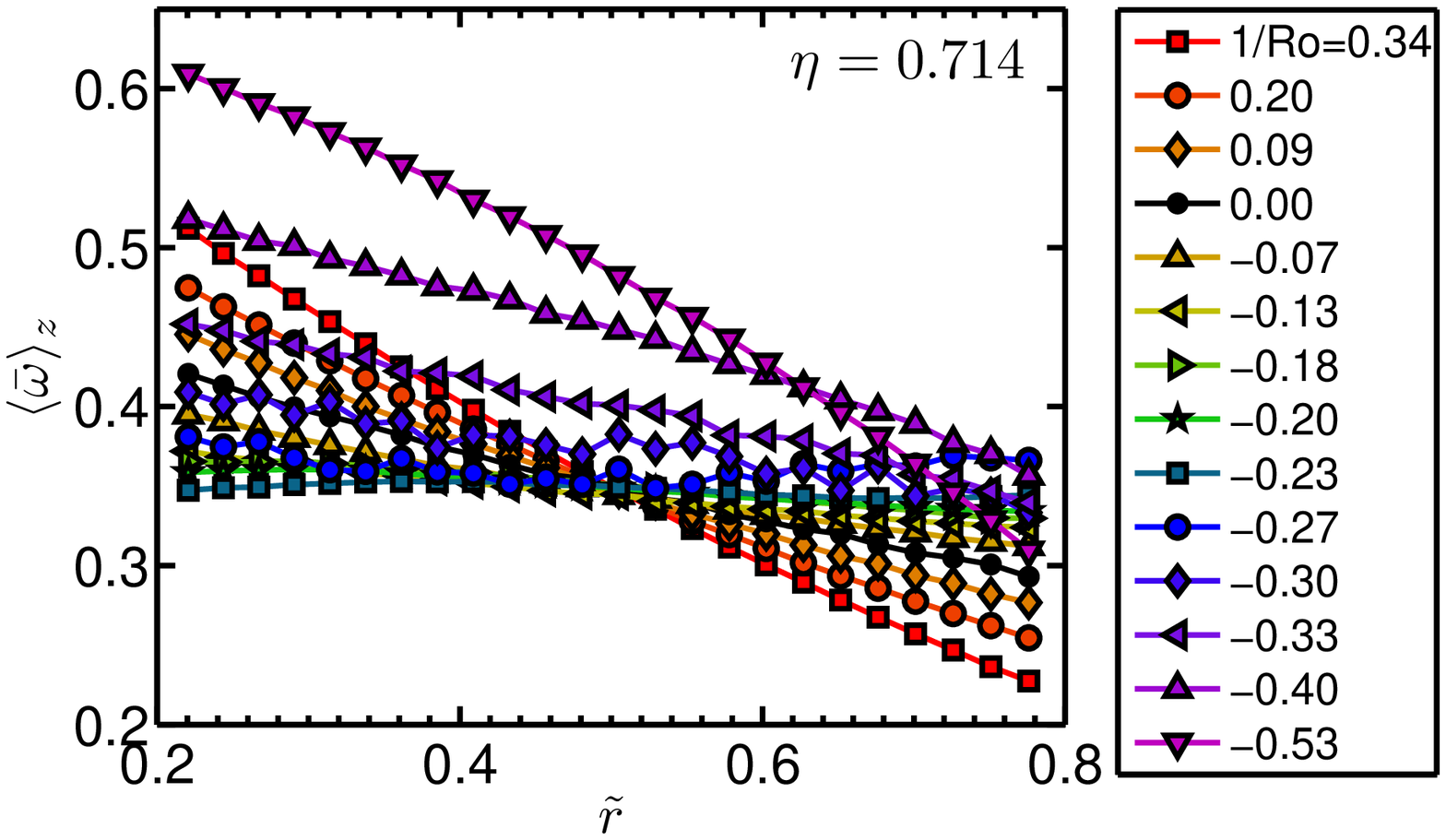}}\\
  \subfloat{\label{fig:lda_Eta0833}\includegraphics[height=0.33\textwidth]{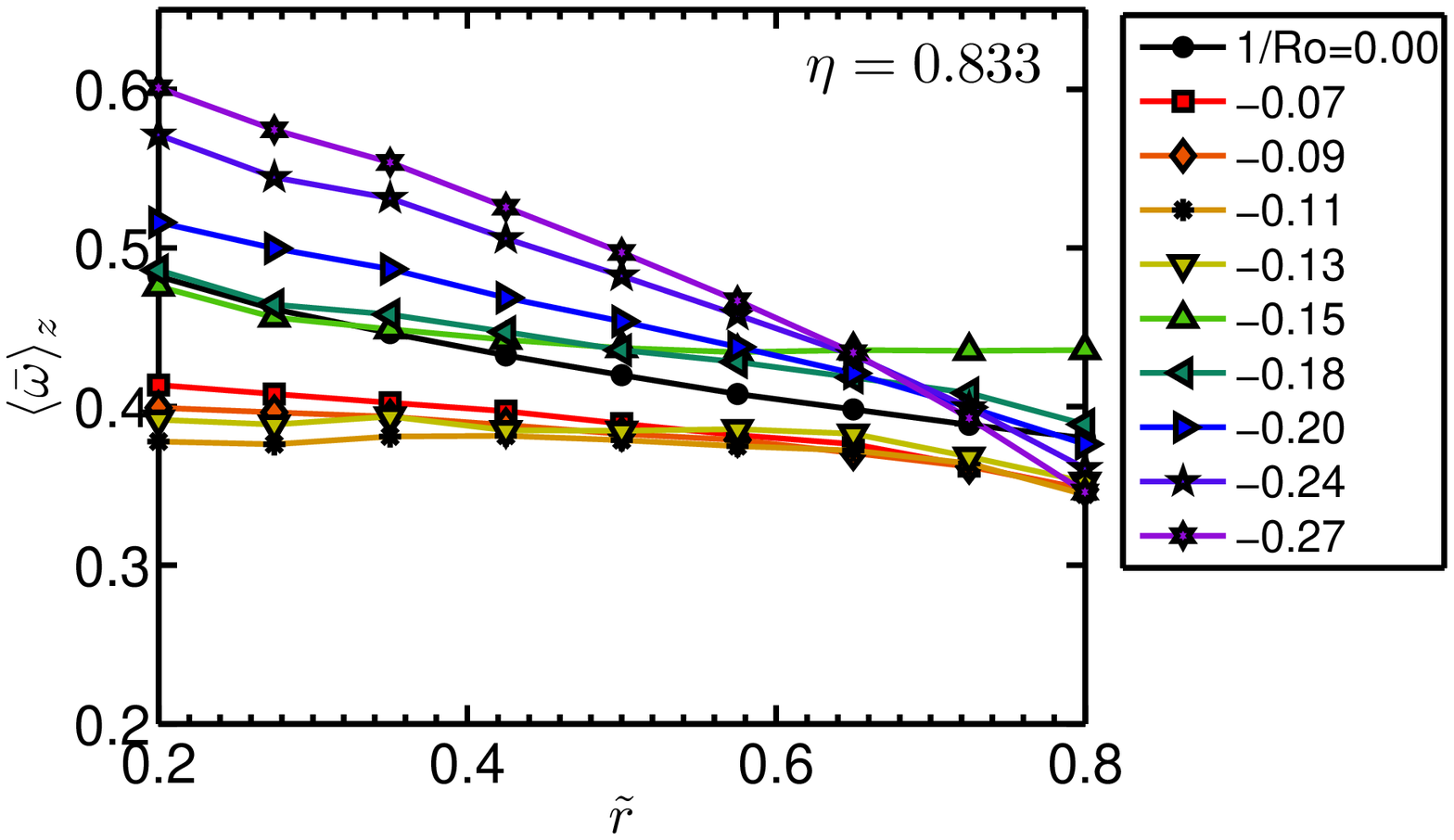}}
  \subfloat{\label{fig:lda_Eta0909}\includegraphics[height=0.33\textwidth]{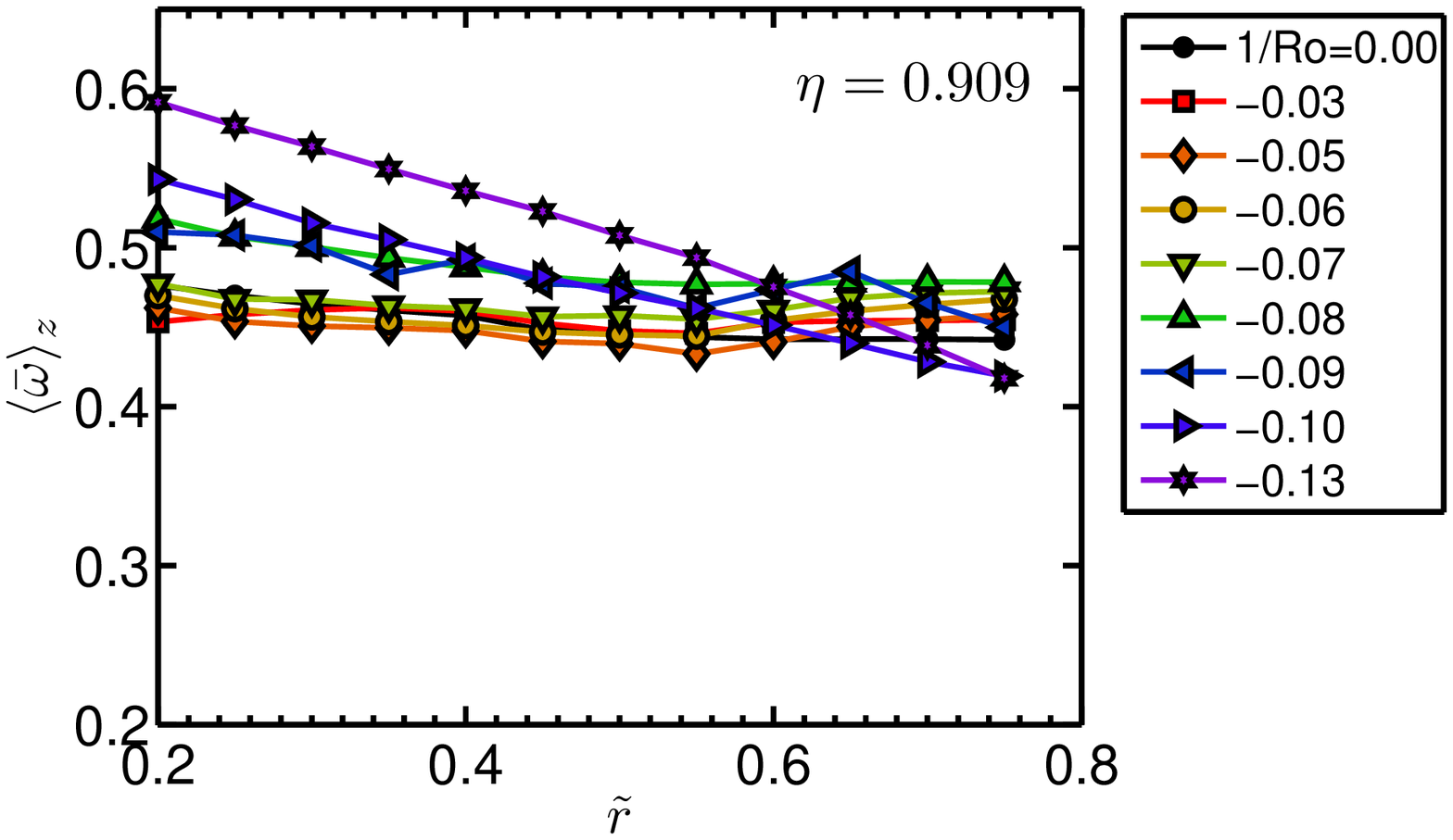}}
  \caption{Angular velocity profiles obtained by LDA for $\eta=0.714$ at either $Re_i-Re_o=10^6$  (top), 
$\eta=0.833$ or at $Ta=5\cdot10^{11}$ (bottom-left) and $\eta=0.909$ at $Ta=1.1\cdot10^{11}$ (bottom-right), to 
explore different dependencies in parameter space. Data is taken at a fixed axial height (ie. the cylinder 
mid-height, $z=L/2$), but as the Taylor number $Ta$ is much larger than in the numerics, 
the axial dependence is much weaker. }
  \label{fig:romegaprof_exp}
 \end{center} 
\end{figure}

Angular velocity $\omega$ profiles obtained from numerics are shown in figure \ref{fig:romegaprof_num}. Results are presented
for four values of $\eta$ and selected values of $\usro$ at $Ta=2.5\cdot10^7$ (and $Ta=2.39\cdot10^7$ for $\eta=0.714$).
Experimental data obtained by using LDA are shown in figure \ref{fig:romegaprof_exp} for three values 
of $\eta$: from top-left to bottom, $\eta=0.714$ for $Re_i-Re_o=10^6$, $\eta=0.833$ for $Ta=5\cdot10^{11}$, 
and $\eta=0.909$ for $Ta=1.1\cdot10^{11}$.

The different radius ratios affect the angular velocity profiles on both boundary layers, as the two boundary layers 
are more asymmetric for the wide gaps; and they affect the bulk, as the bulk angular velocity is smaller for 
wide gaps. These effects will be analyzed in the next sections.

\subsection{Angular velocity profiles in the bulk}

\label{sec:localbulk}

We now analyze the properties of the angular velocity profiles in the bulk. We find that the slope of the profiles 
in the bulk is controlled mainly by $\usro$ and less so by $Ta$. This can be understood as follows: The Taylor number 
$Ta$ acts through the viscous term, dominant in the boundary layers, while $\usro$ acts through the Coriolis force, 
present in the whole domain. These results extend the finding from \cite{ost12} to other values of $\eta$. 

\begin{figure}
 \begin{center}
  \subfloat{\label{fig:blobliex}\includegraphics[width=0.49\textwidth]{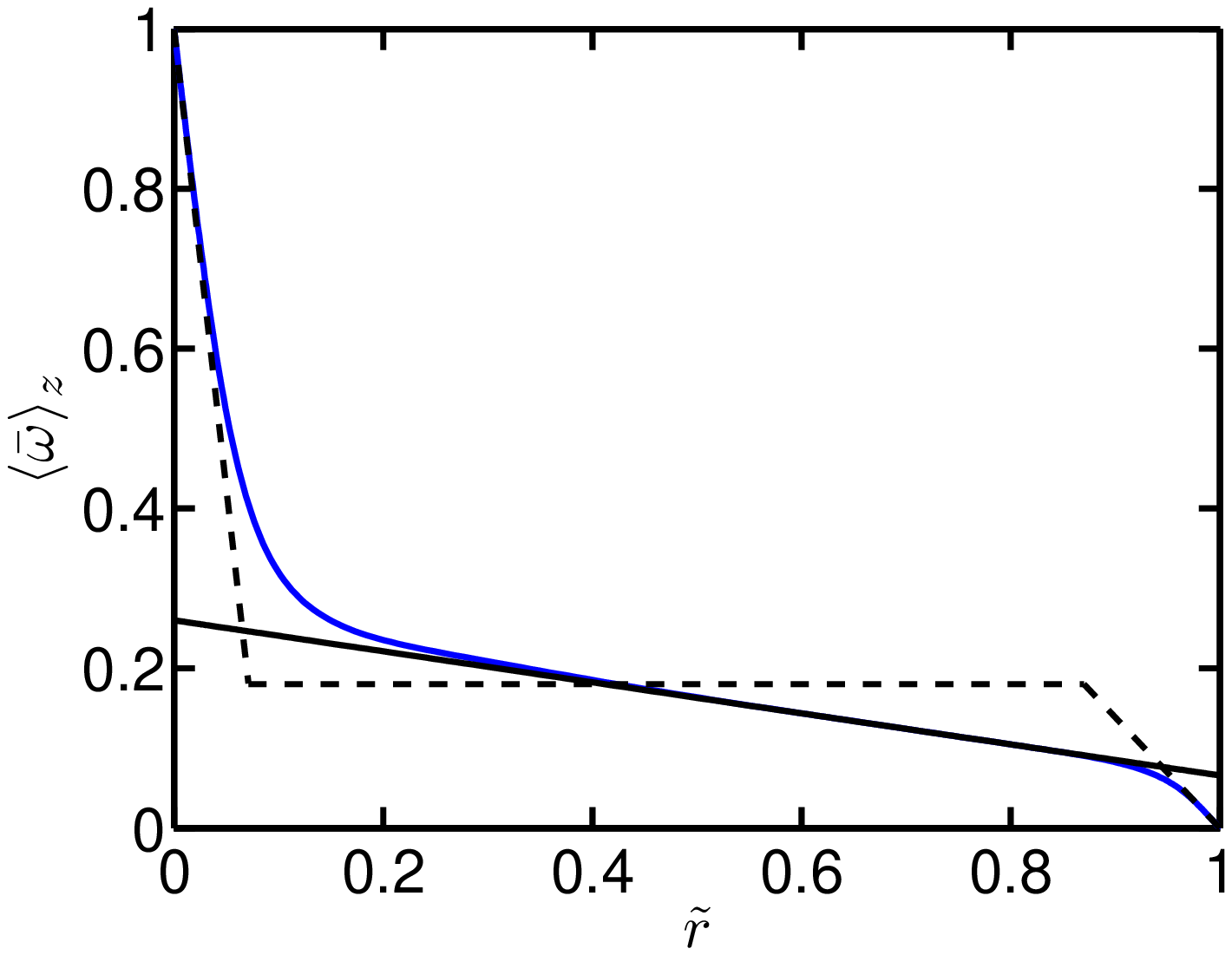}}
  \subfloat{\label{fig:domegadrexample}\includegraphics[width=0.49\textwidth]{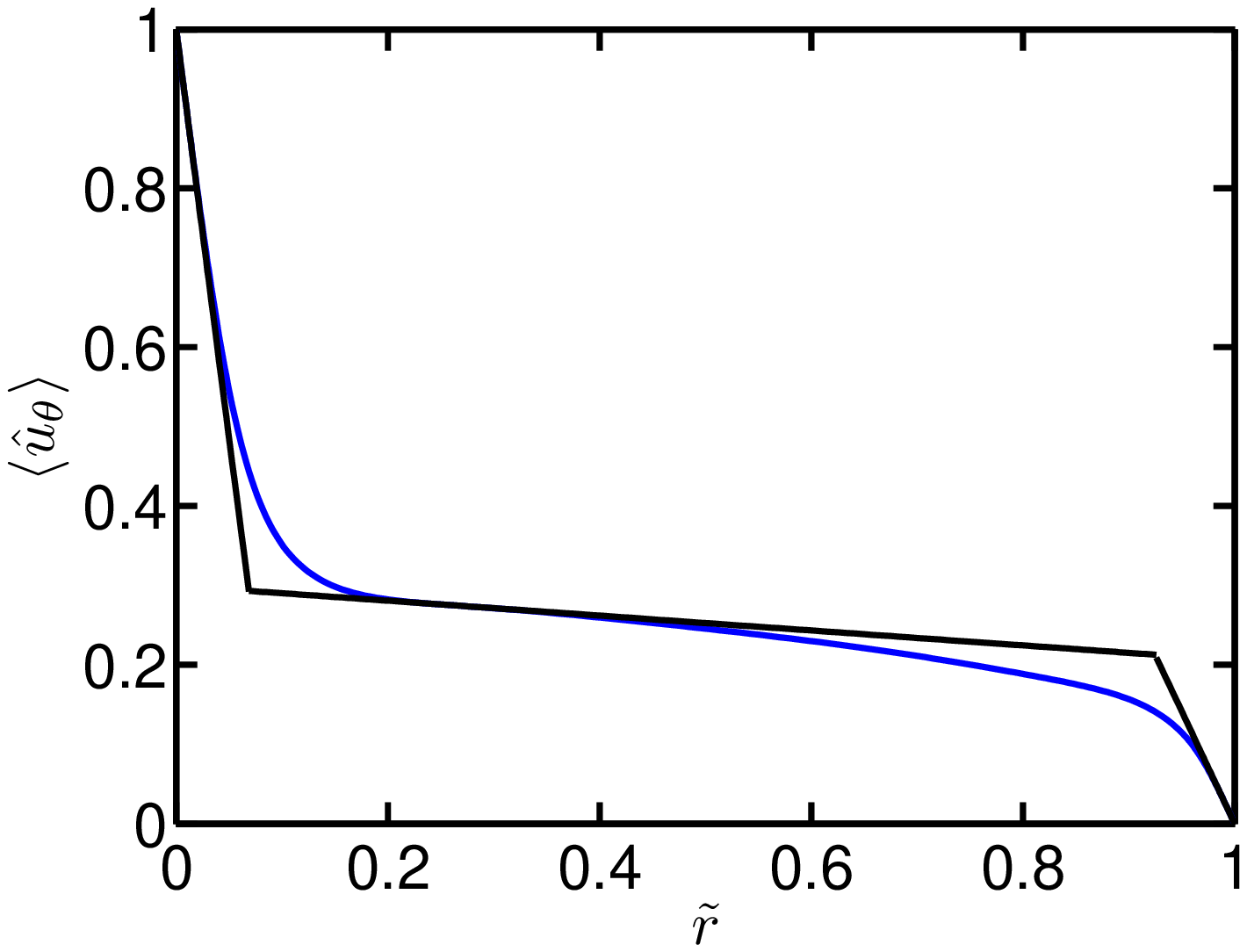}}\\	
  \caption{An example of the two fitting procedures for the bulk angular velocity gradient and for boundary layer thicknesses
done on the DNS data is shown here. Both panels show the $\theta$,$z$, and $t$ averaged azimuthal velocity and angular
velocity for $\eta=0.5$, $Ta$=$1\cdot10^7$, and pure inner cylinder rotation. In the left panel a line
is fitted to the bulk of the angular velocity to obtain the bulk gradient. The dashed lines indicate the EGL 
approximation. The right panel shows
the three-lines-fit to the whole profile to obtain the width of its boundary layers, used in 
Section \ref{sec:blobli}. Both bulk fits are done at the inflection point, but for different
variables ($\bar{\omega}$ or $\bar{u}_\theta$), which gives slightly different slopes (and
intersection points).}
  \label{fig:domegablobliex}
 \end{center} 
\end{figure}

To further quantify the effect of $\usro$ on the bulk profiles, we calculate the gradient of
 $\langle \bar{\omega} \rangle_z$. For the DNS data, this is done by numerically fitting a tangent line 
to the profile at the inflection point using the two neighboring points on both sides (at a distance of $0.01-0.02$ r-units);  
such fit is shown in the left panel of figure \ref{fig:domegablobliex}.

As the spatial resolution of the LDA data is more limited, the fit is done differently. A linear regression to
the $\omega$-profile between $0.2<\tilde{r}<0.8$ is done. The larger range of $\tilde{r}$ is chosen in experiments because:
(i) the boundary layers are small enough due to the high $Ta$ that they are outside of the fitting range, and (ii) the 
fluctuations of the data are much higher in experiments, especially for the LDA of the narrow gaps ($\eta=0.833$ and
$\eta=0.909$).
From this regression, we calculate $\langle \bar{\omega} \rangle_z$, and an error taken from the covariance matrix of the fit.

\begin{figure}
 \begin{center}
  \subfloat{\label{fig:RodwdrEta05}\includegraphics[height=0.3\textwidth]{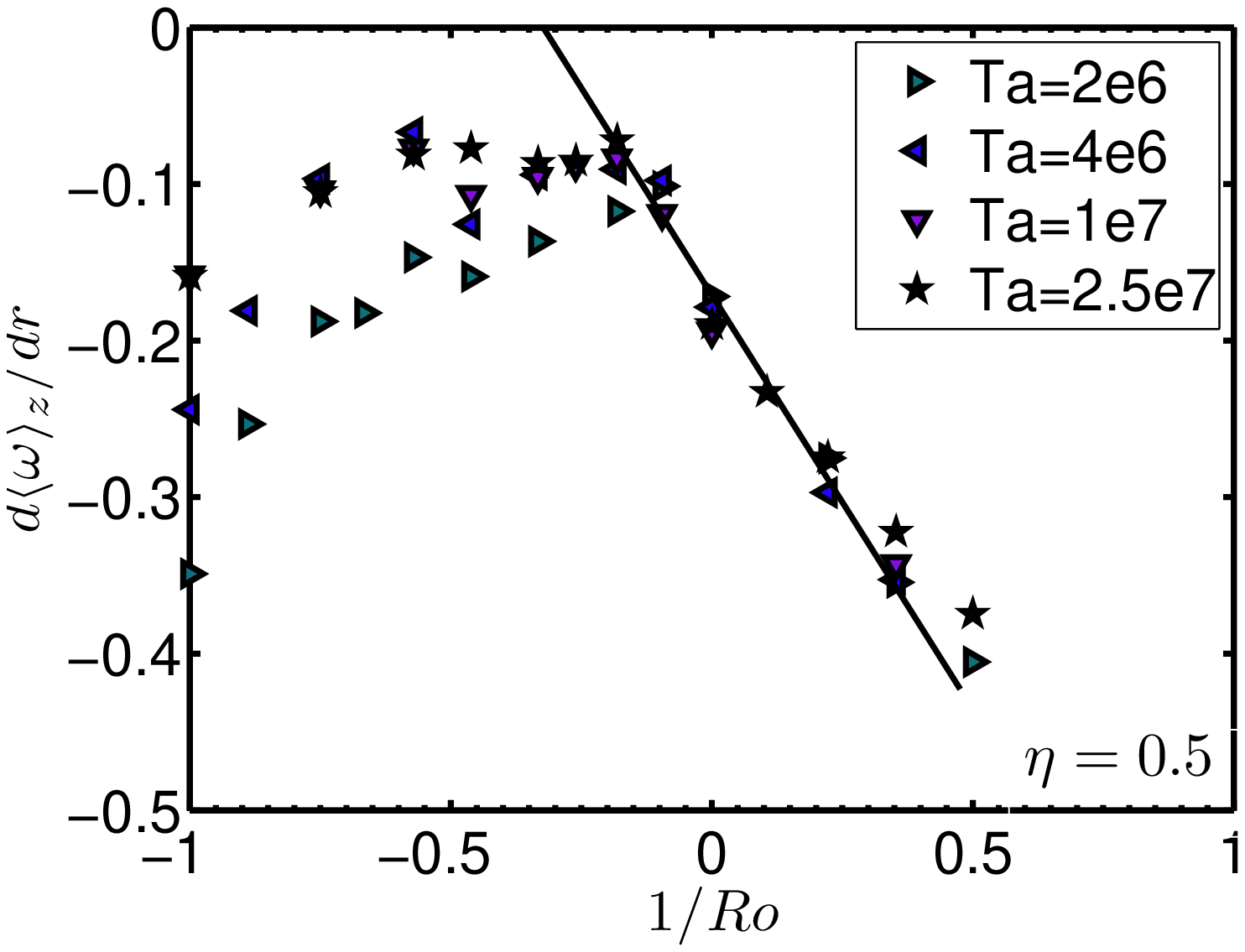}}
  \subfloat{\label{fig:RodwdrEta0714}\includegraphics[height=0.3\textwidth]{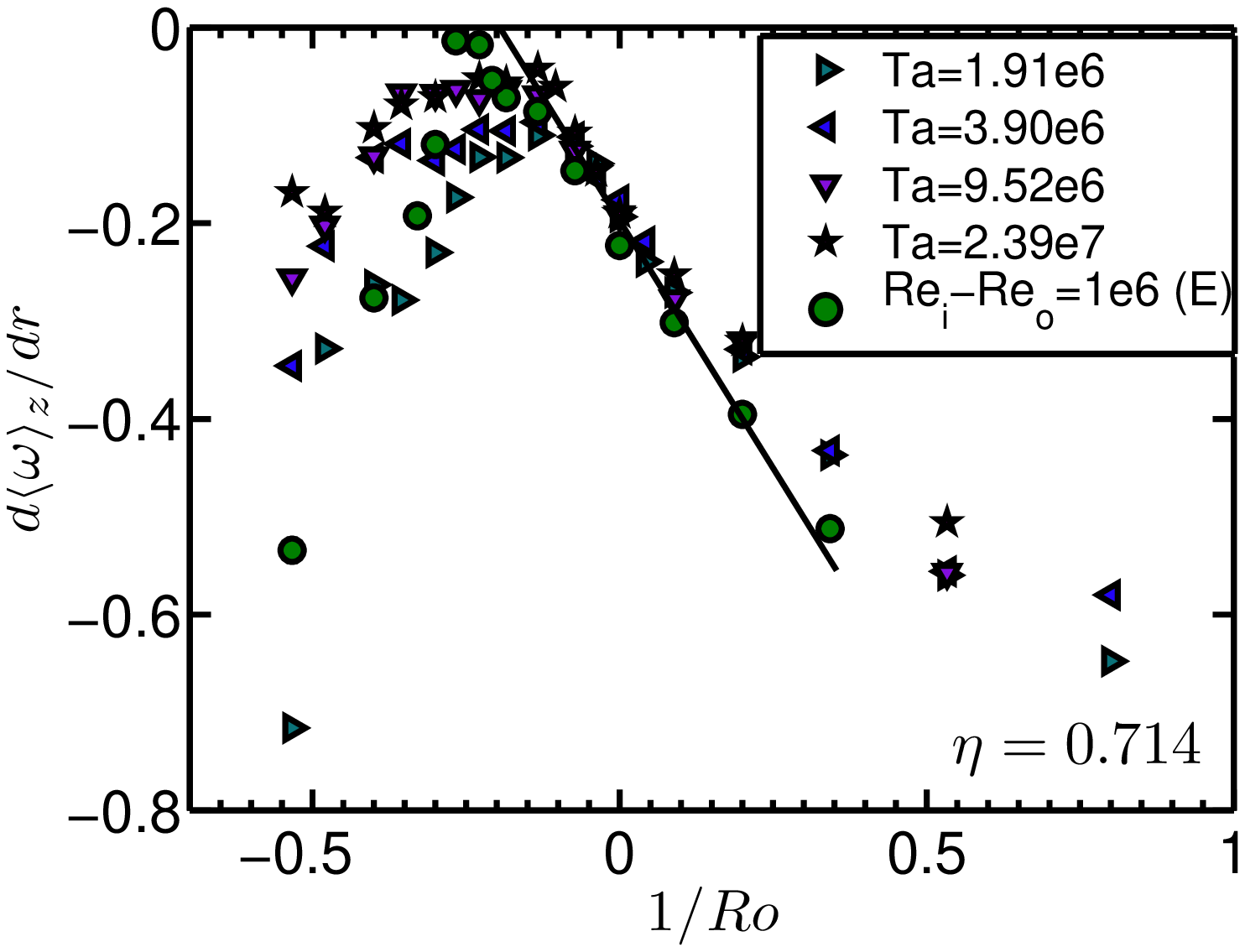}}\\
  \subfloat{\label{fig:RodwdrEta0833}\includegraphics[height=0.3\textwidth]{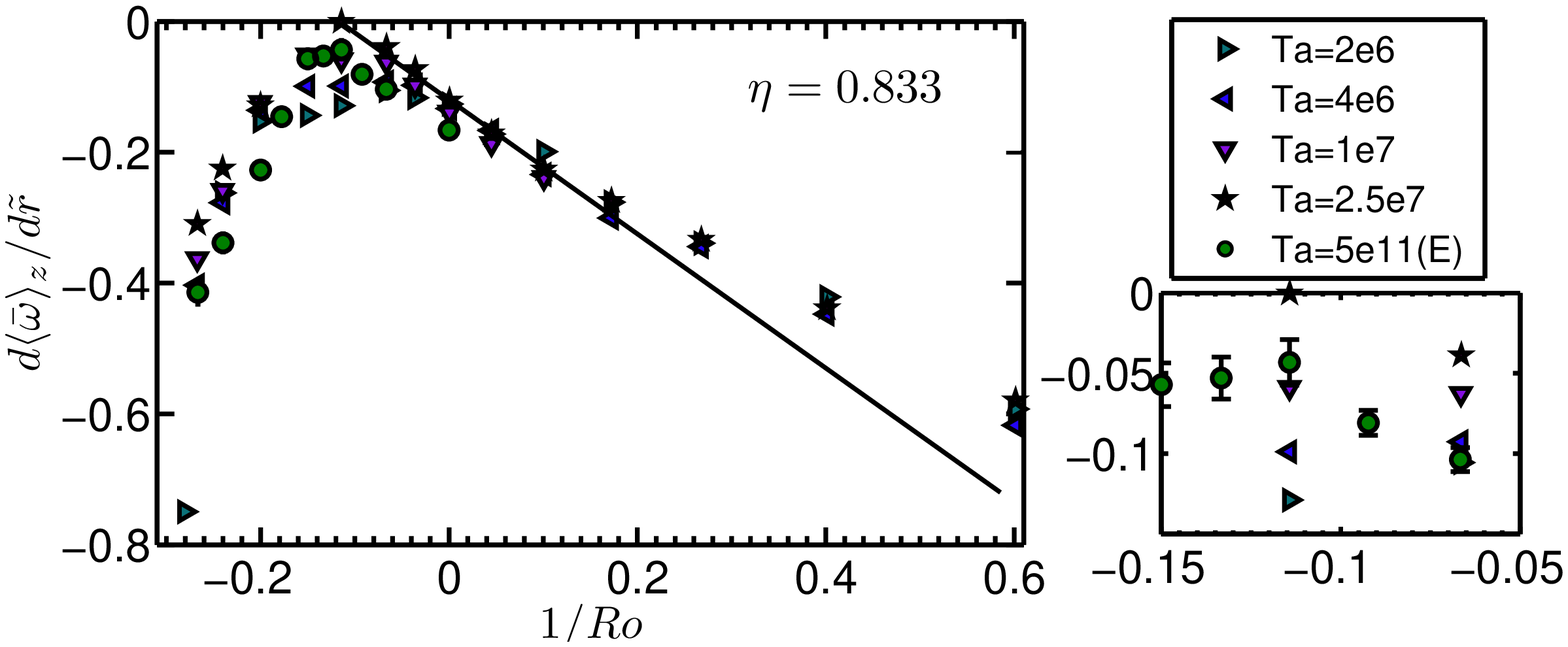}}\\
  \subfloat{\label{fig:RodwdrEta0909}\includegraphics[height=0.3\textwidth]{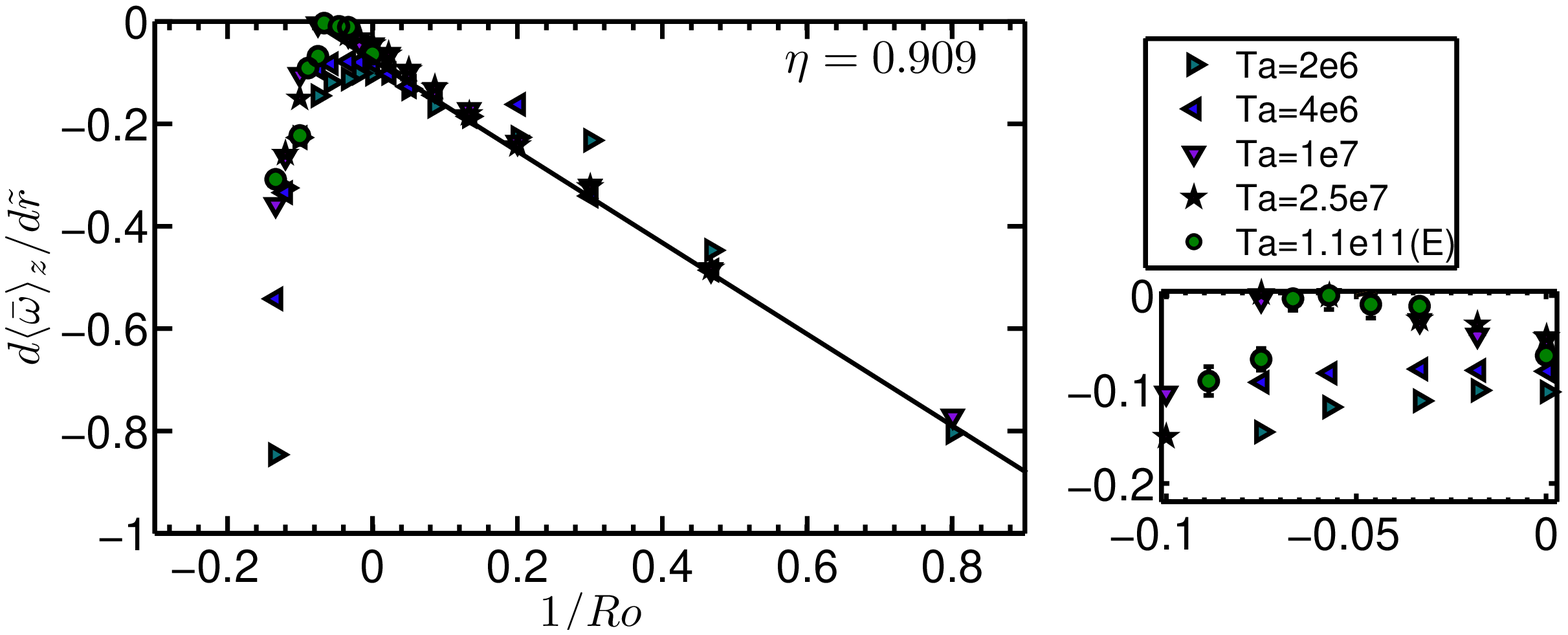}}
  \caption{Bulk angular velocity gradient $d\langle\bar{\omega}\rangle_z/d\tilde{r}$ against $\usro$ for
the four values of $\eta$ explored in simulations, $\eta=0.5$ (top left), $\eta=0.714$, (top right),
$\eta=0.833$, (middle), and $\eta=0.909$ (bottom). Data from experiments obtained by LDA is also plotted for the
three values of $\eta$ for which it was experimentally measured (green circles). For all values of $\eta$ except 
$\eta=0.5$, for co--rotation and slight counter--rotation
there is once again an almost linear relationship between $\usro$ and $d\langle\bar{\omega}\rangle_z/d\tilde{r}$.
A black straight line is added to extrapolate this relationship in order to 
estimate $\usro_{opt}$.  A plateau, in which the radial gradient of $\langle\bar{\omega}\rangle_z$ is small
can be seen around optimal transport, indicating a large convective transport of angular velocity.}
 \label{fig:Rodwdr}
 \end{center} 
\end{figure}

Figure \ref{fig:Rodwdr} shows four panels, each containing
the angular velocity gradient in the bulk from the numerical simulations and
experiments for a given value of $\eta$. We first notice that the angular velocity gradients from experiment
and numerics are in excellent agreement. Next 
the connection between a flat angular velocity profile and optimal transport for the highest
drivings explored in the experiments can now be seen for
other values of $\eta$ and not just for $\eta=0.714$ as reported previously \citep{gil11}. 
Once $\usro<\usro_{opt}$, the large scale balance analyzed in \cite{ost12} breaks
down, and a ``neutral'' surface which reduces the transport appears in the flow.

In simulations, because of resolution requirements, we are unable of driving
the flow strongly enough to see a totally flat bulk profile. 
Also, the influence of the large scale structures causes a small discrepancy
between the flattest profile and the value of $\usro_{opt}$ measured from $\nom$. This is expected to 
slowly dissapear with increasing $Ta$.

In \cite{ost12}, 
a linear extrapolation of the bulk angular velocity
gradient was done to give an estimate for the case when this profile would become horizontal,
i.e., $d\langle\bar{\omega}\rangle_z/d\tilde{r}=0$,
and thus give an estimate of $\usro_{opt}$. For $\eta=0.714$ this estimate
agreed with the numerical result within error bars. Here, we extend this analysis for the other values of $\eta$ and,
as we shall see, successfully. 

As in \cite{ost12}, an almost linear relationship between $\usro$ and $d\langle\bar{\omega}\rangle_z/d\tilde{r}$
can be seen. This linear relationship is extrapolated and plotted in each panel. 
This extrapolation gives an estimate for $\usro_{opt}(Ta\to\infty)$, which we can compare to the experimentally
determined $\usro_{opt}(Ta\to\infty)$. For $\eta=0.833$, 
$\usro(Ta\to\infty)\approx-0.12$ corresponding to $a\approx0.38$ is obtained, and for
$\eta=0.909$, $\usro(Ta\to\infty)\approx-0.05$, corresponding to $a\approx0.31$ is obtained. 
These values are (within error bars) also obtained for $\usro_{opt}$ at the large $Ta$ investigated
in experiments, namely  $\usro_{opt}=-0.10$ and $-0.05$, respectively. 

For $\eta=0.5$, $\usro_{opt}(Ta\to\infty)\approx-0.33$ is obtained, corresponding to 
$a\approx 0.2$. This value is consistent with the numerical results in \cite{bra13}, which report
$a_{opt}\approx 0.2$. However, care must be taken, as fitting straight lines to the $\omega$-profiles
gives higher residuals for $\eta=0.5$ as the profiles deviate from straight lines (cf. top-left panel
of figure \ref{fig:romegaprof_num}). A fit to the ``quarter-Couette'' profile derived from
upper bound theory \citep{bus67}
is much more appropriate for $\eta=0.5$ at the strongest drivings achieved in experiments \citep{mer12}.
This is because the flow feels much more the effect of the curvature at the small $\eta$. On the other end of the scale,
the linear relationship works best for
smallest gaps, i.e. $\eta=0.909$ (cf. the bottom right panel of figure \ref{fig:romegaprof_num}) where
curvature plays a small effect.

\begin{figure}
 \begin{center}
  \includegraphics[width=0.7\textwidth]{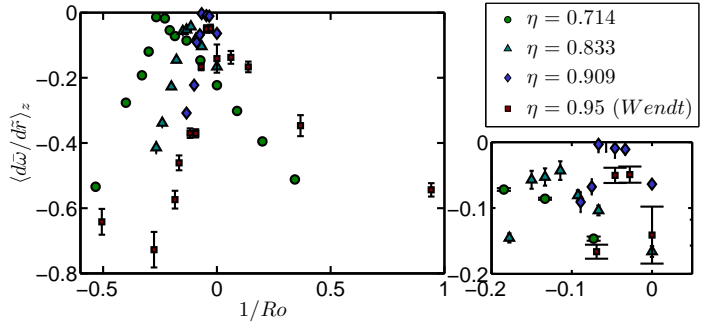}
  \caption{Bulk angular velocity gradient $d\langle\bar{\omega}\rangle_z/d\tilde{r}$ against $\usro$ for
the three values of $\eta$ explored in experiments, and for $\eta=0.95$ (digitized from \cite{wen33}). The error bars 
of Wendt's data are larger due to the quality of the digitization. As seen previously, the flattest
profile occurs around weak counter-rotation, for all values of $\eta$ including $\eta=0.95$.}
 \label{fig:Rodwdrwendt}
 \end{center} 
\end{figure}

To further elaborate the link between $\eta$, flat $\omega$-profiles and 
$\usro$, data for the smallest gap $\eta=0.95$ from \cite{wen33} has been digitized, 
and $d\langle\bar{\omega}\rangle_z/d\tilde{r}$ was been determined for it. This data corresponds to a driving
of $Ta\sim 10^8-10^9$.  Figure \ref{fig:Rodwdrwendt} shows $d\langle\bar{\omega}\rangle_z/d\tilde{r}$ against
$\usro$ for Wendt's data and also for the current experimental data. 
The flattest profile can be seen to occur for increasing (in absolute value) $\usro$ for larger gaps, simular to the
shift of $\usro$.
For $\eta=0.95$, almost no curvature is felt by the flow and a flat profiles can be seen for $-0.05<\usro<0$.
However, adding a Coriolis force (in the form of $\usro$) 
a large $\omega$-gradient is sustained in the bulk. This corroborates
the balance between $\usro$ and the bulk $\omega$-gradients proposed in \cite{ost12}.

\subsection{Angular velocity profiles in the boundary layers in the classical turbulent regime}
\label{sec:blobli}

As the driving is increased, the transport is enhanced. To accommodate for this, the boundary layers (BLs) 
become thinner and therefore the $\omega$-slopes ($\partial_r \omega$) become steeper. Due to the geometry of the TC system
an intrinsic asymmetry in the BL layer widths is present. More precisely, the exact relationship 
$\partial_r \langle \omega \rangle |_o = \eta^3 \partial_r \langle \omega \rangle |_i$ holds for the slopes of the 
boundary layers, due to the $r$ independence of $J^\omega$, cf. EGL 2007 and eq.(\ref{eq:jomega}). 

An analysis of the boundary layers was not possible in the present experiments because the present LDA measurements 
have insufficient spatial resolution to resolve the flow in the near--wall region. Therefore, only DNS results will 
be analyzed here. In simulations the driving is not as large as in experiments, and as a consequence
the shear in the BLs is expected to not be large enough to cause a shear-instability. 
This means that the BLs are expected to be of Prandtl-Blasius (i.e. laminar) type, even if the bulk is turbulent.
On the other hand, in the experiments both boundary layers and bulk are turbulent, i.e. the system is in 
the ``ultimate regime''. 

Using the DNS data, we can compare the ratio of the numerically obtained boundary layer widths with the analytical 
formula for this ratio obtained by EGL 2007 for laminar boundary layers, namely:

\begin{equation}
 \label{eq:thicknessratio}
 \displaystyle\frac{\lambda^o_\omega}{\lambda^i_\omega}=\eta^{-3}\displaystyle\frac{|\omega_o-\omega_{bulk}|}{|\omega_i-\omega_{bulk}|},
\end{equation}

\noindent where the value of $\omega_{bulk}$ is some appropriate value in between for which the angular velocity at the 
inflection point of the profile might be chosen, i.e., the point at which the linear bulk profile 
fit was done to obtain $\lambda^o_\omega$ and 
$\lambda^i_\omega$. The value $\omega_{bulk}$ is taken from the numerics, and may bias the estimate. 

To calculate the boundary layer thicknesses, the profile of the mean azimuthal 
velocity $\langle \bar{u}_\theta\rangle_z$ is approximated by three straight lines, 
one for each boundary layer and one for the bulk. For the boundary layers the slope of the fit is calculated by 
fitting (by least-mean-squares) a line through the first two computational grid points. For the bulk, first
the line is forced to pass through the grid point which is numerically closest to the inflection point
of the profile. Then its slope is taken from a least mean square fit using two grid points on both sides of this 
inflection point. The respective boundary layer line will 
cross with this bulk line at a point which then defines the thickness of that boundary layer. 

The results obtained for $\lambda^o_\omega/\lambda^i_\omega$
both from equation \ref{eq:thicknessratio} and directly from the simulations is shown in figure \ref{fig:Rodwdr}. 
Results are presented for the four values of $\eta$ and only for the highest value of $Ta$ achieved in the simulations.
The boundary layer asymmetry for counter--rotating cylinders (i.e., $\usro<0$) grows with larger gaps.
This is to be expected, as the $\eta^{-3}$ term is much larger ($\approx 8$)
for the largest gap as compared to the smallest gap ($\approx 1.3$). This is consistent with the $\eta$ and thus 
$\sigma$ restriction in EGL 2007 to a range of smaller gap widths.  

As noticed already in \cite{ost12} we find that the fit is not satisfactory for co--rotation (i.e., $\usro>0$) at the lowest
values of $\eta$, but is satisfactory for counter--rotation 
(ie. $\usro<0$).  In EGL 2007,  equation (\ref{eq:thicknessratio}) is obtained 
by approximating the profile by three straight lines, two for the BLs and a constant $\omega$ line for the bulk. Therefore,
we expect the approximation to hold best when the bulk has a flat gradient. 
For co--rotating cylinders and strongly counter--rotating cylinders, the bulk has a steep gradient (see figure \ref{fig:Rodwdr}),
but characteristically different shapes. The only free parameter in equation (\ref{eq:thicknessratio}) is $\omega_{bulk}$, which
is chosen to be $\omega$ at the point of inflection. Due to the different shapes of the $\omega$-profiles, this choice seems
more correct for counter-rotating cylinders, as there is a clear inflection point in the profile. On the other hand, for co-rotating
cylinders, the profile appears to be more convex-like, and there, the choice of $\omega_{bulk}$ as the inflection point induces 
errors in the approximation (cf. \ref{fig:blobliex}).
For $\eta=0.5$, the error from the constant $\omega$ approximation is even more pronounced, and the formula fails.

\begin{figure}
 \begin{center}
  \subfloat{\label{fig:RoBlobliEta05}\includegraphics[width=0.49\textwidth]{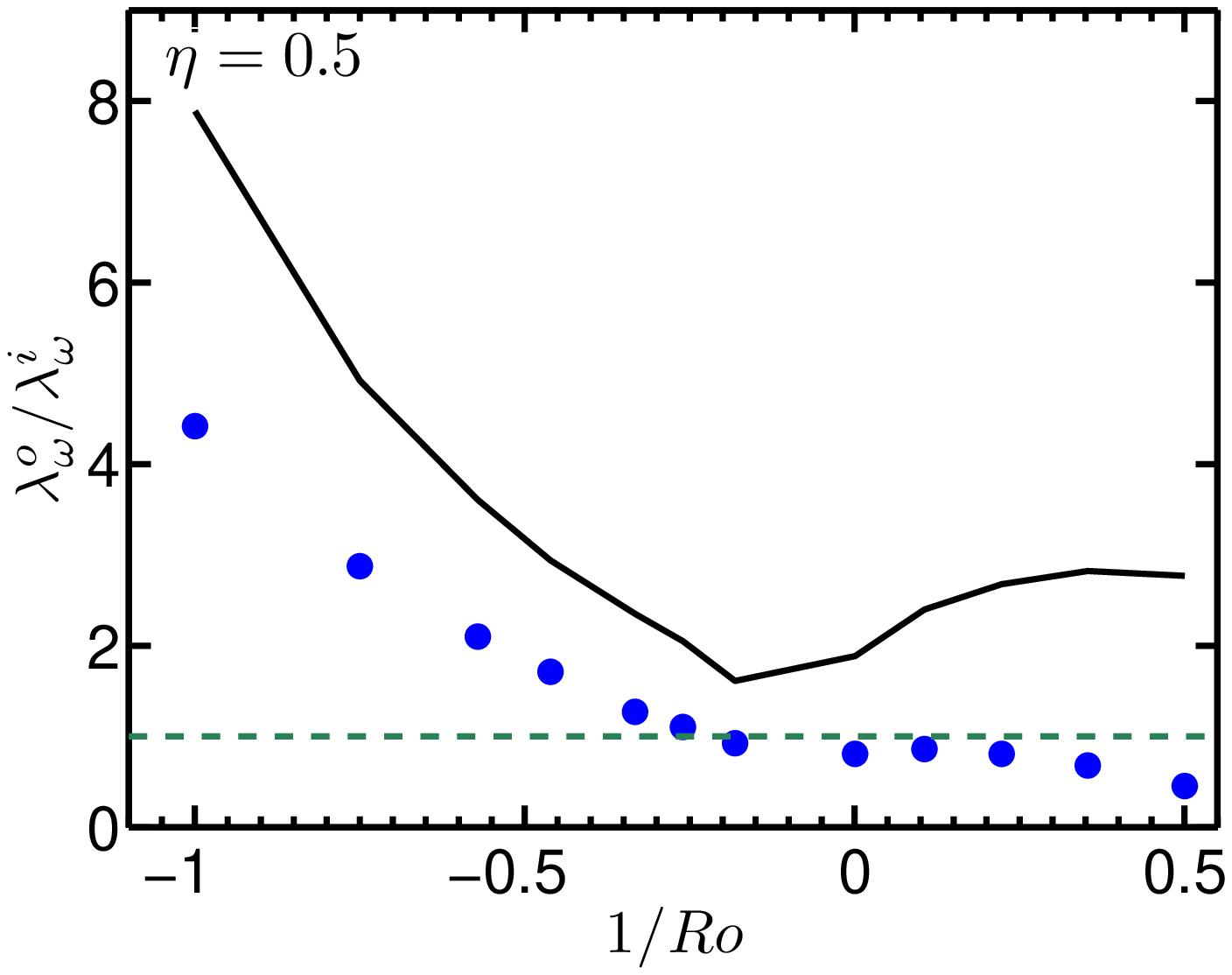}}
  \subfloat{\label{fig:RoBlobliEta0714}\includegraphics[width=0.49\textwidth]{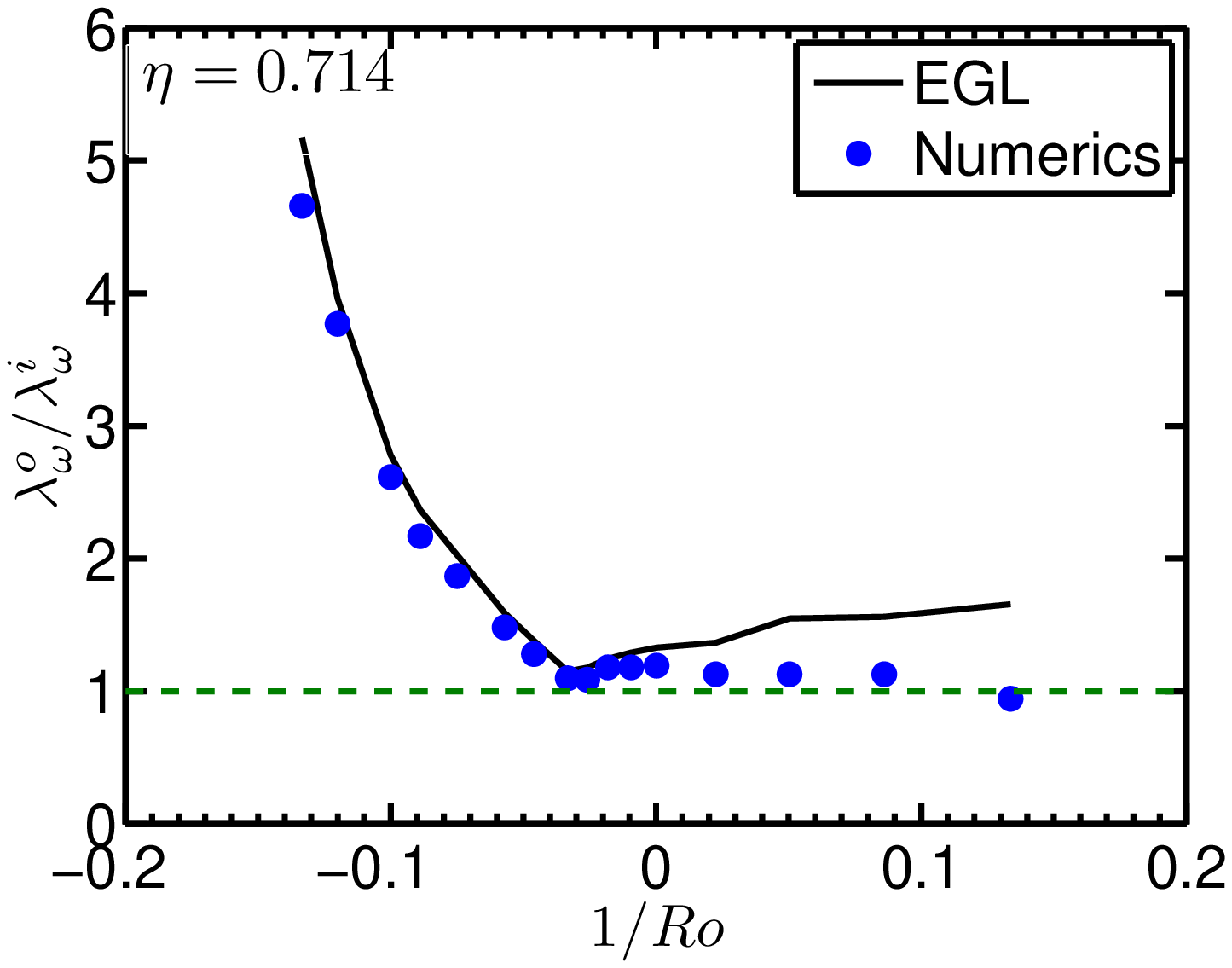}}\\
  \subfloat{\label{fig:RoBlobliEta0833}\includegraphics[width=0.49\textwidth]{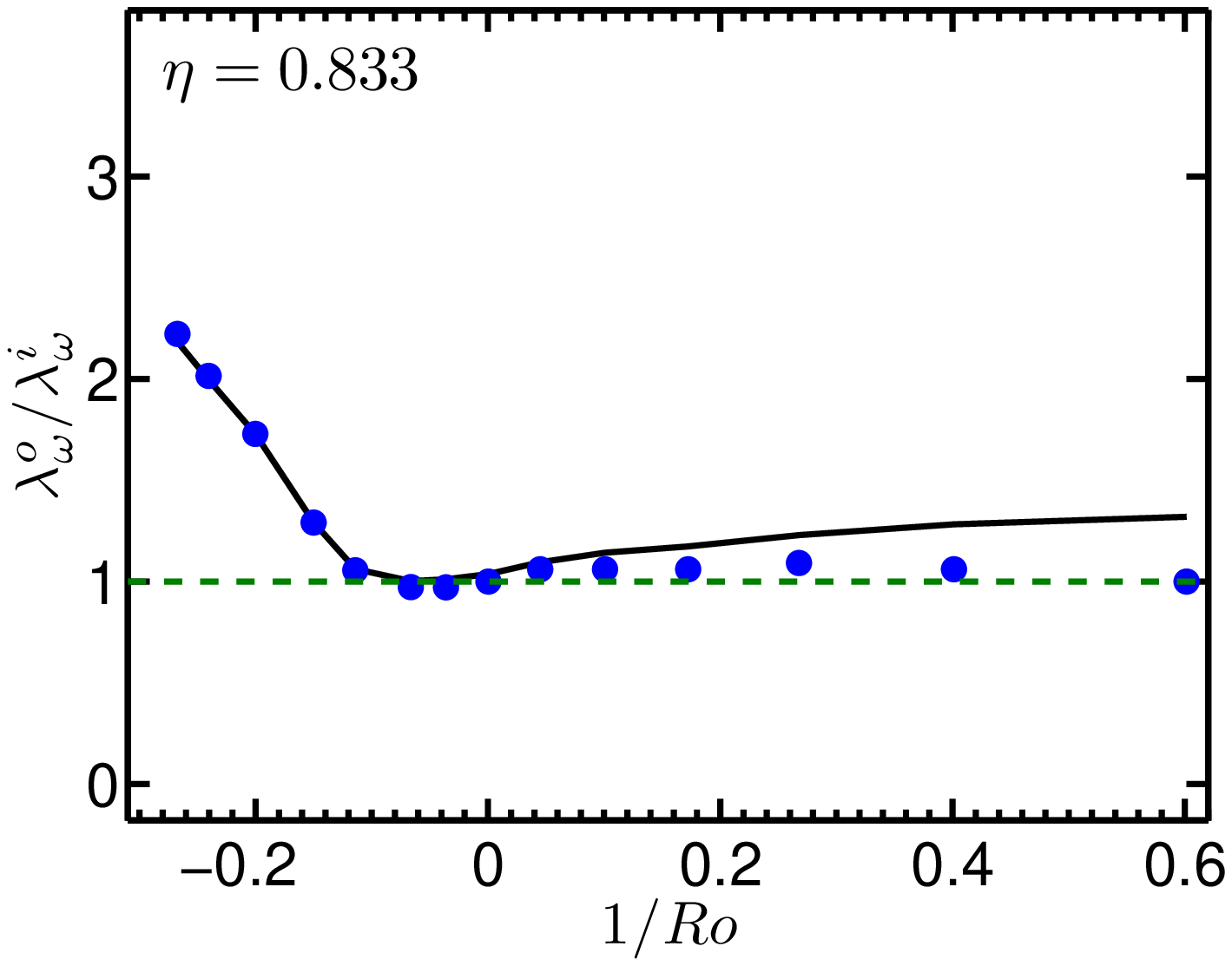}}
  \subfloat{\label{fig:RoBlobliEta0909}\includegraphics[width=0.49\textwidth]{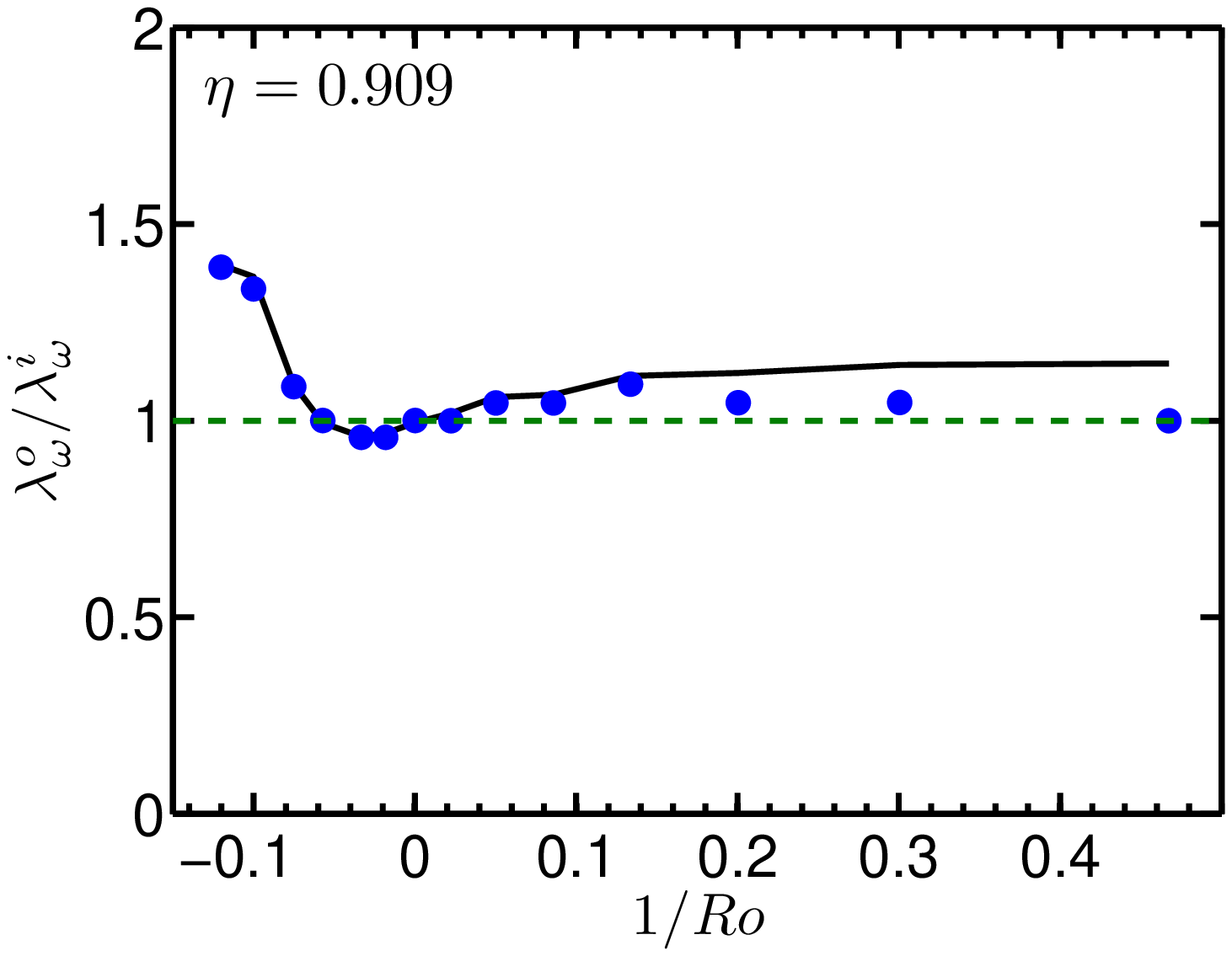}}\\
  \caption{$\lambda^o_\omega/\lambda^i_\omega$ from simulations (dots) and from equation (\ref{eq:thicknessratio})
 (solid lines) versus $\usro$ for the four values of 
 $\eta$ studied numerically, $\eta=0.5$ (top-left), $0.714$ (top-right), $0.833$ (bottom-left) 
 and $0.909$ (bottom-right) at $Ta=2.5\cdot10^7$. The numerical results and the estimate from equation 
(\ref{eq:thicknessratio}) match very well for larger values of $\eta$ and especially for counter--rotating cylinders 
($1/Ro < 0$). The asymmetry between the boundary layers can be seen to be larger for smaller values of $\eta$, which 
is expected as equation (\ref{eq:thicknessratio}) contains the explicit factor $\eta^{-3}$.}
 \label{fig:RoBlobli}
 \end{center} 
\end{figure}

\begin{figure}
 \begin{center}
  \includegraphics[height=0.33\textwidth]{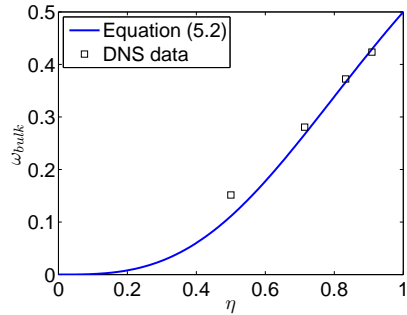}
  \caption{ $\omega_{bulk}$ as a function of $\eta$, taken from both eq. \ref{eq:bulkest}
and from the crosspoints of the $\omega$-profiles in figure \ref{fig:romegaprof_num}. The trend is the 
same in both data sets. A smaller value of $\eta$ decreases the value of the bulk angular velocity. }
 \label{fig:eta_omegabulk}
 \end{center} 
\end{figure}

For co--rotation the boundary layers are approximately of the same size, and the ratio $\lambda^o_\omega/\lambda^i_\omega$
is very close to $1$. If one inverts equation (\ref{eq:thicknessratio}) by aproximating this ratio by $1$, an estimate of what
the angular velocity will be in the bulk due to the boundary layer slope asymmetry is obtained:

\begin{equation}
 \label{eq:bulkest}
 \omega_{bulk}=\displaystyle\frac{\eta^3}{1+\eta^3},
\end{equation}

\noindent corresponding to:

\begin{equation}
 \label{eq:bulkestlab}
 \omega^\ell_{bulk}=\displaystyle\frac{-\omega^\ell_o+\eta^3\omega_i^\ell}{1+\eta^3},
\end{equation}

\noindent in the lab frame. This expression gives an estimate for $\omega_{bulk}$ when the profile is flattest, and
has been represented graphically in figure \ref{fig:romegaprof_num}. Indeed, one can take this
estimate (for example $0.27$ for $\eta=0.714$) and compare it with figures \ref{fig:romegaprof_num} and 
\ref{fig:romegaprof_exp}. We note that the value of $\omega$ in the bulk for 
the flattest profile in the numerics (at $\usro\approx\usro_{opt}(Ta)$) lies around $\omega_{bulk}$. 
We can also notice that the profiles for $\usro>\usro_{opt}$ approximately cross each other at the same point, 
and this point has a value of $\omega\approx\omega_{bulk}$. This effect can only be seen in the numerics, as these 
approximations break down once the boundary layers become turbulent. The cross points of the curves are taken
as an estimate for $\omega_{bulk}$, and this is represented against eq. \ref{eq:bulkest} in figure \ref{fig:eta_omegabulk}.

To understand why the boundary layers are of approximately the same thickness despite the different initial slopes at 
the cylinders one has to go back to equation (\ref{eq:jomega}). The angular velocity current has a diffusive part and 
a convective part. Per definition in the boundary layer the diffusion dominates and in the bulk the convection does. 
Thus the boundary layer ceases when convection becomes significant. But convection is controlled by the wind. Thus in 
essence the boundary layer size is controlled by the wind and not immediately by the initial slope at the wall. Due to 
continuity, if the rolls penetrate the whole domain the wind may be expected to be the same close to the inner and close 
to the outer cylinder. This suggests that the flow organizes itself in a way that the boundary 
layer extensions (or widths) might be similar, even if the initial slopes at the walls are different.

What happens for counter--rotation, or more precisely when $\usro<\usro_{opt}$? For $\usro$ below the optimum $\usro_{opt}$ 
a so-called neutral surface will be present in the flow, which separates the Rayleigh-stable and -unstable areas. 
The wind drastically changes in the Rayleigh-stable 
areas \citep{ost12}, leading to very different wind velocities close to the outer and inner cylinder, respectively. 
The wind at the outer cylinder will be weaker, as the rolls cannot fully penetrate the Rayleigh-stable domain.
This means that the outer cylinder boundary layer will extend deeper into the flow, in accordance
to what is seen in figure \ref{fig:RoBlobli}.

\section{Summary and conclusions}

Experiments and direct numerical simulations (DNS) were analysed to explore the
effects of the radius ratio $\eta$ on turbulent Taylor-Couette flow. Numerical results corresponding
to Taylor numbers in the range of $10^4<Ta<10^8$ alongside with experiments in a Taylor
number range of $10^{10}<Ta<10^{12}$ were presented for four values of the radius ratio $\eta$. 

First the influence of the radius ratio on the global scaling laws $Nu_\omega\sim Ta^\gamma$ was studied. The 
local scaling exponent $\gamma$ describing the response of the torque caused by a Taylor number increase, is barely 
modified by varying the radius ratio $\eta$. Indeed, in experiments a universal exponent of $\gamma \approx 0.39$ is obtained,
independent of radius ratio and outer cylinder rotation.
For the numerical simulations at lower $Ta$ similar universal behavior can be observed. 
The transition associated to the vanishing of coherent
structures can also be appreciated at $Ta\sim10^6$ for all values of $\eta$. Before this transition
local exponents of $\gamma\approx0.33$ are seen and after the transition these decrease to
about $\gamma\approx0.2$.

The radius ratio does play a very important role in optimal transport. At smaller gaps, i.e., for larger $\eta$, 
at the lower end of the $Ta$ range a very large increase in transport for corotating 
cylinders can be seen.  The shift towards the asymptotic optimal transport 
happens in a much slower way for small gaps, but this shift is seen 
for all studied radius ratios. For the largest gap ($\eta=0.5$), 
optimal transport for pure inner cylinder rotation at the lowest drivings is obtained. The shift
towards the asymptotic value happens suddenly, as two peaks can be seen in the $\nom$ versus $\usro$
curve, and one of the peaks becomes larger than the other one as driving increases. This might
point in the direction of different phenomena and transitions in the flow topology happening at larger gaps.
Finally, the asymptotic values of $\usro_{opt}$ obtained in experiments  
were compared to the speculation of \cite{gil12} and the prediction of \cite{bra13}.
Both of the models were found to deviate from experimental and numerical results.

When looking at the local results, as in \cite{ost12} we can link the optimal transport in the
smallest gaps to a balance between Coriolis forces and the inertia terms in the equations of motion.
The flattest profiles in the bulk are linked to optimal transport in experiments. With the numerics
the extrapolation presented in \cite{ost12} for predicting optimal transport was extended to other
radius ratios. It is found to work well for all selected $\eta$ except for $\eta=0.5$. At this $\eta$, i.e., for 
the largest gap considered here, the 
most obvious problem is that the $\omega$ profiles strongly feel the effect of curvature difference 
at the inner and outer cylinders and a 
straight line fit to the bulk is not appropriate. There may be additional reasons
for this discrepancy and optimal transport in large gaps requires more investigation. 
A summary of the results for determining $\usro_{opt}$ using both the experimentally measured
torque maxima from section \ref{sec:globalro} and the numerical extrapolation from
section \ref{sec:localbulk} are presented in table \ref{tbl:final}. 

\begin{table}
  \begin{center} 
  \def~{\hphantom{0}}
  \begin{tabular}{ccccc}
  Radius ratio ($\eta$)  & $\usro_{opt}$/$a_{opt}$ from extrapolation & Measured $\usro_{opt}$/$a_{opt}$\\
  0.5   & -0.33/0.20 & -/-   \\
  0.714 & -0.20/0.33 & -0.20/0.33 \\
  0.769 & -/-        & -0.20/0.36 \\
  0.833 & -0.12/0.41 & -0.10/0.37 \\
  0.909 & -0.05/0.34 & -0.05/0.34 \\
 \end{tabular}
 \caption{ Summary of values obtained for $\usro_{opt}$ and $a_{opt}$ through both the extrapolation of 
$d\langle\bar{\omega}\rangle_z/d\tilde{r}$ (Section \ref{sec:localbulk}) and direct measurement of the torque 
(Section \ref{sec:globalro}) for the various values of $\eta$ studied in this paper.}
 \label{tbl:final}
\end{center}
\end{table}

Finally, the boundary layers have been analyzed. The outer boundary layer is found to be much
thicker than the inner boundary layer when $\usro < \usro_{opt}$. We attribute this to the appearance
of Rayleigh-stable zones in the flow. This prevents the turbulent Taylor vortices from covering the full 
domain between the cylinders. 
As the boundary layer size is essentially determined by the wind,
if the rolls penetrate the whole domain (which is the case for $\usro > \usro_{opt}$), 
both boundary layers are approximately of the same size. If the rolls do not penetrate the full domain,
the outer boundary layer will be much larger than the inner boundary layer, in accordance with the smaller initial
slope of $\omega(r)$ at the cylinder walls. 

In this work, simulations and experiments have been performed on a range of radius ratios 
between $0.5 \le \eta \le 0.909$. Insights for the small gaps seem to be consistent with
what was discussed in \cite{ost12}. However, for $\eta=0.5$ the phenomena of optimal
transport appears to be quite different. Therefore, our ambition is to extend the DNS
towards values of $\eta$ smaller than $0.5$ to improve the understanding of that regime.

Acknowledgements: We would like to thank H. Brauckmann, B. Eckhardt, 
S. Merbold, M. Salewski, E. P. van der Poel and R. C. A. van der Veen for various stimulating discussions during these years,
and G.W. Bruggert, M. Bos and B. Benschop for technical support.
We acknowledge that the numerical results of this research have been achieved 
using the PRACE-2IP project (FP7 RI-283493) resource VIP based in Germany at Garching. We would also like 
to thank the Dutch Supercomputing Consortium SurfSARA for technical support and computing resources. 
We would like to thank FOM, the Simon Stevin Prize of the Technology Foundation 
STW of The Netherlands, COST from the EU and ERC for financial support through an Advanced Grant.

\bibliographystyle{jfm}

\bibliography{literatur}

\begin{thebibliography}{57}
\expandafter\ifx\csname natexlab\endcsname\relax\def\natexlab#1{#1}\fi

\bibitem[Ahlers(1974)]{ahl74}
{\sc Ahlers, G.} 1974 Low temperature studies of the
  {{{{Rayleigh-B{\'e}nard}}}} instability and turbulence. {\em Phys. Rev.
  Lett.\/} {\bf 33}, 1185--1188.

\bibitem[Ahlers {\em et~al.\/}(2009)Ahlers, Grossmann \& Lohse]{ahl09}
{\sc Ahlers, G., Grossmann, S. \& Lohse, D.} 2009 Heat transfer and large scale
  dynamics in turbulent {{{{Rayleigh-B{\'e}nard}}}} convection. {\em Rev. Mod.
  Phys.\/} {\bf 81}, 503.

\bibitem[Andereck {\em et~al.\/}(1986)Andereck, Liu \& Swinney]{and86}
{\sc Andereck, C.~D., Liu, S.~S. \& Swinney, H.~L.} 1986 Flow regimes in a
  circular couette system with independently rotating cylinders. {\em J. Fluid
  Mech.\/} {\bf 164}, 155.

\bibitem[Behringer(1985)]{beh85}
{\sc Behringer, R.~P.} 1985 {{{{Rayleigh-B{\'e}nard}}}} convection and
  turbulence in liquid-helium. {\em Rev. Mod. Phys.\/} {\bf 57}, 657 -- 687.

\bibitem[Benjamin(1978)]{ben78}
{\sc Benjamin, T.~B.} 1978 Bifurcation phenomena in steady flows of a viscous
  liquid. {\em Proc. R. Soc. London A\/} {\bf 359}, 1--43.

\bibitem[Bodenschatz {\em et~al.\/}(2000)Bodenschatz, Pesch \& Ahlers]{bod00}
{\sc Bodenschatz, E., Pesch, W. \& Ahlers, G.} 2000 Recent developments in
  {{{{Rayleigh-B{\'e}nard}}}} convection. {\em Ann. Rev. Fluid Mech.\/} {\bf
  32}, 709--778.

\bibitem[Brauckmann \& Eckhardt(2013{\natexlab{{\em a\/}}})]{bra12}
{\sc Brauckmann, H. \& Eckhardt, B.} 2013{\natexlab{{\em a\/}}} Direct
  numerical simulations of local and global torque in {{{Taylor-Couette}}} flow
  up to {{{Re}}}=30.000. {\em J. Fluid Mech.\/} {\bf 718}, 398--427.

\bibitem[Brauckmann \& Eckhardt(2013{\natexlab{{\em b\/}}})]{bra13}
{\sc Brauckmann, H. \& Eckhardt, B.} 2013{\natexlab{{\em b\/}}} Intermittent
  boundary layers and torque maxima in {{{Taylor-Couette}}} flow. {\em Phys.
  Rev. E\/} {\bf 87}, 033004.

\bibitem[Busse(1967)]{bus67}
{\sc Busse, F.~H.} 1967 The stability of finite amplitude cellular convection
  and its relation to an extremum principle. {\em J. Fluid Mech.\/} {\bf 30},
  625--649.

\bibitem[Chandrasekhar(1981)]{cha81}
{\sc Chandrasekhar, S.} 1981 {\em Hydrodynamic and Hydromagnetic Stability\/}.
  New York: Dover.

\bibitem[Couette(1890)]{cou890}
{\sc Couette, M.} 1890 {\em {\'E}tudes sur le frottement des liquides\/}.
  Gauthier-Villars et fils.

\bibitem[Coughlin \& Marcus(1996)]{cou96}
{\sc Coughlin, K. \& Marcus, P.~S.} 1996 Turbulent bursts in
  {{{Couette-Taylor}}} flow. {\em Phys. Rev. Lett.\/} {\bf 77}~(11), 2214--17.

\bibitem[Cross \& Hohenberg(1993)]{cro93}
{\sc Cross, M.~C. \& Hohenberg, P.~C.} 1993 Pattern formation outside of
  equilibrium. {\em Rev. Mod. Phys.\/} {\bf 65}~(3), 851.

\bibitem[Dominguez-Lerma {\em et~al.\/}(1986)Dominguez-Lerma, Cannell \&
  Ahlers]{dom86}
{\sc Dominguez-Lerma, M.~A., Cannell, D.~S. \& Ahlers, G.} 1986 Eckhaus
  boundary and wavenumber selection in rotating {{Couette-Taylor}} flow. {\em
  Phys. Rev. A\/} {\bf 34}, 4956.

\bibitem[Dong(2007)]{don07}
{\sc Dong, S} 2007 Direct numerical simulation of turbulent taylor-couette
  flow. {\em J. Fluid Mech.\/} {\bf 587}, 373--393.

\bibitem[Dong(2008)]{don08}
{\sc Dong, S} 2008 Turbulent flow between counter-rotating concentric
  cylinders: a direct numerical simulation study. {\em J. Fluid Mech.\/} {\bf
  615}, 371--399.

\bibitem[Donnelly(1991)]{don91b}
{\sc Donnelly, R.} 1991 {{{Taylor-Couette}}} flow: the early days. {\em Physics
  Today\/} pp. 32--39.

\bibitem[Drazin \& Reid(1981)]{dra81}
{\sc Drazin, P.G. \& Reid, W.~H.} 1981 {\em Hydrodynamic stability\/}.
  Cambridge: Cambridge University Press.

\bibitem[Eckhardt {\em et~al.\/}(2007)Eckhardt, Schneider, Hof \&
  Westerweel]{eck07}
{\sc Eckhardt, B., Schneider, T.M., Hof, B. \& Westerweel, J.} 2007 Turbulence
  transition in pipe flow. {\em Annu. Rev. Fluid Mech.\/} {\bf 39}, 447--468.

\bibitem[Esser \& Grossmann(1996)]{ess96}
{\sc Esser, A. \& Grossmann, S.} 1996 Analytic expression for
  {{Taylor-Couette}} stability boundary. {\em Phys. Fluids\/} {\bf 8},
  1814--1819.

\bibitem[Fasel \& Booz(1984)]{fas84}
{\sc Fasel, H. \& Booz, O.} 1984 Numerical investigation of supercritical
  taylor-vortex flow for a wide gap. {\em J. Fluid Mech.\/} {\bf 138}, 21--52.

\bibitem[Gebhardt \& Grossmann(1993)]{geb93b}
{\sc Gebhardt, Th. \& Grossmann, S.} 1993 The {{{Taylor-Couette}}} eigenvalue
  problem with independently rotating cylinders. {\em Z. Phys. B\/} {\bf
  90}~(4), 475--490.

\bibitem[van Gils {\em et~al.\/}(2011{\natexlab{{\em a\/}}})van Gils, Bruggert,
  Lathrop, Sun \& Lohse]{gil11a}
{\sc van Gils, D. P.~M., Bruggert, G.~W., Lathrop, D.~P., Sun, C. \& Lohse, D.}
  2011{\natexlab{{\em a\/}}} The {{Twente Turbulent}} {{Taylor-Couette}}
  ({{$T^3C$}}) facility: strongly turbulent (multi-phase) flow between
  independently rotating cylinders. {\em Rev. Sci. Instr.\/} {\bf 82}, 025105.

\bibitem[van Gils {\em et~al.\/}(2011{\natexlab{{\em b\/}}})van Gils, Huisman,
  Bruggert, Sun \& Lohse]{gil11}
{\sc van Gils, D. P.~M., Huisman, S.~G., Bruggert, G.~W., Sun, C. \& Lohse, D.}
  2011{\natexlab{{\em b\/}}} Torque scaling in turbulent {{Taylor-Couette}}
  flow with co- and counter-rotating cylinders. {\em Phys. Rev. Lett.\/} {\bf
  106}, 024502.

\bibitem[van Gils {\em et~al.\/}(2012)van Gils, Huisman, Grossmann, Sun \&
  Lohse]{gil12}
{\sc van Gils, D. P.~M., Huisman, S.~G., Grossmann, S., Sun, C. \& Lohse, D.}
  2012 Optimal {{Taylor-Couette}} turbulence. {\em J. Fluid Mech.\/} {\bf 706},
  118.

\bibitem[Grossmann \& Lohse(2000)]{gro00}
{\sc Grossmann, S. \& Lohse, D.} 2000 Scaling in thermal convection: A unifying
  view. {\em J. Fluid. Mech.\/} {\bf 407}, 27--56.

\bibitem[Grossmann \& Lohse(2001)]{gro01}
{\sc Grossmann, S. \& Lohse, D.} 2001 Thermal convection for large {{Prandtl}}
  number. {\em Phys. Rev. Lett.\/} {\bf 86}, 3316--3319.

\bibitem[Grossmann \& Lohse(2011)]{gro11}
{\sc Grossmann, S. \& Lohse, D.} 2011 Multiple scaling in the ultimate regime
  of thermal convection. {\em Phys. Fluids\/} {\bf 23}, 045108.

\bibitem[Grossmann \& Lohse(2012)]{gro12}
{\sc Grossmann, S. \& Lohse, D.} 2012 Logarithmic temperature profiles in the
  ultimate regime of thermal convection. {\em Phys. Fluids\/} {\bf 24}, 125103.

\bibitem[Huisman {\em et~al.\/}(2012{\natexlab{{\em a\/}}})Huisman, van Gils \&
  Sun]{hui12b}
{\sc Huisman, S.G., van Gils, D.P.M. \& Sun, C.} 2012{\natexlab{{\em a\/}}}
  Applying laser doppler anemometry inside a taylor-couette geometry -- using a
  ray-tracer to correct for curvature effects. {\em Eur. J. Mech. - B/Fluids\/}
  {\bf 36}, 115--119.

\bibitem[Huisman {\em et~al.\/}(2012{\natexlab{{\em b\/}}})Huisman, van Gils,
  Grossmann, Sun \& Lohse]{hui12}
{\sc Huisman, S.~G., van Gils, D. P.~M., Grossmann, S., Sun, C. \& Lohse, D.}
  2012{\natexlab{{\em b\/}}} Ultimate turbulent {{Taylor-Couette}} flow. {\em
  Phys. Rev. Lett.\/} {\bf 108}, 024501.

\bibitem[Huisman {\em et~al.\/}(2013)Huisman, Scharnowski, Cierpka, Kaehler,
  Lohse \& Sun]{hui13}
{\sc Huisman, S.~G., Scharnowski, S., Cierpka, C., Kaehler, C., Lohse, D. \&
  Sun, C.} 2013 Logarithmic boundary layers in highly turbulent
  {{{Taylor-Couette}}} flow. {\em Phys. Rev. Lett\/} {\bf 110}, 264501.

\bibitem[Kadanoff(2001)]{kad01}
{\sc Kadanoff, L.~P.} 2001 Turbulent heat flow: Structures and scaling. {\em
  Phys. Today\/} {\bf 54}~(8), 34--39.

\bibitem[Lathrop {\em et~al.\/}(1992{\natexlab{{\em a\/}}})Lathrop, Fineberg \&
  Swinney]{lat92a}
{\sc Lathrop, D.~P., Fineberg, Jay \& Swinney, H.~S.} 1992{\natexlab{{\em
  a\/}}} Transition to shear-driven turbulence in {{Couette-Taylor}} flow. {\em
  Phys. Rev. A\/} {\bf 46}, 6390--6405.

\bibitem[Lathrop {\em et~al.\/}(1992{\natexlab{{\em b\/}}})Lathrop, Fineberg \&
  Swinney]{lat92}
{\sc Lathrop, D.~P., Fineberg, Jay \& Swinney, H.~S.} 1992{\natexlab{{\em
  b\/}}} Turbulent flow between concentric rotating cylinders at large
  {{Reynolds}} numbers. {\em Phys. Rev. Lett.\/} {\bf 68}, 1515--1518.

\bibitem[Lewis \& Swinney(1999)]{lew99}
{\sc Lewis, G.~S. \& Swinney, H.~L.} 1999 Velocity structure functions,
  scaling, and transitions in high-{{Reynolds}}-number {{Couette-Taylor}} flow.
  {\em Phys. Rev. E\/} {\bf 59}, 5457--5467.

\bibitem[Lohse \& Xia(2010)]{loh10}
{\sc Lohse, D. \& Xia, K.-Q.} 2010 Small-scale properties of turbulent
  {{Rayleigh-B{\'e}nard}} convection. {\em Ann. Rev. Fluid Mech.\/} {\bf 42},
  335--364.

\bibitem[Lorenz(1963)]{lor63}
{\sc Lorenz, E.~N.} 1963 Deterministic nonperiodic flow. {\em J. Atmos. Sci\/}
  {\bf 20}, 130--141.

\bibitem[Mallock(1896)]{mal896}
{\sc Mallock, A.} 1896 Experiments on fluid viscosity. {\em Phil. Trans. R.
  Soc. Lond. A\/} {\bf 187}, 41--56.

\bibitem[Merbold {\em et~al.\/}(2013)Merbold, Brauckmann \& Egbers]{mer12}
{\sc Merbold, S., Brauckmann, H. \& Egbers, C.} 2013 Torque measurements and
  numerical determination in differentially rotating wide gap
  {{{Taylor-Couette}}} flow. {\em Phys. Rev. E\/} {\bf 87}, 023014.

\bibitem[Ostilla {\em et~al.\/}(2013)Ostilla, Stevens, Grossmann, Verzicco \&
  Lohse]{ost12}
{\sc Ostilla, R., Stevens, R. J. A.~M., Grossmann, S., Verzicco, R. \& Lohse,
  D.} 2013 Optimal {{{Taylor-Couette}}} flow: direct numerical simulations.
  {\em J. Fluid Mech.\/} {\bf 719}, 14--46.

\bibitem[{Ostilla Monico} {\em et~al.\/}(2013){Ostilla Monico}, van~der Poel,
  Verzicco, Grossmann \& Lohse]{ost13}
{\sc {Ostilla Monico}, R., van~der Poel, E., Verzicco, R., Grossmann, S. \&
  Lohse, D.} 2013 Boundary layer dynamics at the transition between the
  classical and the ultimate regime of {{{Taylor-Couette}}} flow. {\em
  submitted to Physics of Fluids\/} .

\bibitem[Paoletti \& Lathrop(2011)]{pao11}
{\sc Paoletti, M.~S. \& Lathrop, D.~P.} 2011 Angular momentum transport in
  turbulent flow between independently rotating cylinders. {\em Phys. Rev.
  Lett.\/} {\bf 106}, 024501.

\bibitem[Pfister \& Rehberg(1981)]{pfi81}
{\sc Pfister, G. \& Rehberg, I.} 1981 Space dependent order parameter in
  circular {{Couette}} flow transitions. {\em Phys. Lett.\/} {\bf 83}, 19--22.

\bibitem[Pfister {\em et~al.\/}(1988)Pfister, Schmidt, Cliffe \& Mullin]{pfi88}
{\sc Pfister, G, Schmidt, H, Cliffe, K~A \& Mullin, T} 1988 Bifurcation
  phenomena in {{Taylor-Couette}} flow in a very short annulus. {\em J. Fluid
  Mech.\/} {\bf 191}, 1--18.

\bibitem[Pirro \& Quadrio({2008})]{pir08}
{\sc Pirro, Davide \& Quadrio, Maurizio} {2008} {Direct numerical simulation of
  turbulent Taylor-Couette flow}. {\em {Eur. J. Mech. B-Fluids}\/} {\bf {27}},
  {552}.

\bibitem[Siggia(1994)]{sig94}
{\sc Siggia, E.~D.} 1994 High {{Rayleigh}} number convection. {\em Annu. Rev.
  Fluid Mech.\/} {\bf 26}, 137--168.

\bibitem[Smith \& Townsend(1982)]{smi82}
{\sc Smith, G.~P. \& Townsend, A.~A.} 1982 Turbulent {{Couette}} flow between
  concentric cylinders at large {{Taylor}} numbers. {\em J. Fluid Mech.\/} {\bf
  123}, 187--217.

\bibitem[Stevens {\em et~al.\/}(2011)Stevens, Lohse \& Verzicco]{ste11}
{\sc Stevens, R. J. A.~M., Lohse, D. \& Verzicco, R.} 2011 Prandtl and
  {{Rayleigh}} number dependence of heat transport in high {{Rayleigh}} number
  thermal convection. {\em J. Fluid Mech.\/} {\bf 688}, 31--43, submitted.

\bibitem[Stevens {\em et~al.\/}(2010)Stevens, Verzicco \& Lohse]{ste10}
{\sc Stevens, R. J. A.~M., Verzicco, R. \& Lohse, D.} 2010 Radial boundary
  layer structure and {{Nusselt}} number in {{Rayleigh-B{\'e}nard}} convection.
  {\em J. Fluid Mech.\/} {\bf 643}, 495--507.

\bibitem[Strogatz(1994)]{str94}
{\sc Strogatz, S.~H.} 1994 {\em Nonlinear dynamics and chaos\/}. Reading:
  Perseus Press.

\bibitem[Swinney \& Gollub(1981)]{swi81}
{\sc Swinney, H.~L. \& Gollub, J.~P.} 1981 {\em Hydrodynamic instabilities and
  the transition to turbulence\/}, , vol. 45 (Topics in Applied Physics).
  Berlin: Springer-Verlag.

\bibitem[Taylor(1936)]{tay36}
{\sc Taylor, G.~I.} 1936 Fluid friction between rotating cylinders. {\em Proc.
  R. Soc. London A\/} {\bf 157}, 546--564.

\bibitem[Tong {\em et~al.\/}(1990)Tong, Goldburg, Huang \& Witten]{ton90}
{\sc Tong, P., Goldburg, W.~I., Huang, J.~S. \& Witten, T.~A.} 1990 Anisotropy
  in turbulent drag reduction. {\em Phys. Rev. Lett.\/} {\bf 65}, 2780--2783.

\bibitem[Verzicco \& Orlandi(1996)]{ver96}
{\sc Verzicco, R. \& Orlandi, P.} 1996 A finite-difference scheme for
  three-dimensional incompressible flow in cylindrical coordinates. {\em J.
  Comput. Phys.\/} {\bf 123}, 402--413.

\bibitem[Wendt(1933)]{wen33}
{\sc Wendt, F.} 1933 Turbulente {{Str{\"o}mungen}} zwischen zwei rotierenden
  {{Zylindern}}. {\em Ingenieurs-Archiv\/} {\bf 4}, 577--595.

\bibitem[Xia {\em et~al.\/}(2002)Xia, Lam \& Zhou]{xia02}
{\sc Xia, K.-Q., Lam, S. \& Zhou, S.~Q.} 2002 Heat-flux measurement in
  high-{{Prandtl}}-number turbulent {{{{Rayleigh-B{\'e}nard}}}} convection.
  {\em Phys. Rev. Lett.\/} {\bf 88}, 064501.

\end{thebibliography}

\end{document}